\documentclass[%
 twocolumn,
 superscriptaddress,
 amsmath,amssymb,
 prb,
]{revtex4-2}

\usepackage{graphicx}
\usepackage{dcolumn}
\usepackage{bm}
\usepackage[all]{xy}
\usepackage{color}
\usepackage{soul}

\begin{document}



\title{One dimensional nexus objects, network of Kibble-Lazarides-Shafi string walls, and their spin dynamic response in polar distorted B-phase of $^3$He}


\author{K.~Zhang}%
\affiliation{Low Temperature Laboratory, Aalto University,  P.O. Box 15100, FI-00076 Aalto, Finland}
\affiliation{University of Helsinki, Department of Mathematics and Statistics, P.O. Box 68 FI-00014, Helsinki, Finland}

\date{\today}

\begin{abstract}
The domain wall problem in the axion solution of strong CP violation has condensed-matter based analogy in nafen-distorted superfluid Helium-3. The Kibble-Lazarides-Shafi (KLS) domain wall, which appears during the temperature of early universe cooling down to QCD scale, attaches on the string defect appeared in the first time symmetry breaking phase transition. Recent experiment in rotating superfluid Helium-3 produced the network of KLS string walls in human controllable system. In this system, the half quantum vortices (HQVs) appear in the first time symmetry breaking from normal phase vacuum to polar phase, while the KLS domain walls appear and attach on the HQVs in the phase transition from polar phase to polar-distorted B-phase. Based on the method of relative homotopy group, the KLS string walls have turned out to be the descendants of HQVs of polar phase.
Here we further show the KLS string walls smoothly connect to spin solitons with length scale around $\xi_{D}$ when the spin orbital coupling is taken into account. This means HQVs are one dimensional (1D) nexuses which connect the spin solitons and the KLS domain walls. This is because the subgroup $G=\pi_{1}(S_{S}^{1},\tilde{R}_{2})$ of relative homotopy group describing the spin solitons is isomorphic to the group describing the half spin vortices -- the textures of spin degree of freedom of KLS string wall. In the nafen-distorted Helium-3 system, 1D nexus objects and the spin solitons with topological invariant $2/4$ have two different types of networks, which are named as pseudo-random lattices of inseparable and separable spin solitons. These two types of pseudo-random lattices correspond to two different representations of $G$. We discuss the condition under which pseudo-random lattices model works. The equilibrium configuration and surface densities of free energies of pseudo-random lattices are calculated by numeric minimization. Based on the equilibrium spin textures, we calculate their transverse spin dynamic response of NMR, the resulted frequency shifts and $\sqrt{\Omega}$-scaling of ratio intensity exactly coincide with the experimental measurements. We also discuss the mirror symmetry in the presence of KLS domain wall and its explicit breaking. Our discussions and considerations can be applied to the composite defects in other condensed matter and cosmological system.           
\end{abstract}

\maketitle


\section{Introductions}
The composite objects formed by topological defects with different dimensions, such as Kibble-Lazarides-Shafi (KLS) string wall \cite{Kibble1982,Kibble2000}, play significant roles in Grand Unified Theories and cosmological models. The KLS string wall  typically appears when two different symmetries with well separated energy scales are spontaneously broken \cite{Vilenkin1982,Everett1982,Zeldovich1974}. An example of $spin(10)$ gauge theory breaking to $H=\{H_{0}, K\}$ was provided in Ref.~\onlinecite{Kibble1982}, where $H_{0}=spin(6){\otimes}spin(4)$ and $K=H_{0}i\sigma_{67}$. Thus the string defects in $spin(10)$ model are described by $\pi_{1}(R)=\pi_{0}(H)=\mathbb{Z}_{2}$, where $R=spin(10)/H$ is vacuum manifold and the nontrivial element of $\pi_{1}(R)$ corresponds to $1$-loops containing charge-conjugated state, which is generated by charge-conjugation transformation $C{\in}H$. This means the string defect is Alice string, around which particle converts to its charge-conjugation \cite{Kibble1982,Schwarz1978,Kiskis1978}. In the second time symmetry breaking from $H$, the Alice strings become boundaries of domain walls because of the spontaneous breaking of charge-conjugation symmetry.

Similar two-step symmetry breaking pattern may happen in different unified gauge theories and cosmological models. Particularly it induces the domain wall problem of the axion solution of the CP violation in QCD \cite{Zeldovich1974}. In the axion solution, two phase transitions successively occur in our universe during its temperature cools down. In the first time transition, the $U(1)_{PQ}$ symmetry of Peccei-Quinn mechanism  spontaneously breaks, then the axion and string defect appear. When the cosmic temperature reaches the QCD temperature, the $U(1)_{PQ}$ symmetry is explicitly broken by QCD instanton to discrete symmetry and then the domain wall appears. As a result, the cosmic strings formed in the first time symmetry breaking attach on the the domain walls formed under QCD temperature \cite{Vilenkin1982}. This string wall system is topologically protected and then stable during the evolution of universe. The universe which has this stable structure will be very different with what we have observed. A lots of ideas have been reported to solve this problem, and the corresponding decay dynamics of the string wall system also be researched \cite{Lazarides1982,Lazarides1985,Sato2018,Chatterjee2019,Andrea2019}. 

On the other side, the similar ideas about string wall system are introduced into condensed matter system and soft matter system. These systems may provide very stable instances of string wall with human controllable methods. For example, the ferroelectric nematic liquid crystal was observed recently \cite{Chen2020}. The molecules of this liquid crystal have big enough dipole moments and show ferroelectric-like polar arrangement of polarization vectors. The formation of the string wall by two successive phase transitions during cooling down in this new system was expected \cite{Oleg2020}. In this paper, we focus on the nafen-distorted Helium-3 superfluid system \cite{Dmitriev2014}. This system generally belongs to nanoconfined superfluid Helium-3. In this kinds of system, the objects with nanometers geometric sizes are immersed into liquid Helium-3. In the low temperature at which the liquid Helium-3 is superfluid, these objects, which geometric sizes are less than the coherent length of $p$-wave triplet cooper paring, will strongly modify the microscopic scatting properties of quisipaticle and then induce new stable phases such as stripe phase \cite{Ikeda2014,Ikeda2019,Vorontsov2007,Levitin2019,Shook2019}. The nafen is one of these kinds of nanostructed material which consists of randomly-distributed-parallel $Al_{2}O_{3}$ strands with diameter $8$~nm. This geometric size is far less than the typical coherent length $\xi_{0}$ ($\sim$~$20$~nm~-~$80$~$nm$). The polar phase, which can never be stable in bulk Helium-3, was predicted be a stable vacuum state in this system \cite{Ikeda2014} and latter be experimentally identified \cite{Askhadullin2012}. Recently, The Anderson-Fomin theorem, which is the extension of Anderson theorem \cite{Anderson1959}, further explains the reason of the domination of polar phase in this uniaxial system \cite{Fomin2018,Fomin2020}.  Moreover, the observation of the $T^{3}$ dependence of gap amplitude of polar phase verified the Anderson-Fomin theorem \cite{Eltsov2019}. In multiorbital superconductor, similar extension of the Anderson theorem was also be discussed \cite{Ramires2018}.      

\begin{figure*}
\centerline{\includegraphics[width=0.73\linewidth]{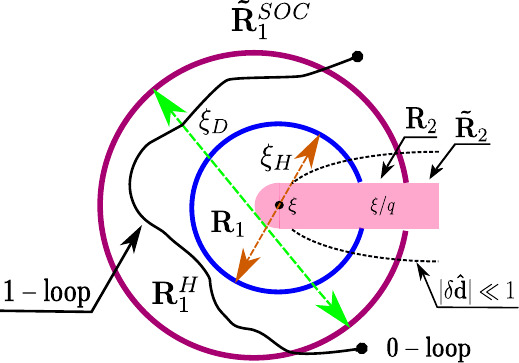}}
\caption{Illustration of vacuum manifolds with length scales $\xi_{H} < r < \xi_{D}$ and $r > \xi_{D}$ in the vicinity of transition from polar phase to PdB phase. As been discussed in Ref.~\onlinecite{volovik2020}, the vacuum manifolds of PdB in the vicinity of phase transition from polar phase to PdB phase are $R_{1}$ and $R_{2}$ in the region with $r<\xi_{H}$. The hierarchy of length scales extends in the presence of magnetic energy and SOC energy. We have known there is KLS string wall described by $\pi_{1}(R_{1},R_{2}) \cong \tilde{\mathbb{Z}}$. In larger region with length scale $\xi_{H} < r < \xi_{D}$, $R_{1}$ reduces to $R_{1}^{H} = S^{1}_{S} \times U(1)^{\Phi}$ by magnetic energy. To minimize the magnetic energy, spin vector $\hat{\mathbf{d}}$ is perpendicular to static magnetic field $\mathbf{H}^{(0)}$, while the $R_{2}$ is unchanged. When taking in account the SOC energy, $R_{1}^{H}$ further reduces to $\tilde{R}_{1}^{SOC} = R_{s}^{SOC} \times U(1)^{\Phi}$ and $R_{2}$ reduces to $\tilde{R}_{2} = \mathbb{Z}_{2}^{S-\Phi}$. As a results, there are linear topological objects described by $\pi_{1}(R_{1}^{H},\tilde{R}_{1}^{SOC})$, in which the relative 1-loop (black solid curve) is mapped to $R_{1}^{H}$ while its end points 0-loop is mapped to $\tilde{R}_{1}^{SOC}$. \label{Degenerate_Spaces}}
\end{figure*}

The observation of stable polar phase provides an ideal platform to research the Alice string i.e., half quantum vortex (HQV). At the 1970s, HQVs were predicted to appear in Helium-3 A-phase \cite{VolovikMineev1976,Cross1977}. Unfortunately, HQVs have higher energy than phase vortices in A-phase, then it actually never be observed in bulk A-phase. Nevertheless, many researches about the structures, spin dynamics and spin polarization of HQV in A-phase were reported in last few decades years because its unusual properties \cite{Salomaa1985,Hu1987,Vakaryuk2009,Volovik1999,ReadGreen2000,Ivanov2001}. Now this novel string defect can be easily observed in polar phase and polar distorted A-phase of nafen-distorted Helium-3 system \cite{Autti2016,Makinen2019}. The fundamental group $\pi_{1}(R_{P})$ of polar phase is isomorphic to $\tilde{\mathbb{Z}}=\{n/2|n\in\mathbb{Z}\}$, where $R_{P}$ is vacuum manifold of polar phase \cite{volovik2020}. The coset $\{n'/2|n'=2n+1\}$ of $\pi_{1}(R_{P})$ characterizes the topological stability of HQVs. The appearance of HQVs in polar phase during cooling down from normal phase is an instance of the formation of cosmic strings by symmetry breaking phase transition in $p$-wave superfluid system. When the temperature of polar phase superfuid reaches the transition temperature of polar distorted B-phase (PdB), the second time symmetry breaking phase transition occurs \cite{Makinen2019}. In some spatial regions of polar phase, which has HQVs generated in the first time transition, the degenerate parameter $\hat{\mathbf{d}}$ of spin degree of freedom  asymptotically trends to be constant. Thus the vacuum manifold $R_{2}$ of PdB, which appears in the second time symmetry breaking in regions with constant $\hat{\mathbf{d}}$, is smaller than the PdB vacuum manifold $R_{1}$ of the whole system. In other word, the inhomogeneous distribution of polar phase degenerate parameter reduces the original vacuum symmetry of normal phase to vacuum symmetry of polar phase in some parts of system \cite{volovik2020}. This mechanism is quite similar with the explicit breaking of the $U(1)_{PQ}$ symmetry by the appearance of QCD instanton in the cosmic domain wall problem \cite{Chatterjee2019,Andrea2019}. In the vicinity of the second time symmetry breaking, it is clear that the HQVs formed in polar phase turn to be string-wall composite topological objects described by relative homotopy group $\pi_{1}(R_{1},R_{2})$ i.e.,
\begin{equation}
\pi_{1}(R_{P}) = \pi_{1}(R_{1},R_{2}),
\end{equation}      
here the disconnected subsets of $R_{2}$ form the KLS domain wall as shown in Fig.~\ref{Degenerate_Spaces} \cite{volovik2020,nash1988}. 

Earlier the non-axialsymmetric core of quantized vortex was suggested be the string-wall system \cite{Thuneberg1986,VolovikSalomaa1985,Kondo1991,Volovik1990}. However, the wall between the separated cores is merely around few coherent lengths \cite{Silaev2015}. Similar string-wall-like double cores elliptic vortex also be proposed in spin-1 BEC \cite{takeuchi2020}. In contrast, the KLS string wall formed by the two-step phase transition in PdB phase has around $10$ to $20$ times of dipole lengths, and the length of wall can be controlled by changing the angular velocity of the system. These perfect properties allow the KLS string wall be experimentally observed in continuous wave NMR experiment \cite{Makinen2019}. The reason, which makes PdB phase has these features, is the pinning effect of HQVs by nafen strands \cite{Makinen2019,Volovik2008}. The HQVs are strongly pinned and never move once they appear, thus the KLS domain walls formed in the second time symmetry breaking do not shrink even they have tensions. Another significant consequence of this strong pinning results from the randomness of distribution of nafen strands. This randomness makes KLS string walls connect to each other randomly and form a random network of composite string-wall system. Because the geometric size of KLS string wall is around dipole length, the spin orbital coupling (SOC) energy further reduces the vacuum manifold of PdB to discrete sets. This gives rise to spin solitons, which are described by relative homotopy group \cite{mineyev1979}. Here in this paper, we show the subgroup of the relative homotopy group of spin solitons is isomorphic to the group, which characterizes the spin degree of freedom of KLS string wall. And then the spin soliton smoothly connect to KLS domain wall via the HQV i.e., HQV is one dimensional (1D) nexus \cite{volovik2020}. As a results, the network of KLS string walls is also the network of 1D nexus objects, in which randomly distributed spin solitons connect to each others by KLS domain wall. We show under the low angular velocity limit, the randomly distributed spin soliton network can be mapped to models of regular lattices consisting of spin solitons. We named these kinds of models as pseudo-random lattices. We calculate the dimensionless frequency shifts of spin dynamic response  of different pseudo-random lattices under continuous wave drive and the results exactly coincide with the experimental measurements in Ref.~\onlinecite{Makinen2019}.

This paper is organized as following sequence. In Sec. \ref{VacuaManifold} we introduce the gradient energy density and all orientation energy densities in our question. The healing length $\xi_{H}$ of magnetic energy and healing length $\xi_{D}$ of SOC energy are introduced \cite{VollhardtWolfle1990}.  Based on these well separated characteristic lengths we describe the reduced vacuum manifolds of degenerate parameters in different length scales. In Sec. \ref{SpinSolitonsAndRelativeHomotopyGroup} we utilize the exact sequences of relative homotopy group of the reduced vacuum manifolds to find out the linear topological defects. We calculate the group which describes the spin degree of freedom of KLS string wall in the region $\xi_{H} < r < \xi_{D}$ and the relative homotopy group of spin solitons when $r > \xi_{D}$. We prove the former is isomorphic to the subgroup of the latter, thus the spin soliton is smoothly connected to KLS domain wall by HQV. This means HQV is 1D nexus. Because this subgroup has two different representations, there are two classes of 1D nexus objects. One is formed by inseparable spin solitons and the other is formed by separable spin solitons. In Sec. \ref{EqulibriumTextures}, we discuss the condition under which the 1D nexus objects and spin solitons form pseudo-random lattices. The equilibrium configurations of pseudo-random lattices and the corresponding surface densities of free energy are calculated with BFGS optimization. In Sec. \ref{SpinDynamics} we calculate the spin dynamic response properties  of different types of pseudo-random lattices of spin solitons. The results are exactly coincide with the experimental observations. In Sec. \ref{DiscreteSymmetry} we discuss the mirror symmetry, which results from the reduction of vacuum manifold in the presence of KLS domain wall, and its explicit breaking. In Sec. \ref{ConclusionAndDiscussions} we summarize our main results and discuss the observing of soliton glasses in the presence of coupling between spin solitons under high angular velocity. We also discuss the possible planar spin solitons attached on string monopole networks in PdB phase.  
\section{\label{VacuaManifold}Vacuum manifolds in the presence of magnetic energy and spin-orbital coupling energy}
The PdB phase achieved by two-step continuous phase transition, which starts from uniaxial anisotropy normal phase vacuum, has two well separated length scales $\xi$ and $\xi/q$ in the vicinity of transition from polar phase to PdB phase \cite{volovik2020}. In the Ref.~\onlinecite{volovik2020}, we discussed the vacua of order parameters of superfluid in the nafen-distorted Helium-3. These vacua have dramatically different characteristic lengths determined by the energy gaps. As a result, the PdB phase in the vicinity of transition from polar phase to PdB phase has several composite topological objects within different dimensions. These novel composite objects are classified by relative homotopy groups $\pi_{n}(R_{1},R_{2})$ between vacua $R_{1}$ and $R_{2}$, where $R_{1}$ and $R_{2}$ are vacuum manifolds of PdB phase achieved from normal phase vacuum and polar phase vacuum respectively. The stable objects of polar phase are stabilized again in PdB phase by forming composite objects described by relative homotopy groups $\pi_{n}(R_{1},R_{2})$.  

More length scales appear additionally if we take into account more orientation energies. In nafen-distorted Helium-3 system, these length scales are magnetic length $\xi_{H}$ and dipole length $\xi_{D}$ \cite{Dmitriev2014,VollhardtWolfle1990}. These two length scales characterize the spatial ranges in which the gradient energy are larger than orientations energies. When the length scale of spatial variations is larger than these characteristic lengths, the vacua of order parameters are reduced to minimize the orientation energies. We discussed the consequence of this kinds of reduction by magnetic energy and magnetic length $\xi_{H}$ i.e., the vortex skyrmions in Ref.~\onlinecite{volovik2020}. We will see there are more interesting results when dipole length $\xi_{D}$ is introduced in addition to $\xi_{H}$ in rest parts of this paper. $\xi_{H}$ is determined by gradient energy density 
\begin{align}
f_{\rm grad} = & \frac{1}{2} K_1 \partial_{i} A_{\alpha j} \partial_{i} A^{*}_{\alpha j} +\frac{1}{2} K_2 \partial_{j} A_{\alpha i} \partial_{i} A^{*}_{\alpha j} \notag \\ & + \frac{1}{2} K_3 \partial_{i} A_{\alpha i} \partial_{j} A^{*}_{\alpha j}
\label{GradientEnergy}
\end{align}    
where 
\begin{equation}
A_{{\alpha}i} \equiv A_{{\alpha}i}^{PdB} = e^{i\Phi} [\Delta_{P}\hat{d}_{\alpha} \hat{z}_{i} + \Delta_{\bot1}\hat{e}^{1}_{\alpha} \hat{x}_{i} + \Delta_{\bot2}\hat{e}^{2}_{\alpha} \hat{y}_{i}]
\label{PdBOrderParameter}
\end{equation}
is the order parameter of PdB phase. $\hat{\mathbf{d}} \equiv \hat{d}_{\alpha}$ and $\hat{\mathbf{e}}^{1(2)} \equiv \hat{e}_{\alpha}^{1(2)}$ are the spin degenerate parameters and they form the triad in spin space. $\Phi$ and $\hat{x}_{i}\equiv\hat{x}$,$\hat{y}_{i}\equiv\hat{y}$,$\hat{z}_{i}\equiv\hat{z}$ are phase and orbital degenerate parameters respectively. Here $|\Delta_{\bot1}|=|\Delta_{\bot2}|=|q|\Delta_{P}$ with $|q| \leq 1$, and $K_{1}=K_{2}=K_{3}$ \cite{VollhardtWolfle1990}. The magnetic energy density is
\begin{equation}
f_{\rm H} = -\frac{1}{2} \chi_{\alpha \beta} H_{\alpha} H_{\beta}=\frac{1}{2}\gamma^{2} S_{a}S_{b}(\chi^{-1})_{ab}- \gamma H_{a}S_{a},
\label{MagneticEnergy}
\end{equation}
here the $\chi_{\alpha \beta}$ is uniaxial tensor of magnetic susceptibility of PdB phase, $H_{\alpha}$ are magnetic field strengths with $\alpha=1,2,3$, $S_{a}$ are spin densities with $a=1,2,3$ and $\gamma$ is gyromagnetic ratio \cite{VollhardtWolfle1990}. With the help of Eq.~(\ref{GradientEnergy}) and Eq.~(\ref{MagneticEnergy}), the magnetic length is given as 
\begin{equation}
\xi_{H}=[\frac{K_{1}\Delta_{P}^{2}}{(\chi_{\bot}-\chi_{\|})H^{2}}]^{\frac{1}{2}},
\label{MagneticLength}
\end{equation}
where $\chi_{\bot}$ and $\chi_{\|}$ are transverse and longitude spin magnetic susceptibilities of PdB phase. In the experiment for PdB phase, a static magnetic field $\mathbf{H}^{(0)}$ with fixed direction is turned on \cite{Makinen2019}. Then the degenerate space of PdB order parameter reduces to 
\begin{equation}
R^{H}_{1} = S^{1}_{S} \times U(1)^{\Phi}
\end{equation}
from $R_{1}$ in the region in which length scale of spatial variation is larger than $\xi_{H}$ \cite{volovik2020}. Because the magnetic energy locks the $\hat{\mathbf{d}}$ vector into the plane perpendicular to $\mathbf{H}^{(0)}$, $R_{2}$ keeps the same form as it is inside the region with length scale $\xi_{H}$. Then we still have $R_{2} = SO(2)_{S-L} \times \mathbb{Z}^{S-\Phi}_{2}$ in the region where condition $|\delta \hat{\mathbf{d}}| \ll 1$ is satisfied. In Fig.~\ref{Degenerate_Spaces}, we illustrate the $R_{1}^{H}$ and $\xi_{D}$ in the presence of KLS string wall. 

Following the same idea, the dipole length $\xi_{D}$ is determined by gradient energy density $f_{grad}$ and SOC energy density
\begin{equation}
f_{\rm soc} = \frac{3}{5}g_D(A^*_{ii}A_{jj}+A^*_{ij}A_{ji} - \frac{2}{3}A^*_{ij}A_{ij}), 
\label{EnergyDensityOfSOC}
\end{equation}
where $g_{D}$ is strength of spin orbital coupling. Then we have
\begin{equation}
\xi_{D}=(\frac{5K_{1}}{6g_{D}})^{\frac{1}{2}}.
\label{DipoleLength}
\end{equation}
When the Spin-Orbit coupling (SOC) is taken into account, degenerate vacuum manifolds of order parameters are  further reduced from $R^{H}_{1}$ and $R_{2}$. In general consideration, the requirement of minimizing SOC energy in region with length scale larger than $\xi_{D}$ fixes the relative directions between spin vectors and orbital vectors. The resulted vacuum manifold always could be represented by spin degree of freedom because the broken symmetry is relative symmetry \cite{VollhardtWolfle1990}. Thus $R_{1}^{H}$ reduces to
\begin{equation}
\tilde{R}_{1}^{SOC} = R_{S}^{SOC}\times U(1)^{\Phi}
\end{equation}
in the region with length scale larger than $\xi_{D}$, where $R_{S}^{SOC}$ is the reduced vacuum manifold of spin degree of freedom. In general case, $R_{S}^{SOC}$ is a complicated space. However $R_{S}^{SOC}$ may be simplified by using parametrization of $\hat{\mathbf{d}}$ and $\hat{\mathbf{e}}^{1(2)}$ vectors of $A_{{\alpha}i}^{PdB}$. To facilitate comparison between experimental observations and our theoretical analysis, the paramentrizations 
\begin{align}
\mathbf{\hat{d}} & =\hat{x}cos\theta-\hat{z}sin\theta, \notag \\ 
\mathbf{\hat{e}}^{1} & =-\hat{x}sin\theta-\hat{z}cos\theta, \label{PARA1} \\ 
\mathbf{\hat{e}}^{2} & =\hat{y}, \,\, \mathbf{H}^{(0)} =H\hat{y} \notag
\end{align}
would be used in this work, where $\theta$ is the angle between $\hat{\mathbf{d}}$ and local orbital-coordinate frame \cite{Makinen2019}. In this case, we find $ R_{S}^{SOC} = \{\theta_{0}, \pi-\theta_{0}, -\theta_{0}, \pi+\theta_{0} \}$, where $\theta_{0}=arcsin[q/(1-|q|)]$. There is a discrete symmetry for free energy of system and this discrete symmetry turns out to be the symmetry between parametrization in Eq.~(\ref{PARA1}) and the alternative in the presence of KLS domain wall. We will discuss the details of this discrete symmetry and its violation in Sec. \ref{DiscreteSymmetry}. Before Sec. \ref{DiscreteSymmetry}, we mainly use the parametrization in Eq.~(\ref{PARA1}). In the region where condition $|\delta \hat{\mathbf{d}}| \ll 1$ is satisfied, SOC energy fixes the relative rotation of $SO(2)_{S-L}$, thus $R_{2}$ reduces to $\tilde{R}_{2} = \mathbb{Z}_{2}^{S-\Phi}$ in the region with length scale larger than $\xi_{D}$. 

From illustrtion of $R_{1}^{H}$, $\tilde{R}_{1}^{SOC}$ and $\tilde{R}_{2}$ in Fig.~\ref{Degenerate_Spaces}, we find again the possibility of utilizing the relative homotopy group to investigate the novel topological objects because of the presence of multiple characteristic length scales \cite{nash1988}. This multilength-scales system belongs to type (i) of the classifications in Ref.~\onlinecite{volovik2020}. Other example of this class is solitons terminated by HQVs observed in spinor Bose condensate with quadratic Zeeman energy \cite{Seji2019,Liu2020}. Both of these systems can be described by the first relative homopoty group. In next section, we discuss this topic.  

\section{\label{SpinSolitonsAndRelativeHomotopyGroup}1D nexus objects and spin solitons classified by relative homotopy groups}

\subsection{\label{RelativeHomotopy}Relative homotopy groups of spin solitons and 1D nexus objects}

\subsubsection{Spin configuration of KLS string wall -- half spin vortices}
In the region with length scale $\xi_{H} \leq r \leq \xi_{D}$, we have the long exact sequence (LES) of homomorphism of $\pi_{1}(R_{1}^{H},R_{2})$
\begin{equation}
\xymatrix@1@R=10pt@C=13pt{
\pi_{1}(R_{2}) \ar@{-}[d] \ar[r]^{i^{*}} & \pi_{1}(R_{1}^{H}) \ar@{-}[d] \ar[r]^{j^{*}} &
\pi_{1}( R_{1}^{H}, R_{2}) \ar@{-}[d] \ar[r]^-{\partial^{*}} & \pi_{0}(R_{2}) \ar@{-}[d] \ar[r]^{k^{*}} & \pi_{0}(R_{1}^{H}) \ar@{-}[d]\\
\mathbb{Z}^{S} \ar[r]^{i^{*}} & \mathbb{Z}^{S} \times \mathbb{Z}^{\Phi} \ar[r]^{j^{*}} & \pi_{1}( R_{1}^{H}, R_{2}) \ar[r]^-{\partial^{*}} & \mathbb{Z}_{2} \ar[r]^{k^{*}} & 0
}\,,
\label{LES1a}
\end{equation}
where $i^{*}$ projects spin vortices of $\pi_{1}(R_{2})$ to the spin vortices of $\pi_{1}(R_{1}^{H})$ \cite{nash1988,suzuki1982}. And boundary homomorphism $\partial^{*}$ maps all relative $1$-loops of $\pi_{1}(R_{1}^{H},R_{2})$ to their $0$-loops of $\pi_{0}(R_{2})$. Because $\pi_{0}(R_{2}) = \mathbb{Z}_{2}$, the end-points of relative $1$-loop may take values from connected or disconnected subsets of $R_{2}$. This LES can be split to the short exact sequence (SES) 
\begin{equation}
\xymatrix@1@R=10pt@C=13pt{
0 \ar[r] & \mathbb{Z}^{\Phi} \ar[r]^-{\iota} &
\pi_{1}( R_{1}^{H},R_{2}) \ar[r]^-{\pi} & \mathbb{Z}_{2} \ar[r] & 0\\
}\,,
\label{SES1}
\end{equation}
where $\iota$ and $\pi$ are inclusion and surjection respectively. Eq.~(\ref{SES1}) suggests $\pi_{1}( R_{1}^{H},R_{2}) \cong \tilde{\mathbb{Z}}$, which is isomorphic to $\pi_{1}(R_{1},R_{2})$ in the region smaller than $\xi_{H}$ \cite{volovik2020}. This means KLS string wall, which determined by two length scales $\xi$ and $\xi/q$ in two-step phase transition, extends into the region with length scale $\xi_{H} \leq r \leq \xi_{D}$.  However Eq.~(\ref{SES1}) only contains degree of freedom (DOF) of phase factor $\Phi$, all information about spin degree of freedom lose because they are trivial elements of $\pi_{1}(R_{1}^{H}, R_{2})$. To understand the spin part of KLS string wall, we should take in to account the continuity of order parameter. The continuity of order parameter $A_{\alpha i}^{PdB}$ requires spin vectors simultaneously change by $(2n+1)\pi$ in the present of KLS string wall \cite{Volovik1990}. This consideration suggests that the spin textures of KLS string wall in the spatial region with length scale $\xi_{H} \leq r \leq \xi_{D}$ are classified by group
\begin{equation}
M \equiv \{ n^{s}/2 | n^{s} \in \mathbb{Z} \},
\end{equation}
such that $M/\pi_{1}(S^{1}_{S}) \cong \mathbb{Z}_{2} = \{ [0], [1/2]\}$. The cosets $[1/2]$ and $[0]$ correspond to the presence or absence of the KLS tring wall in the region $\xi_{H} < r \leq \xi_{D}$ respectively. Coset $[0] \cong 2\mathbb{Z}$ contains all free spin vortices. While  Coset $[1/2] \cong \{n + 1/2 | n \in \mathbb{Z}\}$ contains all spin vortices with half-odd winding number i.e., it is set of half spin vortices. 

\begin{figure*}
\centerline{\includegraphics[width=0.7\linewidth]{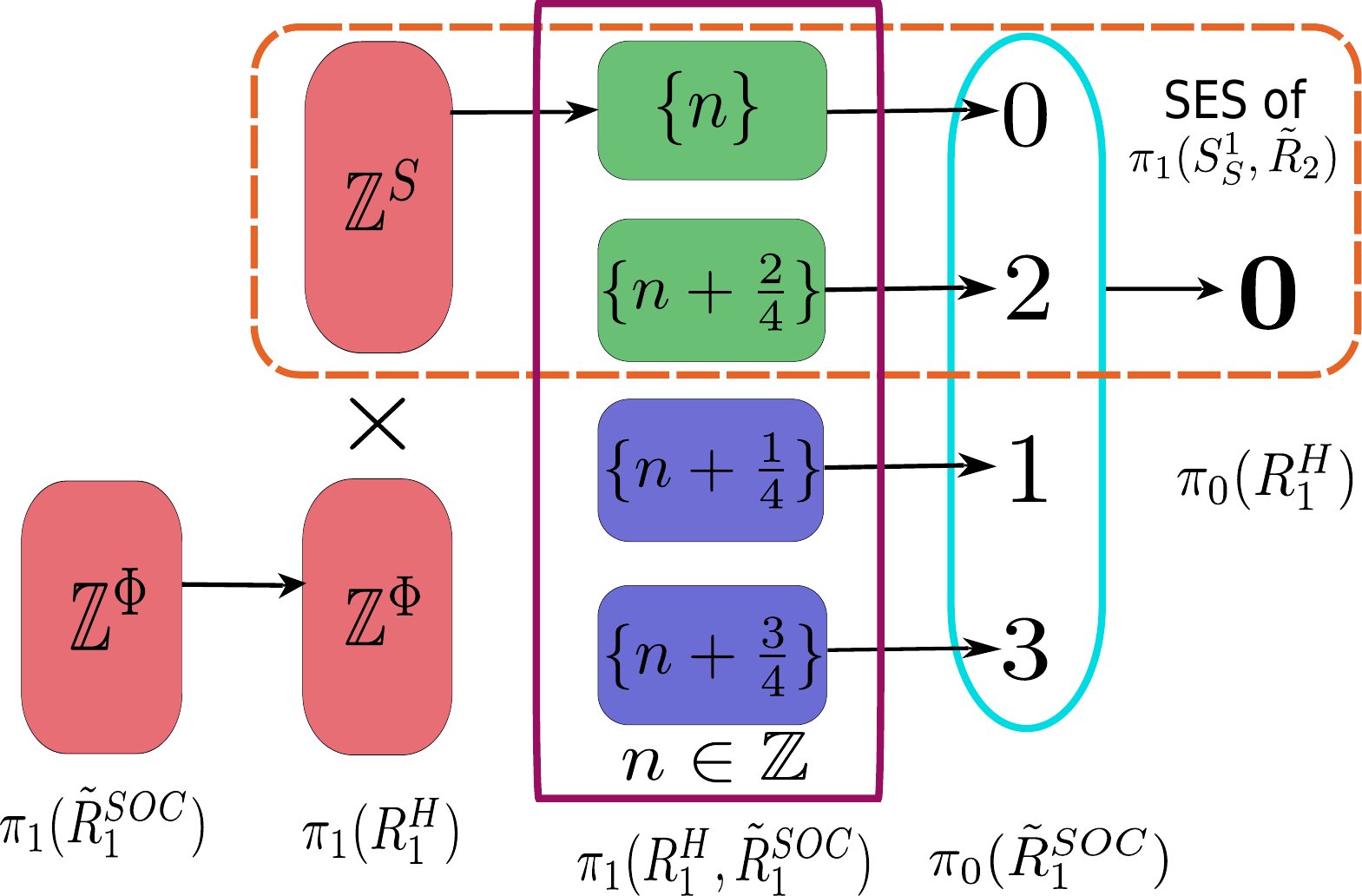}}
\caption{Illustrations of long exact sequence of homomorphism for $\pi_{1}(R_{1}^{H},\tilde{R}_{1}^{SOC})$ and short exact sequence of homomorphism for $\pi_{1}(S^{1}_{S},\tilde{R}_{2})$. The black arrows represent the image of homomorphisms between homotopy groups. This mapping diagram demonstrates the linear objects of $\pi_{1}(R_{1}^{H},\tilde{R}_{1}^{SOC})$ are spin solitons. This is because the mapping between $\pi_{1}(\tilde{R}_{1}^{SOC})$ and $\pi_{1}(R_{1}^{H})$ is projection, the image of homomorphism $i^{*} : \pi_{1}(\tilde{R}_{1}^{SOC}) \rightarrow \pi_{1}(R_{1}^{H}) = \mathbb{Z}^{\Phi}$ i.e., topological invariant of all phase vortices.  As a result, the trivial linear objects of $\pi_{1}(R_{1}^{H},\tilde{R}_{1}^{SOC})$ are all phase vortices because of $\operatorname{im}i^{*} \cong \ker j^{*}$. We found there are one kind of spin vortices and three kinds of spin solitons because $\ker k^{*} \cong \operatorname{im} \partial^{*} = \mathbb{Z}_{4}$ and $j^{*}$ is projection. Moreover, we found form this illustration that the subgroup $G=\{\{n\},\{n+2/4\}\}$ of $\pi_{1}(R_{1}^{H},\tilde{R}_{1}^{SOC})$ is extension of $\mathbb{Z}^{S}$ by $\pi_{0}(\tilde{R}_{2}^{SOC}) = \mathbb{Z}_{2}$ and then isomorphic to $M$. In the orange dash line panel, we shows the corresponding short exact sequence of $G$. As a result, HQV is 1D nexus between spin soliton of coset $[2/4]$ and KLS domain wall in PdB phase. \label{MappingDiagramAndSES}} 
\end{figure*}
\subsubsection{\label{FourSpinSolitons}Spin solition described by $\pi_{1}(R^{H}_{1},\tilde{R}^{SOC}_{1})$}
When taking into account SOC, $R_{1}^{H}$ reduces to $\tilde{R}_{1}^{SOC} = R_{S}^{SOC} \times U(1)$ as mentioned in Sec. \ref{VacuaManifold}. As a result, there are linear objects which classified by $\pi_{1}(R_{1}^{H}, \tilde{R}_{1}^{SOC})$. $\pi_{1}(R_{1}^{H}, \tilde{R}_{1}^{SOC})$ has LES 
\begin{widetext}
\begin{equation}
\xymatrix@1@R=10pt@C=13pt{
\pi_{1}(\tilde{R}_{1}^{SOC}) \ar@{-}[d] \ar[r]^-{i^{*}} & \pi_{1}(R_{1}^{H}) \ar@{-}[d] \ar[r]^-{j^{*}} &
\pi_{1}( R_{1}^{H}, \tilde{R}_{1}^{SOC}) \ar@{-}[d] \ar[r]^-{\partial^{*}} & \pi_{0}(\tilde{R}_{1}^{SOC}) \ar@{-}[d] \ar[r]^-{k^{*}} & \pi_{0}(R_{1}^{H}) \ar@{-}[d]\\
\mathbb{Z}^{\Phi} \ar[r]^-{i^{*}} & \mathbb{Z}^{S} \times \mathbb{Z}^{\Phi} \ar[r]^-{j^{*}} & \pi_{1}( R_{1}^{H}, \tilde{R}_{1}^{SOC}) \ar[r]^-{\partial^{*}} & \mathbb{Z}_{4} \ar[r]^-{k^{*}} & 0
}\,,
\label{LES1}
\end{equation}
\end{widetext}
where $i^{*}$ is projection and $\partial^{*}$ is boundary homomorphism \cite{volovik2020,nash1988,suzuki1982}. Figure~\ref{MappingDiagramAndSES} depicts the mapping relation of Eq.~(\ref{LES1}). The relative 1-loop of $\pi_{1}(R_{1}^{H}, \tilde{R}_{1}^{SOC})$ and the boundary 0-loop are shown in Fig.~\ref{Degenerate_Spaces}. Because $\operatorname{im} {\partial}^{*} \cong {\ker} k^{*} = \mathbb{Z}_{4}$, the boundary 0-loop (two end points) of 1-loop takes values from four disconnected subsets of $\tilde{R}_{1}^{SOC}$. For every element of $\tilde{R}_{1}^{SOC}$, there are four possible combinations of elements of $\tilde{R}_{1}^{SOC}$ for 0-loop because of $\pi_{0}(\tilde{R}_{1}^{SOC}) = \mathbb{Z}_{4}$. As a result, we found there are four kinds of linear objects in general, which might be distinguished by four boundary homotopy classes of $\pi_{0}(\tilde{R}_{1}^{SOC})$.  Moreover Eq.~(\ref{LES1}) can be split into SES 
\begin{equation}
\xymatrix@1@R=10pt@C=13pt{
0 \ar[r] & \mathbb{Z}^{S} \ar[r]^-{\iota} &
\pi_{1}( R_{1}^{H},\tilde{R}_{1}^{SOC}) \ar[r]^-{{\partial}^{*}} & \mathbb{Z}_{4} \ar[r] & 0\\
}\,.
\label{SES2}
\end{equation}
Then we find $\pi_{1}( R_{1}^{H},\tilde{R}_{1}^{SOC}) = \{n^{S}/4 | n^{S} \in \mathbb{Z}\} \cong \mathbb{Z}$, such that $\pi_{1}( R_{1}^{H},\tilde{R}_{1}^{SOC})/\mathbb{Z}^{S} \cong \mathbb{Z}_{4}$. Because Eq.~(\ref{SES2}) is merely determined by $ \mathbb{Z}^{S} = \pi_{1}(S^{1}_{S})$ and $\mathbb{Z}_{4} = \pi_{0}(R_{S}^{SOC})$, $\pi_{1}(R_{1}^{H}, \tilde{R}_{1}^{SOC})$  actually is isomorphic to $\pi_{1}(S^{1}_{S}, R_{S}^{SOC})$ i.e.,
\begin{equation}
\pi_{1}(R_{1}^{H}, \tilde{R}_{1}^{SOC}) \cong \pi_{1}(S^{1}_{S}, R_{S}^{SOC}).
\label{isomorphism1}
\end{equation}
\begin{figure*}
\centerline{\includegraphics[width=1.0\linewidth]{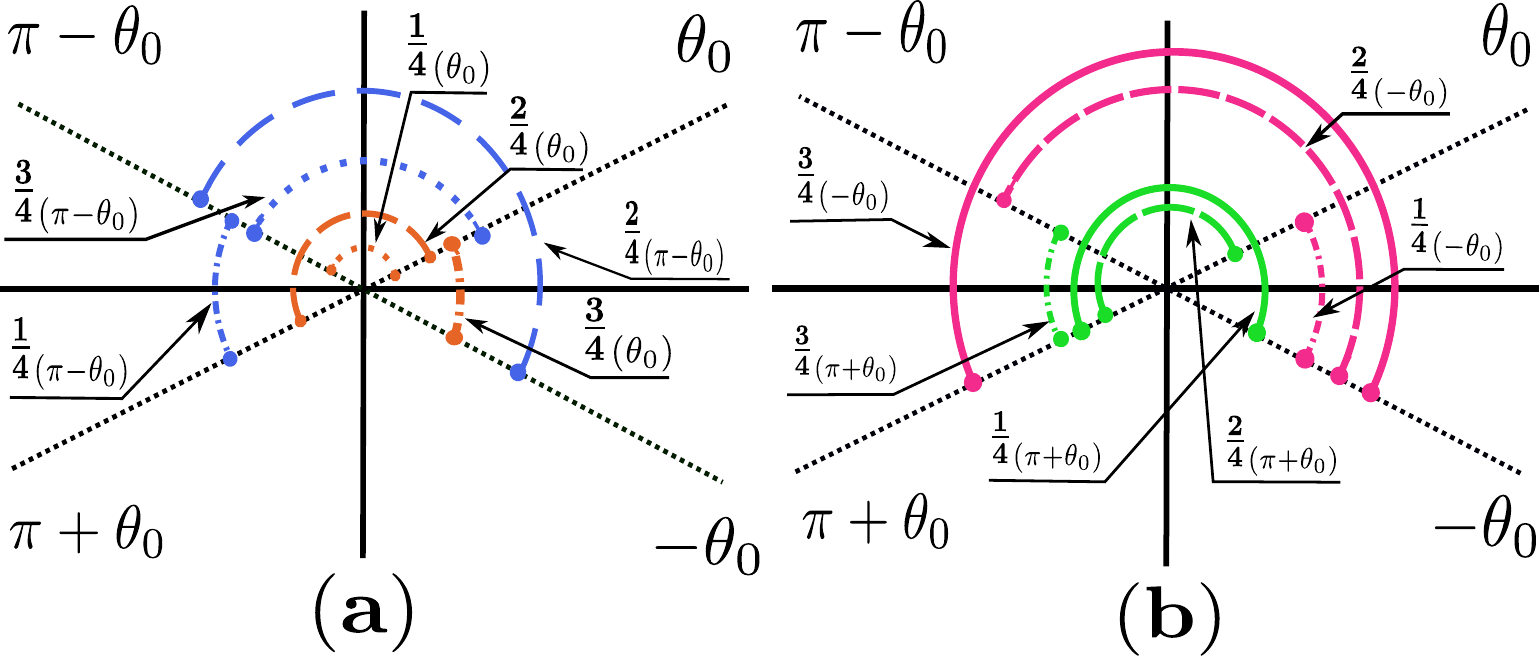}}
\caption{Illustrations of three kinds of spin solitons described by $\pi_{1}(S^{1}_{S},R_{S}^{SOC})$. The black dot lines represent the four elements of $R_{S}^{SOC}$ i.e., $\pm \theta_{0}$ and $\pi \pm \theta_{0}$. The dash line, dot line, dash-dot line and solid line correspond to $\pi$-soliton, solition, KLS-Solition and big-solition respectively. (a) Spin solitons with topological invariants $1/4$, $2/4$ and $3/4$ for $\theta_{0}$ (orange) and $\pi-\theta_{0}$ (blue) respective. (b) Spin solitons with topological invariants $1/4$, $2/4$ and $3/4$ for $-\theta_{0}$ (pink) and $\pi+\theta_{0}$ (green) respective. \label{ClassesOfSolitons}}
\end{figure*}
This means the linear objects classified by $\pi_{1}(R_{1}^{H}, \tilde{R}_{1}^{SOC})$ are spin solitons \cite{mineyev1979}. The four cosets of $\pi_{1}(S^{1}_{S}, R_{S}^{SOC})$ by $\mathbb{Z}^{S}$ are 
\begin{align}
[0] = & \{n^{S} \} \,\,, [\frac{1}{4}] = \{n^{S}+\frac{1}{4} \}, \notag \\ [\frac{2}{4}]  = & \{n^{S}+\frac{2}{4} \} \,\,, [\frac{3}{4}] = \{n^{S}+\frac{3}{4} \}.
\label{cosets}
\end{align} 
The cosets in Eq.~(\ref{cosets}) give out the topological invariants of the four different kinds of linear  objects distinguished by homotopy classes of boundary 0-loop of $\pi_{0}(R_{S}^{SOC})$. They correspond to free spin vortices and three kinds of spin solitons respectively. Figure~\ref{ClassesOfSolitons} shows the representatives of these three classes of spin solitons for every element of $R_{S}^{SOC}$. We omit the spin vortices of $[0]$ from now on because it is not energy-favored stable spin textures. From  Fig.~\ref{ClassesOfSolitons}, we found there are four types of spin solitons distinguished by $|\Delta \theta|$. Following the terminologies in Ref.~\onlinecite{Makinen2019}, they are big-solition ($|\Delta \theta| = \pi +2\theta_{0}$), solition ($|\Delta \theta_{0}|=\pi-2\theta_{0}$), KLS-soliton ($|\Delta \theta_{0}|=2\theta_{0}$) and $\pi$-soliton ($|\Delta \theta|=\pi$). To avoid terminological confusion, we claim here that we use phrase "spin soliton" to denote spin textures of $\pi_{1}(S^{1}_{S},R_{S}^{SOC})$ in rest of this paper, while use phrases "solitons", "big-solitons", "KLS-solitons" and "$\pi$-solitons" to denote particular spin textures with different $|\Delta \theta|$.  

\subsubsection{\label{RelativeHomotopyGroupof1DNexus}Short exact sequence of $\pi_{1}(S^{1}_{S},\tilde{R}_{2})$ and 1D nexus}
A significant property of $\pi_{1}(S^{1}_{S},R_{S}^{SOC})$ is that it has a subgroup $G \equiv \{[0],[2/4]\}$ such that $G/\mathbb{Z}^{S} \cong \mathbb{Z}_{2}$. The SES of $G$ is given as  
\begin{equation}
\xymatrix@1@R=10pt@C=13pt{
0 \ar[r] & \mathbb{Z}^{S} \ar[r] &
G \ar[r]^{\partial^{*}} & \mathbb{Z}_{2} \ar[r] & 0\\
}\,
\label{SES3}
\end{equation}
by Eq.~(\ref{SES2}). The mapping diagram of Eq.~(\ref{SES3}) is shown in the dash panel of Fig.~\ref{MappingDiagramAndSES}. Because $\pi_{0}(\tilde{R}_{2}) \cong \mathbb{Z}_{2}$, Eq.~(\ref{SES3}) can be written as
\begin{equation}
\xymatrix@1@R=10pt@C=13pt{
\pi_{1}(\tilde{R}_{2}) \ar[r] & \pi_{1}(S_{S}^{1}) \ar[r] &
G  \ar[r]^{\partial^{*}} & \pi_{0}(\tilde{R}_{2})  \ar[r] & 0\\
}\,.
\label{SES4}
\end{equation}
This LES suggests 
\begin{equation}
G = \pi_{1}(S^{1}_{S}, \tilde{R}_{2}) \cong \hat{\mathbb{Z}} = M,
\label{isomorphism2}
\end{equation}
here $\hat{\mathbb{Z}} \equiv \{n^{S}/2 | n^{S} \in \mathbb{Z}\}$. Eq.~(\ref{isomorphism2}) is one of main results of this paper. This relation means spin solitons, which are classified by coset $[2/4]$ of $\pi_{1}(S^{1}_{S}, \tilde{R}_{2})$ can continuously transform to half spin vortices of $M$. In  other word, KLS domain wall smoothly connects to $[2/4]$ spin soliton via HQV. Similar with 2D nexus which connects string monopole and vortex skyrmion, the HQV is 1D nexus which connects KLS domain wall and $[2/4]$ spin soliton \cite{volovik2020}. The composite object formed by $[2/4]$ spin soliton and KLS domain wall is then named as 1D nexus object. In Sec. \ref{TwoDifferentSolitonConfiguretions}, we will see there are two possible configurations for $[2/4]$ spin solitons i.e., one $\pi$-solitons or a combination between KLS-soliton and soliton. As a result, there are two different types of 1D nexus objects. 

\subsection{Two different configurations of $[2/4]$ spin soliton of 1D nexus object  -- separable and inseparable }
\label{TwoDifferentSolitonConfiguretions}
Because $\pi_{1}(S^{1}_{S},R^{SOC}_{S})/\mathbb{Z}^{S} \cong \mathbb{Z}_{4}$, we have $[2/4] = [1/4] + [1/4]$. Thus $\pi_{1}(S^{1}_{S},\tilde{R}_{2})$ could also be represented as $\{[0],[1/4]+[1/4]\}$ besides $\pi_{1}(S^{1}_{S},\tilde{R}_{2}) \cong \{[0],[2/4]\}$. This means there are two kinds of spin soliton configurations connecting with KLS domain wall via HQV for a given element of $\pi_{1}(S^{1}_{S},\tilde{R}_{2})$. When the topological invariant is literally $2/4$, the spin soliton is spatially inseparable $\pi$-soliton as shown in Fig.~\ref{ClassesOfSolitons}. When the topological invariant is $1/4+1/4$, the spin soliton is combination of two spatially separable spin solitons with topological invariant $1/4$. To identify these two spatially separable spin solitons, we take in account the requirement of continuity of the order parameters. This requirement is equivalent to the requirement of single-value and continuity of $\theta$. Then the accumulation of $|\Delta\theta|$ of those two spin solitons must equal to $\pi$. Based on the discussions of Sec. \ref{FourSpinSolitons} and Fig.~\ref{ClassesOfSolitons}, These two spin solitons are KLS-soliton and soliton.  

We will see these two dramatically different spin textures of 1D nexus objects have different equilibrium free energies, different spin dynamic response properties and different NMR frequency shifts in Sec. \ref{EqulibriumTextures} and Sec. \ref{SpinDynamics}. These properties help us to identify the objects which be observed in experiment. 

\section{\label{EqulibriumTextures}Equilibrium textures of pseudo-random lattices consisting of spin solitons}
For the PdB phase results from symmetry breaking of nonuniform polar phase, we can use the Ginzburg-Landau model to describe the system when $|q|$ is small enough. The Ginzburg-Landau free energy consists of gradient energy and orientation energies \cite{VollhardtWolfle1990}. In order to quantitatively analyze the equilibrium configurations of 1D nexus objects containing spin solitons with length scale around $\xi_{D}$, we must find out the extreme point of Ginzburg-Landau free energy under given external parameters. Because $\xi_{D} \gg \xi_{0}$ and the strongly uniaxial anisotropy in the presence of nafen strands, we actually did this procedure under London limit \cite{Ikeda2019,VollhardtWolfle1990,volovik1992}. In London limit, all gap parameters attain equilibrium structures and then their magnitudes are constants over whole calculations. When the static magnetic field $\mathbf{H}^{(0)}$ is big enough, the magnetic length $\xi_{H}$ is far smaller than the dipole length $\xi_{D}$, then the magnetic energy has achieved equilibrium over the PdB superfluid. In this situation the Ginzburg-Landau free energy in London limit is 
\begin{equation}
F_{London} =\int\nolimits_{\Sigma} (f_{\rm soc} + f_{\rm grad} ) d\Sigma,
\label{GinzburgLandauFreeEnergyTheta}
\end{equation} 
where $\Sigma$ is the volume of the PdB phase sample. 

Plunging $A_{\alpha i}^{PdB}$ into Eq.~(\ref{GinzburgLandauFreeEnergyTheta}) and substituting $\hat{\mathbf{d}}$, $\hat{\mathbf{e}}^{1}$ and $\hat{\mathbf{e}}^{2}$ with their parametrizations in Eq.~(\ref{PARA1}), we get the gradient energy density and SOC energy density in term of $\theta$ and $\Phi$
\begin{widetext}
\begin{align}
f_{\rm grad}(\Phi,\theta) = 
& \frac{K_{1}}{2}(\Delta_{P}^{2}+\Delta_{\bot1}^{2}+\Delta_{\bot2}^{2})\partial_{i}\Phi \partial_{i}\Phi + \frac{K_{1}}{2}(\Delta_{P}^{2}+\Delta_{\bot1}^{2})\partial_{i} \theta \partial_{i} \theta + \frac{1}{2} (K_{2}+ K_{3}) (\Delta_{P}^{2} \partial_{z} \Phi \partial_{z} \Phi \notag \\ 
& + \Delta_{\bot1}^{2} \partial_{x} \Phi \partial_{x} \Phi + \Delta_{\bot2}^{2} \partial_{y} \Phi \partial_{y} \Phi + \Delta_{P}^{2}\partial_{z}\theta \partial_{z}\theta + \Delta_{\bot1}^{2} \partial_{x}\theta \partial_{x}\theta),\\
f_{\rm soc}(\theta) = & \frac{g_{D}}{5}(\Delta_{P}^{2} + \Delta_{\bot1}^{2} + \Delta_{\bot2}^{2}) -\frac{3g_{D}}{5}(\Delta_{P}+\Delta_{\bot1})^{2}cos2\theta-\frac{6g_{D}}{5}(\Delta_{P}+\Delta_{\bot1})\Delta_{\bot2}sin\theta, \notag
\end{align}
\end{widetext}
where $i=1,2,3$ are the summation indexes of spatial coordinates. In London limit, the term ($g_{D}/5)(\Delta_{P}^{2} + \Delta_{\bot1}^{2} + \Delta_{\bot2}^{2})$ is constant over the sample, thus we omit it in the rest of this paper. Because spin degree of freedom does not couple with phase degree of freedom, $f_{grad}(\Phi,\theta)$ is simply the summation of $f_{grad}(\Phi)$ and $f_{grad}(\theta)$, where $f_{grad}(\Phi)$ and $f_{grad}(\theta)$ are the gradient energy densities of phase and spin vectors respectively. Then we assume $f_{grad}(\Phi)$ has achieved equilibrium and drop it in the rest part of this work. Moreover, because the HQVs are pinned by nafen strands, the system is translation invariant along the direction of nafen strands, thus all $\partial_{z}\theta$ terms vanish. Finally the free energy, which determines the equilibrium textures in London limit is
\begin{equation}
F(\theta)_{London} =\int\nolimits_{\Sigma} [f_{soc}(\theta) + f_{grad}(\theta)] d\Sigma ,
\label{FreeEnergyTheta}
\end{equation}
where $f_{\rm grad}(\theta)$ and $f_{\rm soc}(\theta)$ are given as
\begin{align}
f_{\rm grad}(\theta) = & \frac{K_{1}}{2}(\Delta_{P}^{2}+\Delta_{\bot1}^{2})(\partial_{x} \theta \partial_{x}\theta + \partial_{y} \theta \partial_{y} \theta) \notag \\ & + \frac{1}{2} (K_{2}+K_{3}) \Delta_{\bot1}^{2} \partial_{x}\theta \partial_{x}\theta, \notag \\
f_{\rm soc}(\theta)= & -\frac{3g_{D}}{5}(\Delta_{P}+\Delta_{\bot1})^{2}cos2\theta \\ & -\frac{6g_{D}}{5}(\Delta_{P}+\Delta_{\bot1})\Delta_{\bot2}sin\theta.\notag 
\end{align} 
In this section, we utilize the nonlinear optimization BFGS algorithm to minimize the free energy functional Eq.~(\ref{FreeEnergyTheta}) \cite{jorge2006}.  The saddle points $\theta$ of free energy under different parameters are the equilibrium textures of spin solitons of 1D nexus objects. To facilitate minimization of free energy with nonlinear optimization algorithm, we reduce Eq.~(\ref{FreeEnergyTheta}) to 
\begin{align}
\tilde{F}(\theta)_{London} 
 = & {\frac{1}{\xi_{D}}}\int\nolimits_{\Sigma} [\frac{1}{2} (\gamma_{1}+2\gamma_{2}) \partial_{x}\theta \partial_{x}\theta \notag  + \frac{1}{2}\gamma_{1}\partial_{y}\theta \partial_{y}\theta \\ & + \frac{1}{\xi_{D}^{2}} 
(-\frac{1}{2}{\gamma_{4}}cos2\theta-\gamma_{3}sin\theta)] d\Sigma \notag \\
 = & \frac{1}{\xi_{D}} \int_{\Sigma} (\tilde{f}_{grad} + \tilde{f}_{soc}) d\Sigma
\label{FreeEnergyThetaDimensionless}
\end{align}
by multiplying $(\xi_{D}K_{1}\Delta_{P}^{2})^{-1}$, where 
\begin{align}
q & = \frac{\Delta_{\bot2}}{\Delta_{P}},\,\,
 \gamma_{1}=1+|q|^{2},\,\, 
\gamma_{2}=|q|^{2},\,\, \notag \\
& \gamma_{3}=q(1+|q|),\,\,
\gamma_{4}=(1+|q|)^{2},
\end{align}
and 
\begin{align}
\tilde{f}_{grad} & = \frac{1}{2} (\gamma_{1}+2\gamma_{2}) \partial_{x}\theta \partial_{x}\theta + \frac{1}{2}\gamma_{1}\partial_{y}\theta \partial_{y}\theta \notag \\ 
\tilde{f}_{soc} & = \frac{1}{\xi_{D}^{2}} (-\frac{1}{2}{\gamma_{4}}cos2\theta-\gamma_{3}sin\theta).
\label{ReducedEnergyDensity}
\end{align}
$\xi_{D}K_{1}\Delta_{P}^{2}$ also be used as the characteristic unit of London limit free energy in this paper. Before talking about those numeric results and analyzing the corresponding physics, we discuss the random lattice of HQVs and $2/4$ spin solitons formed by the random pinning effect of nafen strands \cite{Makinen2019,Volovik2008}. We analyze the condition under which the effects of coupling between spin solitons induced by random distributions of HQVs can be neglected. The random lattice of spin solitons is pseudo-random lattices as long as this condition is satisfied. This allows us to understand the network of 1D nexus objects consisting of $2/4$ spin solitons and KLS string walls by calculating and analyzing unit cell of pseudo-random lattices consisting of spin solitons.      
\begin{figure*}
\centerline{\includegraphics[width=0.72\linewidth]{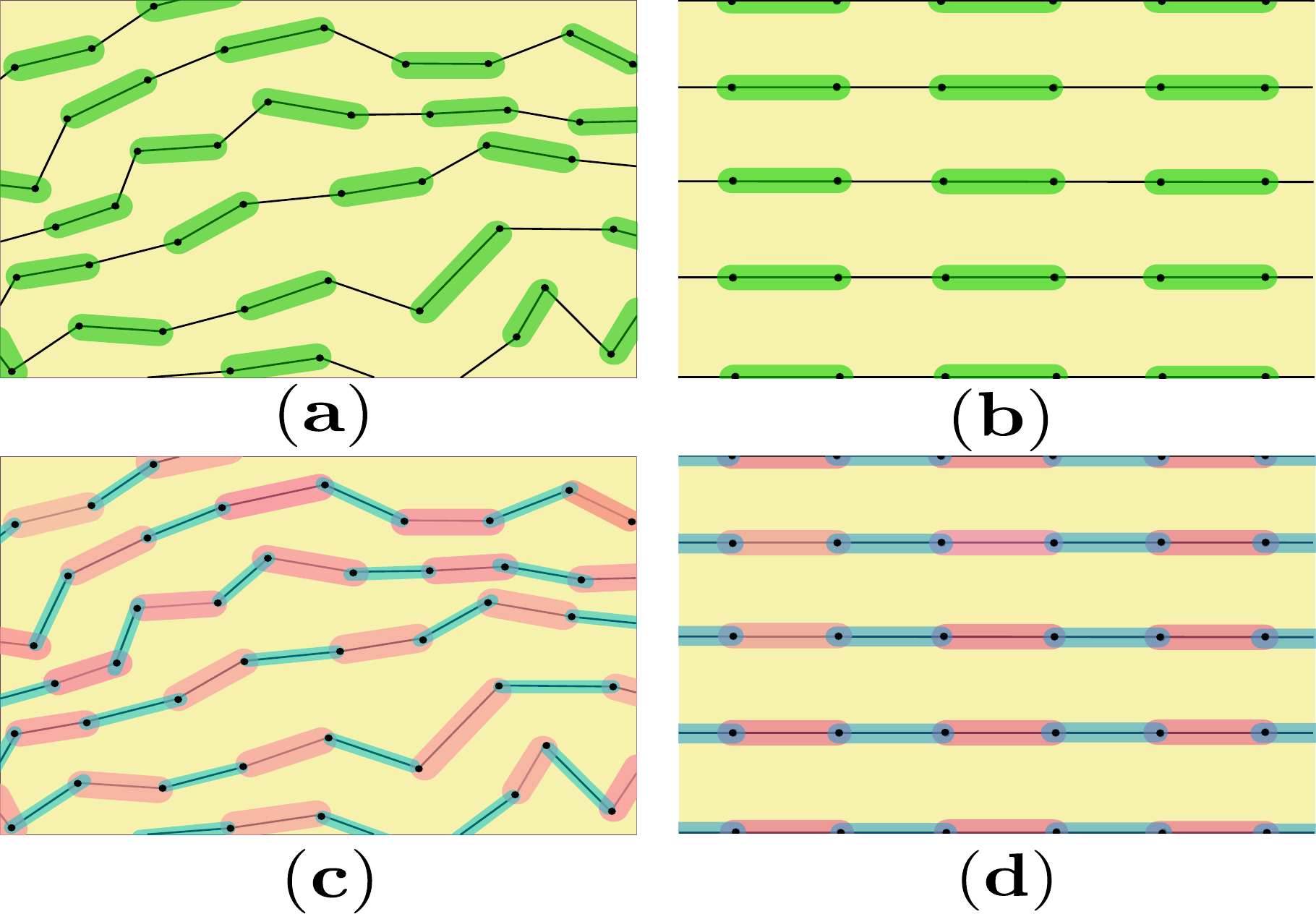}}
\caption{Illustrations of pseudo-random lattices consisting of inseparable and separable spin solitons and their equivalent regular lattices. The black dots represent the HQVs and the black solid lines represent the KLS domain walls. Because every HQV is 1D nexus, two HQVs connect with each others via separable or inseparable $2/4$ spin solitons. These topological objects with different characteristic lengths and spatial dimensions give rise to the network of 1D nexus objects with complex hierarchy of length scales. Thus this network is an instance of the interplay between different homotopy groups of topological objects with dramatically different length scales.
The small green filleted rectangles represent the inseparable spin solitons, while pink and blue filleted rectangles represent separable spin solitons. (a) pseudo-random lattice of inseparable spin solitons ($\pi$-solitons) when $\Omega \ll \Omega_{c}$. The spin solitons are almost identical and well spatially separated with each others. The spin dynamic response properties of pseudo-random lattice is equivalent to (b) 2D regular lattice of $\pi$-solitons. Similarly, (c) pseudo-random lattice of separable spin solitons (KLS-solitons and solitons) has same spin dynamic response with (d) 2D regular lattice consisting of KLS-solitons and solitons. \label{PsudoRamdomLatticeOfSolitons}} 
\end{figure*}

\subsection{Pseudo-random lattices consisting of spin solitons}
In the experiment of polar distorted B-phase, the HQVs are pinned by nafen strands when they appear during cooling down. Hence the 
HQVs and KLS string walls randomly distribute in the PdB sample and form network.
The statistic distribution of HQVs is uniform because there is no reason which provides preferable location for HQV. This means the number of HQVs in unit area is constant for rotating PdB superfluid with angular velocity $\Omega$. Then the average area occupied by one HQV is constant as well. We denote the average area occupied by HQV as $A=D(\Omega)^{2}$, where $D(\Omega)$ is the average distance between two HQVs and $D(\Omega)$ depends on the angular velocity as 
\begin{equation}
D(\Omega)=\sqrt{A}=\sqrt{\frac{\kappa_{0}}{4\Omega}},
\label{DOmega}
\end{equation}
where $\kappa_{0}=h/2m$ is the circulation quantum of HQV and $m$ is mass of Helium-3 atom \cite{Salomaa1985,Autti2016}. In Fig.~\ref{PsudoRamdomLatticeOfSolitons} (a) and (c), we illustrate the uniformly distributed HQVs with given $\Omega$. These HQVs, as we have known at Sec. \ref{RelativeHomotopy} and \ref{TwoDifferentSolitonConfiguretions}, are 1D nexuses which connect $2/4$ spin solitons and KLS domain walls. Because the random distribution of HQVs, the $2/4$ spin solitons are also randomly distributed over the PdB superfluid. Therefore the HQVs and spin solitons form a 2D random lattice \cite{volovik2019}. These spin solitons have almost identical spin configuration and geometric size determined by gradient energy and SOC energy. Their spin dynamic response under weak magnetic drive are almost identical as well. As a result, the spin dynamic response of these spin solitons under weak drive is independent to the distribution of HQVs and spin solitons. The NMR frequency shift under weak magnetic drive is merely determined by the configuration of one spin soliton, and the total ratio intensity of system is the summation of ratio intensities of all spin solitons. We call this kind of random lattice of HQVs and spin solitons as pseudo-random lattice. This means the spin dynamic response properties of pseudo-random lattice of $2/4$ spin solitons are equivalent to the spin dynamic properties of regular lattice of $2/4$ spin solitons. There are two types of regular lattices as shown in Fig.~\ref{PsudoRamdomLatticeOfSolitons} (b) and (d), which correspond to inseparable and separable $2/4$ spin solitons respectively.   
\begin{figure*}
\centerline{\includegraphics[width=0.57\linewidth]{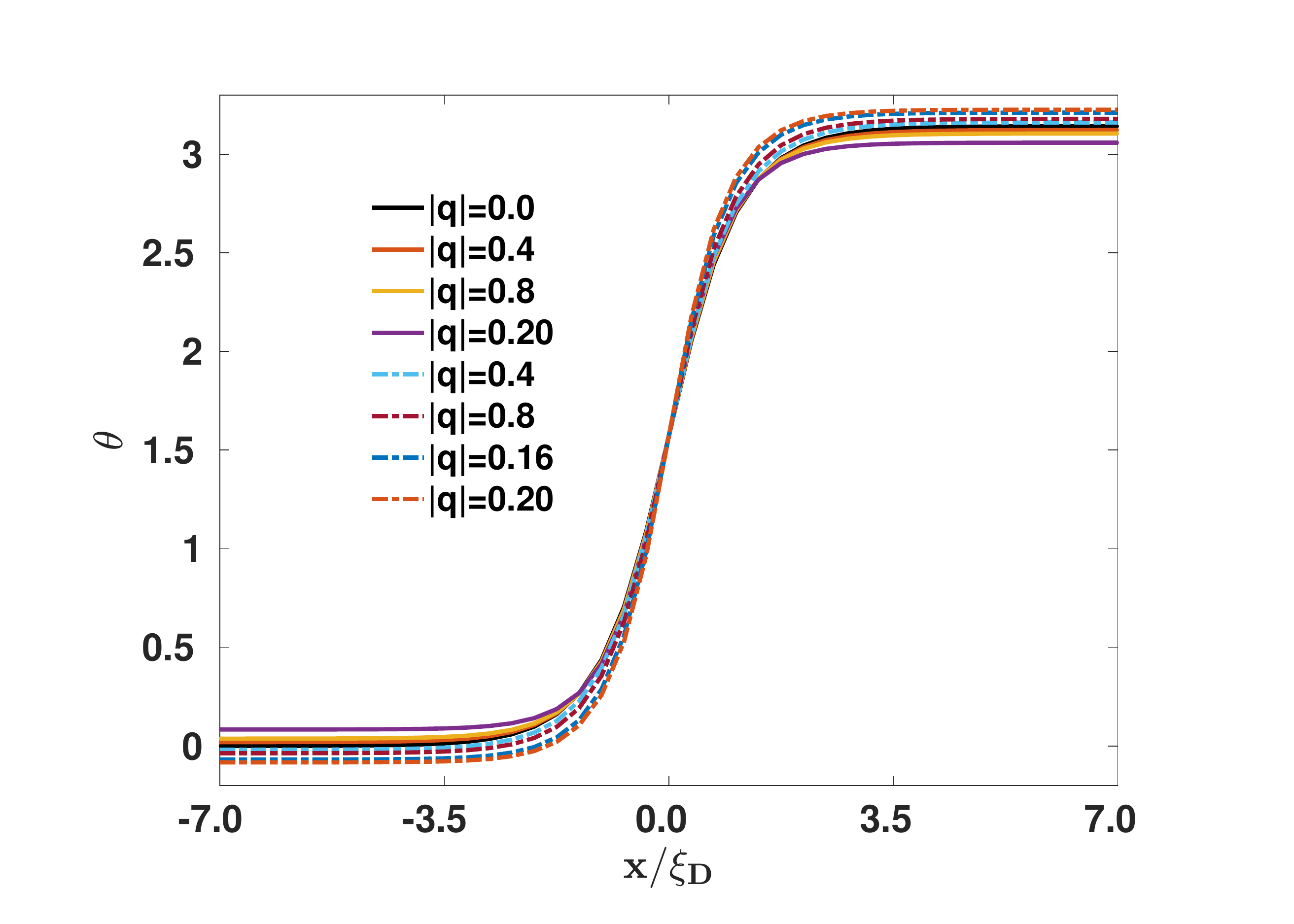}}
\caption{Equlibrium spin configurations of solitons ($|\Delta{\theta}|=2\pi-\theta_{0}$) and big-solitons ($|\Delta{\theta}|=2\pi+\theta_{0}$) in uniform domains. Black solid line represents spin soliton of polar phase ($|q|=0$). The colored dash lines represent big-solitons of PdB phase with $q<0$ from $|q|=0.04$ until $|q|=0.2$. The colored solid lines represent solitons of PdB phase with $q>0$ from $q=0.04$ until $q=0.2$. The spin vectors of all solitons and big-solitons have same relative direction respect to orbital frame because $\theta=\pi/2$ is the stationary point of $\partial_{x} {\theta}$. \label{SolitonAndBigSoliton}} 
\end{figure*}

However, the pseudo-random lattice model is not correct any more when the angular velocity of PdB system increase around a critical value $\Omega_{c}$. To understand this, we notice that the average distance $D(\Omega)$ between two HQVs is proportional to $1/\sqrt{\Omega}$ in Eq.~(\ref{DOmega}). This means the configurations between spin solitons overlap and couple with each others when $\Omega$ big enough. This is because the characteristic thickness of spin soliton i.e., $\sim2\xi_{D}$ is constant under given external parameters. The independence of the spin solitons between two 1D nexus objects loses when $D(\omega) \sim 2\xi_{D}$ and the static textures of spin solitons strongly depend on the distribution of HQVs. As a result, the spin dynamic response of the random lattice of spin solitons under weak magnetic drive strongly depends on the distribution of HQVs as well. Thus the upper limit of $\Omega$ under which pseudo-random lattice model works is determined by $\sqrt{\kappa_{0}/4\Omega_{c}} \sim 2 \xi_{D}$ and then
\begin{equation}
\Omega_{c} \sim \frac{\kappa_{0}}{16 \xi_{D}^{2}}.
\label{OmegaC}
\end{equation}   
For PdB system with $\kappa_{0}=6.62 \times 10^{-8} m^{2}/s$ and $ \xi_{D} \sim 10^{-6}m$ to $\sim 10^{-5}m$, Eq.~(\ref{OmegaC}) suggests $\Omega_{c} \sim 10^{1} rad/s$ to $ \sim 10^{3} rad/s$. These values is larger enough than the angular velocity of PdB system in the experiment of Ref.~\onlinecite{Makinen2019}, then pseudo-random lattice model is good enough and we keep working with it in the rest parts of this paper. 

\subsection{\label{AbsenceOfHQVs}Spin solitons in the absence of KLS string walls -- solitons and big-solitons}
In order to understand the 1D nexus object consisting of $2/4$ spin soliton and KLS string wall, we start from the simpler situation in which there is absence of KLS string wall. We omit the spin solitons with topological invariant larger than $1$ because those kinds of spin solitons cost more energy induced by the existences of spin vortices. In this case $\Delta_{\bot2}$ is single valued over the sample of superfluid, then only solitons ($|\Delta \theta|=\pi-2\theta_{0}$) with topological invariant $1/4_{(\theta_{0})}$, $3/4_{(\pi-\theta_{0})}$ and big-solitons ($|\Delta \theta|=\pi+2\theta_{0}$) with topological invariant $1/4_{(\pi+\theta_{0})}$, $3/4_{(-\theta_{0})}$ are possible in the system. These two different cases correspond to spin solitons in uniform domain with $\Delta_{\bot2}=+|q|\Delta_{P}$ or $\Delta_{\bot2}=-|q|\Delta_{P}$ respectively. Moreover, the spin textures have translation symmetry along transverse direction of spin solitons, then the question reduces to one dimensional question. As mentioned before, we use the BFGS non-linear optimization algorithm on Eq.~(\ref{FreeEnergyThetaDimensionless}) to get the equilibrium configuration of spin solitons \cite{jorge2006} . 

In Fig.~\ref{SolitonAndBigSoliton}, we show the equilibrium configuration of solitons and big-solitons from $|q|=0$ to $|q|=0.2$. The spin textures with $q>0$ are solitons, while the spin textures with $q<0$ are big-solitons. We find that the spin vectors of all solitons and big-solitons have common direction $\theta=\pi/2$. This is because $\theta=\pi/2$ is stationary point of $\partial_{x} \theta$, then $\partial_{x}\partial_{x} \theta|_{\theta=\pi/2}=0$ for all solitons and big-solitons. We will soon see this important feature helps us to set appropriate boundary condition for searching equilibrium textures of pseudo-random lattices consisting of $\pi$-solitons.

\subsection{\label{EquilibriumTexturesOfInseparableAndSeparableSolitons}Spin solitons in the presence of KLS string walls -- inseparable and separable spin solitons }
As we have discussed in Sec. \ref{RelativeHomotopyGroupof1DNexus} and Sec. \ref{TwoDifferentSolitonConfiguretions}, the HQV is 1D nexus connecting KLS domain wall and $2/4$ spin solitons. In London limit, the free energy of network of 1D nexus objects is free energy of pseudo-random lattices consisting of $2/4$ spin solitons. The equilibrium configuration of pseudo-random lattices is the saddle point of Eq.~(\ref{FreeEnergyThetaDimensionless}). The complexity here is the topological invariant $2/4$ has two different representations i.e., literal $2/4$ or $1/4 + 1/4$. Based on the topological analysis, we have known these two cases correspond to inseparable $\pi$-soliton configuration and separable configurations of KLS-soliton and soliton. 

\subsubsection{Boundary conditions on the KLS domain wall}
To quantitatively get the equilibrium spin textures for both configurations of $2/4$ spin solitons, we minimize the London limit free energy Eq.~(\ref{FreeEnergyThetaDimensionless}) in the presence of KLS string wall. For parametriztion Eq.~(\ref{PARA1}), KLS string wall separates two domains with oppsite $\Delta_{\bot2}$ in an unit cell of pseudo-random lattice of spin solitons. 

However, different from the situation with uniform domain for soliton and big-soliton in Sec.~\ref{AbsenceOfHQVs}, the existence of KLS domain wall induces a singularity of the London limit free energy $\tilde{F}(\theta)$. That is because the order parameter $A_{\alpha i}^{PdB}$ in the London limit is ill-defined on the KLS domain wall. As a result, the free energy Eq.~(\ref{FreeEnergyThetaDimensionless}) and corresponding Lagrangian equation of $\theta$ are also ill-defined on the KLS domain wall. On the other hand, we know $\theta$ is a continuous function everywhere for $2/4$ spin soliton because the relative 1-loop of $\pi_{1}(S_{S}^{1},\tilde{R}_{2})$ is continuous mapping. Then $\theta$ keeps single-valued and continuous on the KLS domian wall. These facts require us to set a proper boundary condition of $\theta$ on the KLS domain wall. The London limit free energy Eq.~(\ref{FreeEnergyThetaDimensionless}) can be minimized with this boundary condition.

\begin{figure*}
\centerline{\includegraphics[width=0.7\linewidth]{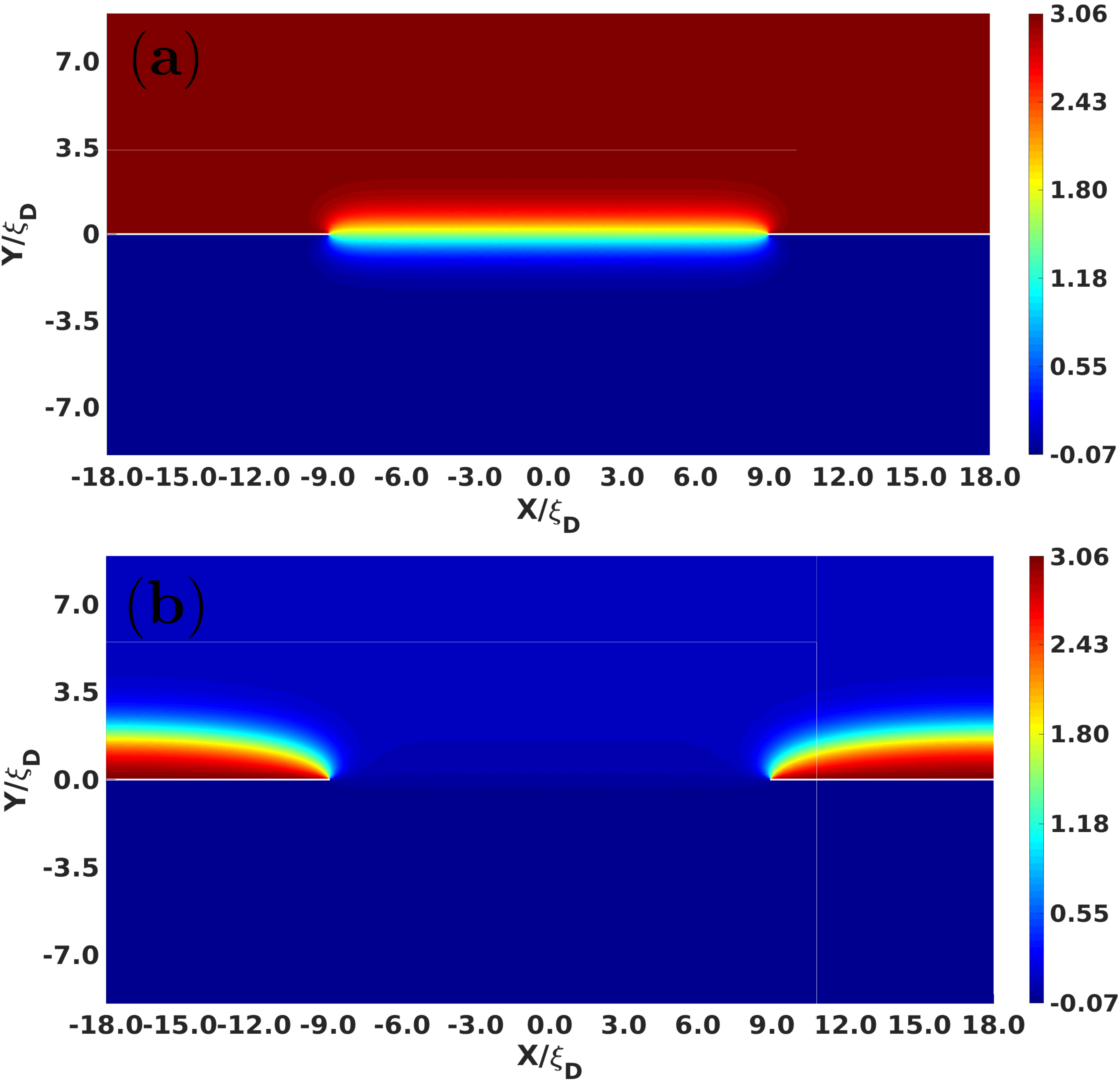}}
\caption{Equilibrium configurations of unit cell of pseudo-random lattices consisting of inseparable spin soliton ($\pi$-solitons) and separable spin solitons (KLS-soliton and soltion) respective. These equilibrium spin textures are gotten by minimizing the reduced London limit free energy $\tilde{F}(\theta)_{London}$ in Eq.~(\ref{FreeEnergyThetaDimensionless}) by BFGS algorithm. The resulted equilibrium distributions of $\theta$ depict the equilibrium textures of spin vectors in a unit cell of pseudo-random lattices.
(a) depicts the unit cell of pseudo-random lattice consisting of inseparable $2/4$ spin solitons ($\pi$-solitons) for $|q|=0.18$ and $D=18\xi_{D}$. (b) depicts the unit cell of pseudo-random lattice consisting of separable spin solitons (KLS-Solitons and solitons) with same parameters of (a) but its topological invariant is $1/4+1/4$. \label{SeparableAndInSeparableSolitons}} 
\end{figure*}
In order to find out this boundary condition properly, we review the fact that the free energy and Lagrangian equation of $\theta$ is ill-defined on the KLS domain wall. This means $\theta$ of different domains in the vicinity of the KLS domain wall does not relate to each other by Lagrangian equation of $\theta$. Then $\theta$ in two different domains, which are separated by KLS domain wall, are determined independently in two uniform domains with opposite $\Delta_{\bot2}$. In this situation, to keep the continuity of $\theta$ on the KLS domain wall, the boundary condition of $\theta$ must be a common value of spin solitons in both two domains with opposite $\Delta_{\bot2}$. For the inseparable spin soliton with literally topological invariant $2/4$, the natural choice is the stationary point of big-soliton and soliton i.e., $\theta_{KLS}=\pi/2$. This boundary condition indicates the $\pi$-soliton may be understood as a hybrid of big-soliton and soliton in London limit. As for the separable spin soliton with topological invariant $1/4+1/4$, because all KLS-solitons have common values $\theta=0$ or $\theta=\pi$ on the KLS domain wall, there are two options of boundary condition \cite{Makinen2019}. However, these two options are identical, they give rise to same spin textures of pseudo-random lattices consisting of separable spin solitons, see details in appendices Sec. \ref{SeparableSolitionTwoBoundryConditions}. Thus in the rest of this paper, we only use $\theta_{KLS} = 0$ for all calculations about separable spin solitons in main text.

\begin{figure*}
\centerline{\includegraphics[width=0.87\linewidth]{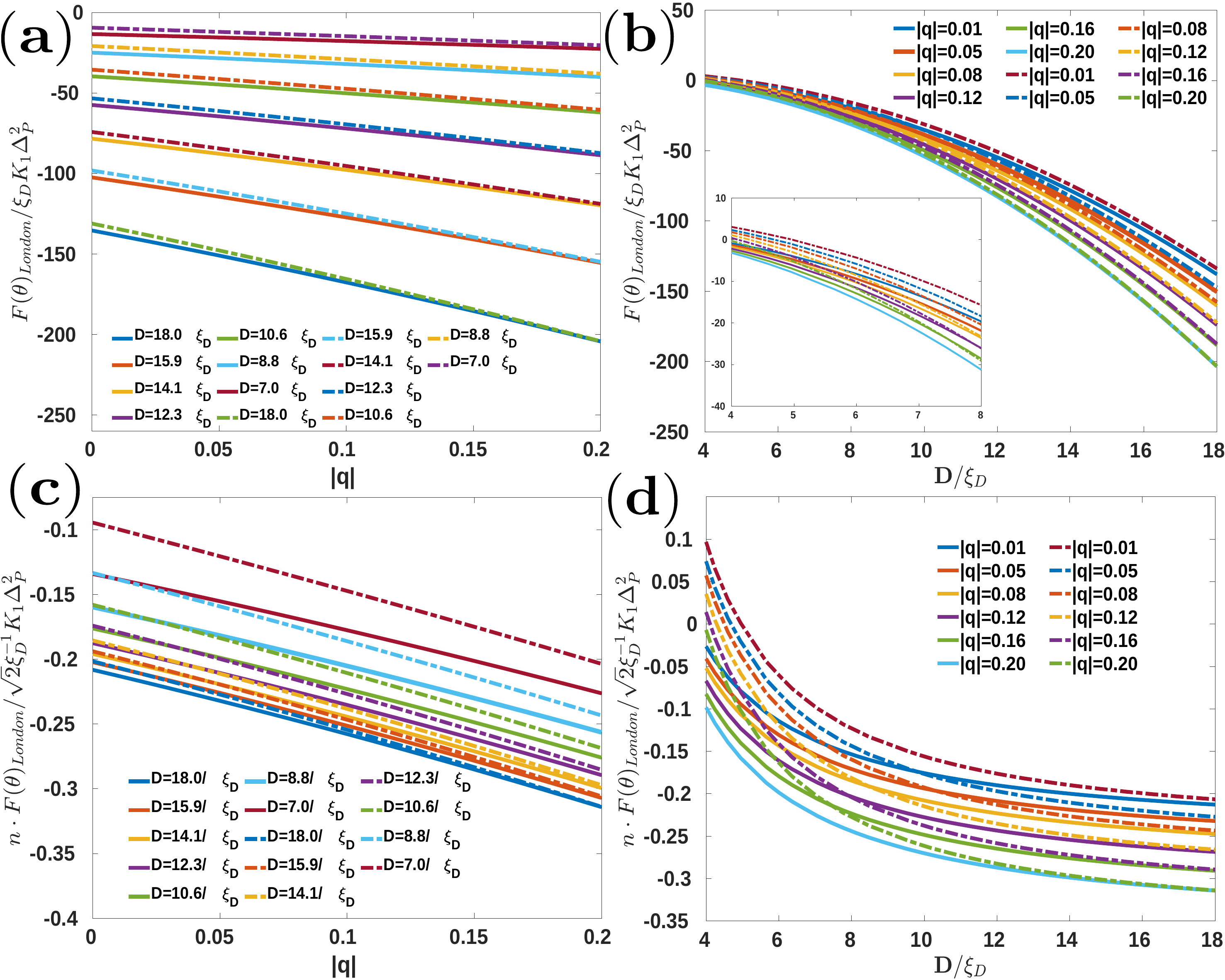}}
\caption{Free energies $F(\theta)_{London}$ of one-half unit cell and surface densities of free energies $n \cdot F(\theta)_{London}$ of pseudo-random lattices consisting of $2/4$ spin solitons. The external parameters $|q|$ are from $0.01$ to $0.20$ and $D$ are from $4\xi_{D}$ to $18\xi_{D}$.
The reduced free energies $\tilde{F}(\theta)_{London}$ and reduced densities of free energies $n \cdot \tilde{F}(\theta)_{London}$ are calculated based on the equilibrium spin textures of one-half unit cell of the lattices when the pseudo-random lattice model works i.e., $\Omega \ll \Omega_{c}$. These results depict the equilibrium free energies and energy densities of spin degree of freedom of 1D nexus objects network in PdB system. The solid lines represent the inseparable spin solitons, and the dash-dot lines represent the separable spin solitons. (a) shows the free energies $F(\theta)_{London}$ as functions of $|q|$ with different average distances $D$. When SOC energy dominates the system in big unit cell, $F(\theta)_{London}$ is monotonically decreasing respect to $|q|$. While $F(\theta)_{London}$ does not show remarkably change as $|q|$ changes when gradient energy is competitive to SOC energy in a small unit cell. (b) shows the free energies $F(\theta)_{London}$ are monotonically decreasing  functions of $D$. The zooming plot between $4\xi_{D}$ and $8\xi_{D}$ in (b) demonstrates this monotonicity is held even when the gradient energy is competitive with the SOC energy. Similarly, (c) depicts the London limit free energy density $n \cdot \tilde{F}(\theta)_{London}$ is monotonically decreasing function of $|q|$. However, (d) demonstrates the London limit free energy densities $n \cdot \tilde{F}(\theta)_{London}$ asymptotically trend to constants determined by SOC energy when SOC energy is the dominating energy in big unit cell. When $D$ is small enough ($D<6\xi_{D}$) and gradient energy becomes to the dominating energy, the free energy densities increase rapidly as $D$ decrease because $n \cdot \tilde{F}(\theta)_{London}|_{D<6\xi_{D}}\propto{1/D}$. All these results show the equilibrium free energies of pseudo-random lattice consisting of inseparable spin solitons ($\pi$-solitons) are lower than those of separable spin solitons (KLS solitons and solitons). \label{FreeEnergyBoxAndDensity}   
} 
\end{figure*}
\subsubsection{\label{EquilibriumSpinTexturesAndFreeEnergy}Equilibrium spin textures and free energies of pseudo-random lattices consisting of inseparable and separable $2/4$ spin solitons }
In Fig.~\ref{SeparableAndInSeparableSolitons}, we show the equilibrium textures of pseudo-random lattices consisting of inseparable and separable $2/4$ spin solitons with $|q|=0.18$ and $D=18\xi_{D}$. These two equilibrium configurations of a pair of 1D nexus objects are unit cells of pseudo-random lattices of inseparable and separable spin solitons respectively. To collect enough data which could be used to calculate spin dynamic response and compare with experiment, we calculated spin textures with parameters $|q|$ from $0$ to $0.2$ and $D$ from $4\xi_{D}$ to $18\xi_{D}$. Based on these data, we further calculated the reduced London limit free energy Eq.~(\ref{FreeEnergyThetaDimensionless}) of these two types of pseudo-random lattices, the results are shown in Fig.~\ref{FreeEnergyBoxAndDensity}. Before we discussing these numeric results, we first evaluate the Eq.~(\ref{FreeEnergyThetaDimensionless}) for one-half of unit cell when $D \geq 10\xi_{D}$. In this case, 
\begin{align}
\int_{\Sigma} \tilde{f}_{grad} \, d\Sigma\,\, & \sim \frac{1}{2}(1+|q|^{2})\frac{\pi^{2}}{4\xi_{D}^{2}}\int_{\Sigma'}d\Sigma, \notag \\ \int_{\Sigma} \tilde{f}_{soc} \, d\Sigma\,\, & \sim \int_{\Sigma'} \tilde{f}_{soc} \, d\Sigma\,\, + \int_{\Sigma-\Sigma'} \tilde{f}_{soc} \, d\Sigma,\,\,
\label{EvaluationOfFreeEnergy1} 
\end{align} 
where $\Sigma'$ is the region which spin solitons occupy and its area in $x$-$y$ plane is around $D\xi_{D}$. Then the integral of $\tilde{f}_{soc}$ in Eq.~(\ref{EvaluationOfFreeEnergy1}) can be evaluated as
\begin{align}
\int_{\Sigma'} \tilde{f}_{soc} \, d\Sigma\,\, & \sim 0, \notag \\ 
\int_{\Sigma-\Sigma'} \tilde{f}_{soc} \, d\Sigma\,\, & \sim \int_{\Sigma-\Sigma'} \tilde{f}_{soc}|_{y>0} \, d\Sigma\,\, + \int_{\Sigma-\Sigma'} \tilde{f}_{soc}|_{y<0} \, d\Sigma.\,\,
\label{EvaluationOfFreeEnergy2} 
\end{align} 
The first integral in Eq.~(\ref{EvaluationOfFreeEnergy2}) vanishes because $\tilde{f}_{soc}$ is not negative-definite function in $\Sigma'$. In contrary, $\tilde{f}_{soc}$ has negative-definite equilibrium values in regions $(\Sigma-\Sigma')_{y>0}$ and  $(\Sigma-\Sigma')_{y<0}$. Hence 
\begin{equation}
\int_{\Sigma-\Sigma'} \tilde{f}_{soc} \, d\Sigma\,\, \sim  \frac{{\xi_{D}}D(D-\xi_{D})}{2} [\tilde{f}_{soc}|_{y>0} \,\, + \tilde{f}_{soc}|_{y<0}]_{q} .\,\,
\label{EvaluationOfFreeEnergy3}
\end{equation}
As a result, the reduced London limit free energy is evaluated as
\begin{align}
\tilde{F}(\theta)_{London} & \, \sim  \notag \\
& \frac{1}{8}(1+|q|^{2})\frac{\pi^{2}D}{\xi_{D}} \, \notag \\ 
& + \, \frac{D(D-\xi_{D})}{2} [\tilde{f}_{soc}|_{y>0} \,\, + \tilde{f}_{soc}|_{y<0}]_{q} \notag \\
& \sim \frac{1}{8}(1+|q|^{2})\frac{\pi^{2}D}{\xi_{D}} \notag \\
& + \, \frac{D(D-\xi_{D})}{2{\xi_{D}^{2}}} \notag \\ 
& \times [-(1+|q|)^{2}cos2|\theta_{0}| -2(1+|q|)|q|sin|\theta_{0}|]. \label{EvaluationOfFreeEnergy4}
\end{align}
Eq.~(\ref{EvaluationOfFreeEnergy4}) immediately suggests SOC energy is dominating energy of London limit free energy when the average distance $D$ between 1D nexuses is big enough and the $\tilde{F}(\theta)_{London}<0$ because $(\tilde{f}_{soc}|_{y>0} \,\, + \tilde{f}_{soc}|_{y<0})<0$ over $\Sigma-\Sigma'$. For $|q|\in[0,0.2]$, $\tilde{F}(\theta)_{London}$ in Eq.~(\ref{EvaluationOfFreeEnergy4}) is around $-130$ to $-200$ with $D=18\xi_{D}$. This is exactly what the numeric results show in Fig.~\ref{FreeEnergyBoxAndDensity}(b).  When $D$ decreases during the angular velocity $\Omega$ of PdB system increases, Eq.~(\ref{EvaluationOfFreeEnergy4}) increases monotonically as shown in Fig.~\ref{FreeEnergyBoxAndDensity} (a) and (b). Other information which Eq.~(\ref{EvaluationOfFreeEnergy4}) indicates is the London limit free energy of unit cell of pseudo-random lattice is decreasing function for $|q|$ as long as SOC energy is dominating energy. This is because $\tilde{f}_{soc}|_{y>0} \,\, + \tilde{f}_{soc}|_{y<0}$ is decreasing function of $|q|$. However, this is not true any more when $D$ is small. Because SOC energy is not dominating energy in this case, the positive-definite gradient energy is competitive with SOC energy. As a result, we can find from Fig.~\ref{FreeEnergyBoxAndDensity} (a) and (b) that the $\tilde{F}(\theta)_{London}$ of one-half unit cell does not change remarkably for different $|q|$ in small unit cell with $D \sim [4\xi_{D},8\xi_{D}]$. The free energy density of per unit area of equlibrium pseudo-random lattices can be evaluated by multiplying the surface density of 1D nexues $n=D^{-2}$ to the Eq.~(\ref{EvaluationOfFreeEnergy4}),
\begin{align}
n \cdot & \tilde{F}(\theta)_{London}\, \sim \notag \\
& \frac{1}{8}(1+|q|^{2})\frac{\pi^{2}}{\xi_{D}D} \notag \\ 
& + \, \frac{1}{2{\xi_{D}^{2}}} [-(1+|q|)^{2}cos2|\theta_{0}| -2(1+|q|)|q|sin|\theta_{0}|].
\label{FreeEnergyDensity}
\end{align}
Then we find the London limit free energy density of pseudo-random lattices trends to be a constant determined by $q$ when SOC energy is dominating with large $D$. We can clearly see this form Fig.~\ref{FreeEnergyBoxAndDensity}(d) when $D$ is larger than $10\xi_{D}$. From Eq.~(\ref{FreeEnergyDensity}), we find the magnitude of $n \cdot F(\theta)_{London}$ is around $10^{-1}{\sqrt{2}}{\xi_{D}^{-1}}K_{1}{\Delta_{P}^{2}}$ for $|q|\in[0,0.2]$ when $D>10\xi_{D}$. This coincides with the numerical results in Fig.~\ref{FreeEnergyBoxAndDensity} (c) and (d). When the system is dominated by gradient energy if $D$ is small enough, the free energy density increases rapidly as shown in Fig.~\ref{FreeEnergyBoxAndDensity} (d). If the angular velocity increase successively, the system will go into a parameters region in which pseudo-random lattice model violates.

In all cases, we find the equilibrium free energies of one-half unit cell of separable spin solitons (KLS soliotns and solitons) are slightly higher than those of inseparable spin solitons ($\pi$-solitons). As a result, the equilibrium free energy densities of pseudo-random lattices consisting of separable spin solitons (KLS-soliotns and solitons) are also slightly higher than those of inseparable spin solitons ($\pi$-solitons). This significant fact suggests that the equilibrium states which was observed in experiment of rotating PdB system is the pseudo-random lattice of inseparable $2/4$ spin solitons ($\pi$-solitons) of 1D nexus objects. We will see this is true in next section by calculating the spin dynamic response under weak magnetic drive.
 
\section{\label{SpinDynamics}Spin dynamic response and NMR of pseudo-random lattices consisting of $2/4$ spin solitons}
We have talked the topological origin of 1D nexus objects as well as the inseparable and separable spin solitons with topological invariant $2/4$ in previous sections. These two kinds of spin solitons connecting with KLS string wall have different equilibrium free energies. Thus the pseudo-random lattices consisting of them have different equilibrium free energy densities. To compare with the experiments and check  the theories, we must calculate the spin dynamic response of system under continuous wave magnetic drive. Under weak enough magnetic drive, the nuclear spin magnetization of PdB superfluid responds a nuclear magnetic resonance (NMR) when the frequency of magnetic drive matches the transverse spin dynamic mode. Because the spin dynamics of symmetry breaking states of Helium-3 is strongly influenced by SOC energy which is determined by the relative orientations between spin and orbital degenerate parameters, the NMR of continuous wave drive is a perfect tool, which can be used to detect the pseudo-random lattice of spin solitons of 1D nexus objects network \cite{VollhardtWolfle1990}.   

When the PdB superfluid is equilibrium, the spin density has equilibrium value $\mathbf{S}^{(0)}$ over the system. If the weak homogeneous magnetic drive is turned on, the spin density gets a tiny variation ${\delta}\mathbf{S}(\mathbf{r},t)$, where $\mathbf{r}$ and $t$ are spatial and time coordinates respectively. In this perturbed system, the transverse spin density ${\delta}S_{+}$ may be expanded as 
\begin{equation}
{\delta}S_{+}({\mathbf{r}},t)={\int}{d{\sigma}'}{\int}{dt'}{\frac{{\delta}{S_{+}}}{{\delta}{H_{a}}}}(\mathbf{r},t,\mathbf{r}',t'){\delta}{H_{a}}({\mathbf{r}}',t')+O({\delta}{H_{a}}^{2}),
\label{LinearResponse}
\end{equation}
where ${\delta}H_{a} \equiv {\delta}\mathbf{H}$ is the homogeneous weak magnetic drive and $a=1,2,3$ are spatial coordinate indexes. Thus the PdB superfluid under magnetic drive is a linear response system if $|{\delta}\mathbf{H}| \ll |\mathbf{H}^{(0)}|$ \cite{Alxander2010}. The poles of the transverse spin dynamic response function ${\delta}S_{+}/{\delta}H_{a}$ correspond to eigenmodes of the NMR. We calculate these eigenmodes for pseudo-random lattices of inseparable and separable spin solitons with topological invariant $2/4$ in this sections.  
  
\begin{figure*}
\centerline{\includegraphics[width=0.66\linewidth]{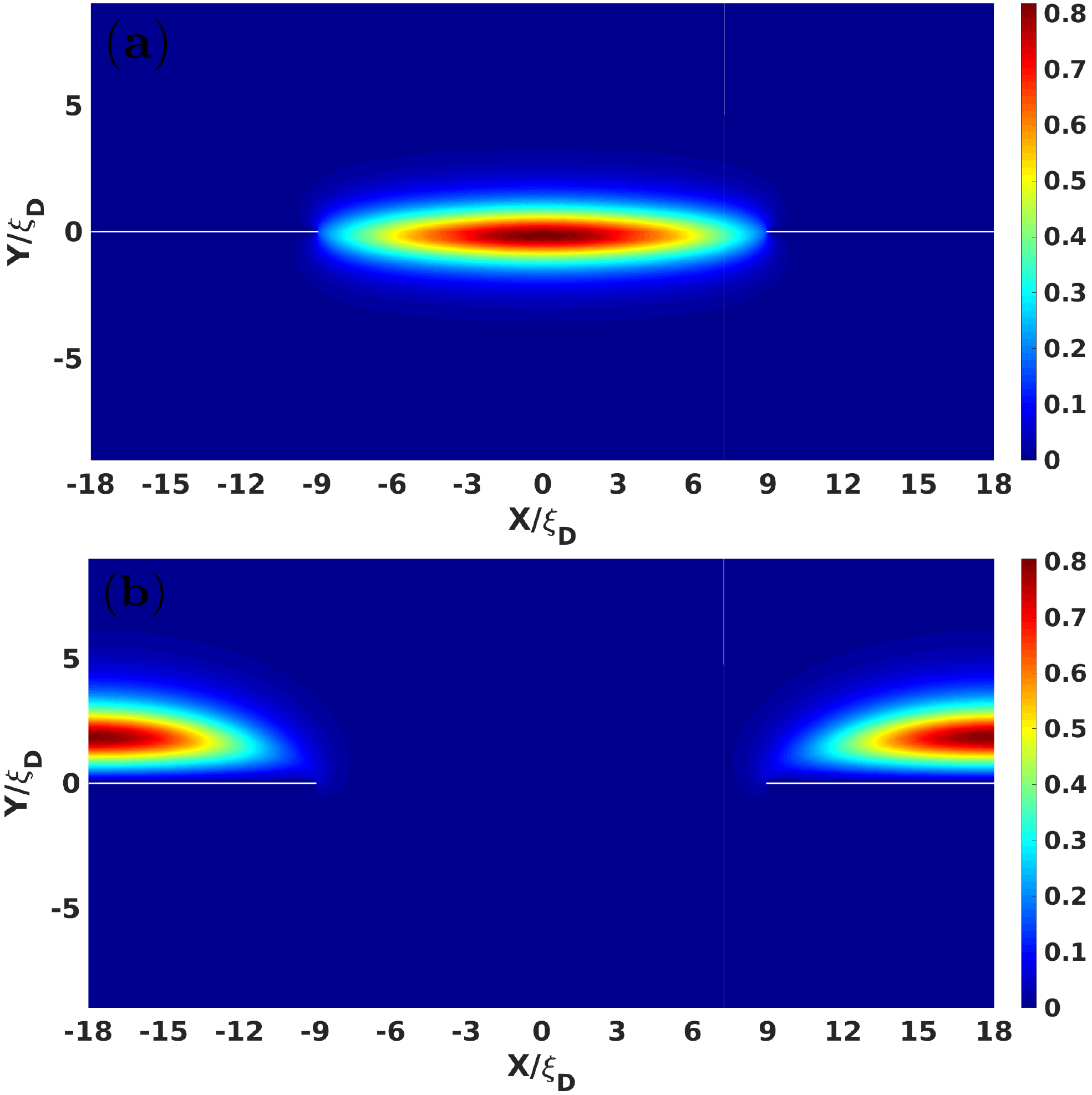}}
\caption{The modulus $|\delta{S_{+}}(\omega)|$ of the lowest transverse spin dynamic response modes located in the unit cells of pseudo-random lattices of inseparable and separable $2/4$ spin solitons. In both cases, we depict the results with parameters $|q|=0.18$ and $D=18\xi_{D}$. (a) $|\delta{S_{+}}(\omega)|$ in  unit cell consisting of inseparable spin solitons ($\pi$-solitons); (b) $|\delta{S_{+}}(\omega)|$ in unit cell consisting of separable spin solitons (KLS-solitons and Solitons). \label{SpinDynamicResponseModes}}
\end{figure*} 
\subsection{Equations of spin dynamic response under homogeneous continuous-wave drive}
Spin-orbit coupling plays an important role in the NMR measurements of significant properties of different superfluid phases in Helium-3  system. This is because the coherence of superfluid states, which breaks relative symmetry between spin and orbital degree of freedom of order parameters in superfluid Helium-3, strengthens the SOC energy \cite{Anderson1973,AndersonVarma1973}. This gives rise to the observable NMR frequency shift of nuclear spin magnetization. In our case, the SOC energy takes into account all the information and effects of spin vectors in spin solitons which connect to the KLS domain wall via 1D nexus. Then the existence of spin solitons could lead to observable frequency shifts in nuclear magnetic resonance spectrum. Thus, what we need to calculate is the spin dynamic response function $\delta{S_{+}}/\delta{H_{a}}$ dominated by SOC energy. 

In this subsection, we utilize the spin dynamic equations dominated by SOC energy to get $\delta{S_{+}}/\delta{H_{a}}$ as well as corresponding eigenequations of poles \cite{chaikin1995}. Because the SOC energy is much smaller than the microscopic energy scales of PdB superfluid i.e., $\Delta_{P}$,  the characteristic time scales of spin dynamic response function $\delta{S_{+}}/\delta{H_{a}}$ is much longer than the time scales of microscopic processes which are proportional to $\Delta_{P}^{-1}$. All the microscopic processes with time scales $\Delta_{P}^{-1}$ are equilibrium in the spin dynamic processes under weak magnetic drive. This means the spin dynamic equations are a system of hydrodynamic equations of spin densities $\delta{S_{a}}$ and spin vectors of order parameter \cite{VollhardtWolfle1990, chaikin1995}.     

In the limit of hydrodynamics, the system of dynamic equations of spin densities $S_{\alpha}$ and spin vectors are system of Liouville equations 
\begin{align}
{\frac{\partial{S_{\alpha}}}{\partial{t}}} & =\{F_{hydrodynamics},S_{\alpha}\}, \notag \\
 {\frac{\partial{V_{\alpha}^{a}}}{\partial{t}}} & =\{F_{hydrodynamics},V_{\alpha}^{a}\},\,\,
V_{\alpha}^{a}={\hat{e}}_{\alpha}^{1}, {\hat{e}}_{\alpha}^{2}, \hat{d}_{\alpha},
\label{LiouvilleEquations}
\end{align}
where $\alpha=1,2,3$ are the indexes of spatial coordinates. And $V_{\alpha}^{a}$ denote the three spin vectors of order parameter i.e.,  $V_{\alpha}^{1}={\hat{e}}_{\alpha}^{1},$ $V_{\alpha}^{2}={\hat{e}}_{\alpha}^{2}$, $V_{\alpha}^{3}=\hat{d}_{\alpha}$. The hydrodynamic free energy of PdB superfluid dominated by SOC energy is   
\begin{equation}
F_{hydrodynamics}=\int\nolimits_{\Sigma} (f_{\rm H} + f_{\rm soc} + f_{\rm grad})d\Sigma.
\label{FreeEnergyOfHydrodynamics}
\end{equation} 
Thus Eq.~(\ref{LiouvilleEquations}) can be further written as
\begin{align}
{\frac{\partial{S_{\alpha}}}{\partial{t}}} & = {\int_{\Sigma}}{{d^{3}}r'}\frac{{\delta}F_{hydrodynamics}}{{\delta}{S_{\beta}}}(r')\{S_{\beta}(r'),{S_{\alpha}}(r)\} \notag \\ & + {\int_{\Sigma}}{{d^{3}}r'}\frac{{\delta}F_{hydrodynamics}}{{\delta}{V_{\beta}^{a}}}(r')\{V_{\beta}^{a}(r'),{S_{\alpha}}(r)\},
\label{LiouvilleEquations1}
\end{align}
and
\begin{equation}
{\frac{\partial{V_{\alpha}^{a}}}{\partial{t}}}={\int_{\Sigma}}{{d^{3}}r'}\frac{{\delta}F_{hydrodynamics}}{{\delta}{S_{\beta}}}(r')\{S_{\beta}(r'),{V_{\alpha}^{a}}(r)\},
\label{LiouvilleEquations2}
\end{equation}
where $\beta=1,2,3$ are indexes of spatial components of hydrodynamic variables. The Poisson brackets between $S_{\alpha}$ and $V_{\alpha}^{a}$ can be gotten by the commutators-based methods in Ref.~\onlinecite{Dzyaloshinskii1980} as 
\begin{align}
\{{S_{\alpha}}(r_{1}),{{S_{\beta}}(r_{2})}\} & ={\epsilon}_{{\alpha}{\beta}{\gamma}}{S_{\gamma}}{\delta}(r_{1}-r_{2}), \notag \\
\{{S_{\alpha}}(r_{1}),{{V_{\beta}^{a}}(r_{2})}\} & ={\epsilon}_{{\alpha}{\beta}{\gamma}}{V_{\gamma}^{a}}{\delta}(r_{1}-r_{2}), \label{PoissonBrackets1}
\end{align}
where $r_{1}$ and $r_{2}$ are the spatial coordinates and ${\epsilon}_{{\alpha}{\beta}{\gamma}}$ is the Levi-Civita symbol.
After plugging Eq.~(\ref{PoissonBrackets1}) into Eq.~(\ref{LiouvilleEquations1}) and Eq.~(\ref{LiouvilleEquations2}), the coupled first order dynamic equations of spin densities $S_{\alpha}$ and $V_{\alpha}^{a}$ are given as
\begin{align}
\frac{{\partial}{S_{\alpha}}}{{\partial}{t}} & ={\gamma}{H_{\beta}}{\epsilon_{{\alpha}{\beta}{\gamma}}}{S_{\gamma}} \notag  \\ 
& -{\frac{6}{5}}{g_{D}}{V_{j}^{d}}{V_{\gamma}^{b}}{{\epsilon}_{{\alpha}{\beta}{\gamma}}}{Q_{{\beta}j}^{bd}}+({{\partial}_{i}}{{\partial}_{j}}{V_{\beta}^{b}}){V_{\gamma}^{a}}{\epsilon_{{\alpha}{\beta}{\gamma}}}{K_{ij}^{ba}}, 
\label{1stOderEquationsA}
\end{align}
\begin{align}
\frac{{\partial}{V_{\alpha}^{a}}}{{\partial}t} & = {\gamma}{H_{\beta}}{\epsilon_{{\alpha}{\beta}{\gamma}}}{V_{\gamma}^{a}} \notag  \\ 
& -{\delta}{\gamma^{2}}{\chi_{\perp}^{-1}}{S_{\eta}V_{\eta}^{3}}{V_{\beta}^{3}}{\epsilon_{{\alpha}{\beta}{\gamma}}}{V_{\gamma}^{a}}-{\gamma^{2}}{\chi_{\perp}^{-1}}{S_{\beta}}{\epsilon_{{\alpha}{\beta}{\gamma}}}{V_{\gamma}^{a}},
\label{1stOderEquationsB}
\end{align}
where $\delta  = (\chi_{\bot} -\chi_{\|})/\chi_{\|}$ in which $\chi_{\bot}$ and $\chi_{\|}$ are the transverse magnetic susceptibility and the longitude magnetic susceptibility of PdB phase respectively.
\begin{align}
K_{ij}^{ba} & =K_{1}{\delta}_{ij}{X_{m}^{b}}{X_{m}^{a}}+K_{2}{X_{j}^{a}}{X_{i}^{b}}+K_{3}{X_{j}^{b}}{X_{i}^{a}},\notag \\
Q_{{\beta}{j}}^{bd} & ={X_{\beta}^{b}}{X_{j}^{d}}+{X_{\beta}^{d}}{X_{j}^{b}}
\end{align}
with
\begin{align}
{X_{i}^{1}}={{\Delta}_{{\perp}1}}{\hat{x}_{i}},\,\, {X_{i}^{2}}={{\Delta}_{{\perp}2}}{\hat{y}_{i}},\,\, {X_{i}^{3}}={{\Delta}_{{\parallel}}}{\hat{z}_{i}}.
\label{2ndEQ}
\end{align}
The details of calculation from Eq.~(\ref{LiouvilleEquations1}) to Eq.~(\ref{1stOderEquationsB}) are shown in appendixes Sec. \ref{For1stOrderEqations}.

Based on the first order equations of spin densities and degenerate parameters in Eq.~(\ref{1stOderEquationsA}) and  Eq.~(\ref{1stOderEquationsB}), we can further derive the second order spin dynamic response equations of $\delta{S_{\alpha}}$ under weak magnetic drive $\delta{H_{\alpha}}$. This was done by plugging $S_{\alpha}=S_{\alpha}^{(0)}+{\delta}{S_{\alpha}(\mathbf{r},t)}$, $V_{\alpha}^{a}=V_{\alpha}^{a(0)}+{\delta}{V_{\alpha}^{a}(\mathbf{r},t)}$ and $H_{\alpha}=H_{\alpha}^{(0)}+{\delta}{H_{\alpha}(t)}$ into Eq.~(\ref{1stOderEquationsA}) and Eq.~(\ref{1stOderEquationsB}). Here the $S_{\alpha}^{(0)}$ and $V_{\alpha}^{a(0)}$ are the equilibrium spin densities and equilibrium degenerate parameters respectively. While the ${\delta}{S_{\alpha}(\mathbf{r},t)}$ and ${\delta}{V_{\alpha}^{a}(\mathbf{r},t)}$ are the dynamic parts of the perturbed spin densities and degenerate parameters. The $H_{\alpha}^{(0)}$ is the static magnetic field and ${\delta}{H_{\alpha}(t)} = |\delta\mathbf{H}|\hat{x}e^{-i{\omega}t}$ is the homogeneous RF continuous-wave drive. We put the details of calculations in appendixes Sec.~\ref{For2stOrderEqations} and the derived spin dynamic response equations within frequency form is 
\begin{align}
i{\omega}{{\delta}{S_{\alpha}}({\omega})} & ={\gamma}{\epsilon_{{\alpha}{\beta}{\gamma}}}{H_{\beta}^{(0)}}{{\delta}S_{\gamma}(\omega)} +{\gamma}{\epsilon_{{\alpha}{\beta}{\gamma}}}S_{\gamma}^{(0)}{{\delta}{H_{\beta}}}(\omega) \notag \\ 
& +{\frac{\Xi_{{\alpha}{\lambda}}}{i\omega}}{\delta}{S_{\lambda}}(\omega) +{\frac{C_{{\alpha}{\eta}}}{i{\omega}}}{\delta}{H_{{\eta}}}(\omega)
\label{2ndOderEquations}
\end{align}
and
\begin{align}
{\Xi_{{\alpha}{\lambda}}} & ={\frac{\gamma^{2}}{\chi_{\perp}}}{K_{ij}^{ba}}{\Lambda_{ij{\alpha}{\lambda}}^{ba}} \notag \\ 
&+{\frac{6{g_{D}}{\gamma^{2}}}{5{\chi_{\perp}}}}{R_{j{\lambda}{\alpha}{\beta}}^{db}}{Q_{{\beta}j}^{bd}}+{\frac{6{g_{D}}{\gamma}^{2}}{5\chi_{\perp}}}{V_{\zeta}^{d{(0)}}}{V_{\gamma}^{b(0)}}{\epsilon_{j{\lambda}{\zeta}}}{\epsilon_{{\alpha}{\beta}{\gamma}}}Q_{{\beta}{j}}^{bd}, \notag \\
C_{{\alpha}{\eta}} & = {\gamma}G_{{i}{j}{\alpha}{\eta}}^{ba}{K_{{i}{j}}^{ba}} \notag \\ & -{\frac{6{g_{D}}{\gamma}}{5}}{R_{j{\eta}{\alpha}{\beta}}^{db}}{Q_{{\beta}j}^{bd}}-{\frac{6{g_{D}}{\gamma}}{5}}{V_{\zeta}^{d(0)}}{V_{\gamma}^{b(0)}}{\epsilon_{j{\eta}{\zeta}}}{\epsilon_{{\alpha}{\beta}{\gamma}}}Q_{{\beta}{j}}^{bd}, 
\label{XiAndC}
\end{align}
where
\begin{align}
R_{j{\eta}{\alpha}{\beta}}^{db} & = {V_{j}^{d(0)}}{V_{{\beta}}^{b(0)}}{\delta_{{\eta}{\alpha}}}-{V_{j}^{d(0)}}{V_{{\alpha}}^{b(0)}}{\delta_{{\eta}{\beta}}}, \notag \\
G_{ij{\alpha}{\gamma}}^{ba} & = ({\partial_{i}}{\partial_{j}}{V_{{\alpha}}^{b(0)}}){V_{{\gamma}}^{a(0)}}-({\partial_{i}}{\partial_{j}}{V_{{\beta}}^{b(0)}}){\delta_{{\beta}{\gamma}}}{V_{{\alpha}}^{a(0)}}, \notag \\
{\Lambda_{ij{\alpha}{\lambda}}^{ba}} & = ({\partial_{i}}{\partial_{j}}{V_{{\beta}}^{b(0)}}){{\delta}_{{\beta}{\lambda}}}{V_{{\alpha}}^{a(0)}} \notag \\ & + ({V_{{\gamma}}^{b(0)}}{V_{{\gamma}}^{a(0)}}{\delta_{{\alpha}{\lambda}}}-{\delta_{{\gamma}{\lambda}}}{V_{{\alpha}}^{b(0)}}{V_{{\gamma}}^{a(0)}}){\partial_{i}}{\partial_{j}} \notag \\ & + [({\partial_{i}}{V_{{\gamma}}^{b(0)}}){V_{{\gamma}}^{a(0)}}{\delta_{{\alpha}{\lambda}}}-({\partial_{i}}{V_{{\alpha}}^{b(0)}}){V_{{\gamma}}^{a(0)}}{\delta_{{\gamma}{\lambda}}}]{\partial_{j}} \notag \\ 
& + [({\partial_{j}}{V_{{\gamma}}^{b(0)}}){V_{{\gamma}}^{a(0)}}{\delta_{{\alpha}{\lambda}}}-({\partial_{j}}{V_{{\alpha}}^{b(0)}}){V_{{\gamma}}^{a(0)}}{\delta_{{\gamma}{\lambda}}}]{\partial}_{i} \notag \\ 
& - {\delta_{{\gamma}{\lambda}}({\partial_{i}}{\partial_{j}}{V_{{\alpha}}^{b(0)}}){V_{{\gamma}}^{a(0)}}}. 
\label{RGandLambda}
\end{align}
The first two terms of Eq.~(\ref{2ndOderEquations}) correspond to the NMR response of Larmor precession of $\delta{S_{\alpha}}$ with frequency $\omega_{L}=\gamma H^{(0)}$. While the last two terms of Eq.~(\ref{2ndOderEquations}) induce the NMR frequency shift. From Eq.~(\ref{XiAndC}) and Eq.~(\ref{RGandLambda}), we found all the NMR frequency shifts are induced by the equilibrium textures of spin vectors. In our case with pseudo-random lattices of $2/4$ spin solitons, the NMR frequency shifts are totally induced by equilibrium textures of spin solitons in 1D nexus objects. That's why the transverse NMR spectrum is perfect tool to observe the network of 1D nexus objects and network of KLS string wall.
Taking into account the static magnetic field $\mathbf{H}^{(0)}=|\mathbf{H}^{(0)}|{\hat{\mathbf{y}}}$ and the parametrization Eq.~(\ref{PARA1}), we can derive the dynamic response equations of transverse spin density ${\delta}S_{+}=[{\delta}S_{1}(\omega)+i{\delta}S_{3}(\omega)]/{\sqrt{2}}$ under weak magnetic drive $\delta{\mathbf{H}(t)}$, see the detail of calculation in appendices Sec.~\ref{TransverseNMRResponseEquation}. This calculation gives
\begin{align}
({\omega}^{2} & -{\omega}_{L}^{2}) {\delta}S_{+}(\omega) \notag \\ & =  ({\Xi}_{11}+{\Xi}_{33}){\delta}S_{+}(\omega)+i({\Xi}_{13}-{\Xi}_{31}){\delta}S_{+}(\omega) \notag \\ 
& -[{\frac{1}{2}}({C_{11}}+{C_{31}})-{\frac{{\chi}_{\perp}}{\sqrt{2}{\gamma}}}({\Xi_{33}}+i{\Xi_{13}}-i{\Xi_{31}})]{\delta}{H_{1}}(\omega).
\label{TransverseResponseEquation}
\end{align}
Thus 
\begin{equation}
\frac{{\delta}S_{+}(\omega)}{{\delta}H_{1}(\omega)} \propto \frac{1}{{\omega}^{2}-{\omega}_{L}^{2}-({\Xi}_{11}+{\Xi}_{33})-i({\Xi}_{13}-{\Xi}_{31})}.
\label{ResponsFunction}
\end{equation}
The poles of spin dynamic response function ${\delta}S_{+}/{\delta}H_{1}$, which are determined by eigenequation
\begin{equation}
({\omega}^{2}-{\omega}_{L}^{2}){\delta}S_{+}(\omega) =({\Xi}_{11}+{\Xi}_{33})+i({\Xi}_{13}-{\Xi}_{31}){\delta}S_{+}(\omega),
\label{NMREigenEquation}
\end{equation} 
correspond to the eigenmodes of transverse NMR spectrum in the presence of pseudo-random lattices of spin solitons. We numerically solve this eigenequation in next subsection with different $D$ and $|q|$.

\begin{figure*}
\centerline{\includegraphics[width=0.89\linewidth]{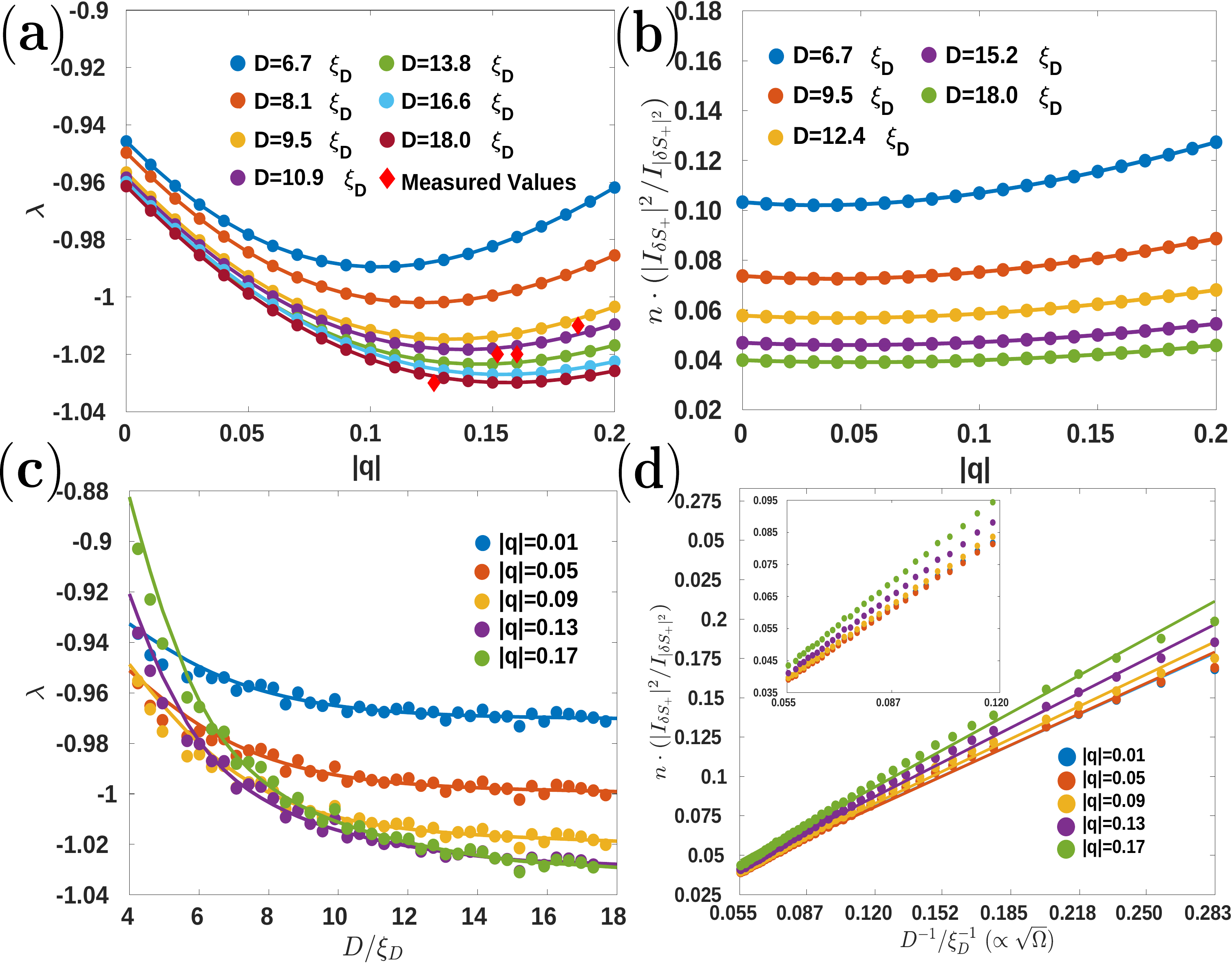}}
\caption{Transverse NMR frequency shifts $\lambda$ and surface densities of ratio intensity $n \cdot (|I_{\delta{S_{+}}}|^{2}/I_{|\delta{S_{+}}|^{2}})$ of pseudo-random lattices consisting of inseparable spin solitons ($\pi$-solitons). The frequency shifts $\lambda$ are eigenvalues of Eq.~(\ref{NMREigenEquationDimensonless}) with equilibrium textures of $\pi$-solitons in London limit. The surface densities of NMR ratio intensity are calculated by using Eq.~(\ref{DensityOfRatioIntensity}). All scattering dots represent the original numeric results, while colored lines are smoothing spline fittings of these original numeric results. (a) Transverse NMR frequency shifts $\lambda$ as functions of $|q|$ with different $D$. For large enough unit cells ($D>10\xi_{D}$), we found $\lambda$ decreases when $|q|$ increases as long as $|q|\leq0.16$. The typical values of $\lambda$ are around $-1.015$ to $-1.03$ when pseudo-random lattices model is good enough. This exactly coincides with the region of $\lambda$ which was observed in experiment of Ref.~\onlinecite{Makinen2019}, as shown via red diamonds. (b) depicts the surface densities of ratio intensity $n \cdot (|I_{\delta{S_{+}}}|^{2}/I_{|\delta{S_{+}}|^{2}})$ of the eigenmodes. (c) depicts the transverse frequency shifts $\lambda$ as function of $D$. (d) The surface densities of ratio intensity $n \cdot (|I_{\delta{S_{+}}}|^{2}/I_{|\delta{S_{+}}|^{2}})$ as function of $1/D \propto \sqrt{\Omega}$. We found $n \cdot (|I_{\delta{S_{+}}}|^{2}/I_{|\delta{S_{+}}|^{2}})$ increases linearly if $\sqrt{\Omega}$ increases. This coincides with the results of experimental observation in Ref.~\onlinecite{Makinen2019}. The inset is the magnified plot between $1/D=0.055$ till $1/D=0.120$. \label{NMRLambdaAndRatioIntensityInseparableSoliton}}   
\end{figure*}
\subsection{NMR frequency shifts and ratio intensities of pseudo-random lattices consisting of inseparable and separable $2/4$ spin solitons}
First we transform the eigenequation of transverse NMR modes into dimensionless form, which is suitable for the numeric calculation. All $\Xi_{{\alpha}{\lambda}}$ operators in Eq.~(\ref{NMREigenEquation}) must be calculated with prarametrization Eq.~(\ref{PARA1}), See the details in appendices Sec. \ref{SimplifyTheNMREigenEquation}. This gives
\begin{align}
\lambda {\delta}S_{+}(\omega) = & \xi^{2}_{D}[(6\rho_{2}^{2}+\rho_{1}^{2} + 1)\partial_{y}\partial_{y} + (3\rho_{1}^{2}+2 \rho_{2}^{2} + 1)\partial_{x}\partial_{x} \notag \\ 
& -2iV]\delta{S_{+}}(\omega) + U\delta{S_{+}}(\omega)
\label{NMREigenEquationDimensonless}
\end{align}
with
\begin{align}
V = & (1+3{\rho_{1}^{2}}cos2\theta){\partial_{x}}\theta{\partial_{x}} + (1+\rho_{1}^{2}){\partial_{y}}\theta{\partial_{y}} \notag \\ 
& +\frac{1}{2 \xi_{D}^{2}} [(1 + \rho_{1})^{2}sin2{\theta} - (1 + \rho_{1})\rho_{2}cos{\theta}], \\
U = & (1+\rho_{1})[-(1+\rho_{1})cos2\theta-5\rho_{2}sin\theta]+1+\rho_{1}^{2}+4\rho_{2}^{2},
\end{align}
where $\rho_{1}=\Delta_{\bot1}/\Delta_{P}$ and  $\rho_{2}=\Delta_{\bot2}/\Delta_{P}$. Here the dimensionless eigenvalue 
\begin{equation}
\lambda=\frac{(\omega^{2}-\omega_{L}^{2})}{\tilde{\Omega}^{2}}
\end{equation}
is the transverse NMR frequency shift under weak magnetic drive and
\begin{equation}
\tilde{\Omega}^{2}=(\frac{5\chi_{\bot}}{6\gamma^{2}{\Delta_{P}^{2}}g_{D}})^{-1}.
\end{equation}
We use the Galerkin strategy under finite-element partition to solve Eq.~(\ref{NMREigenEquationDimensonless}) \cite{ciarlet1978}. The solving regions are the unit cells of pseudo-random lattices of spin solitons. The equilibrium spin textures of pseudo-random lattices of inseparable and separable spin solitons, which we got in Sec. \ref{EquilibriumTexturesOfInseparableAndSeparableSolitons}, are directly used to solve Eq.~(\ref{NMREigenEquationDimensonless}). Because $\delta{H_{\alpha}}$ is low energy drive, we just consider the spin dynamic response mode with the lowest $\lambda$ of Eq.~(\ref{NMREigenEquationDimensonless}). Moreover, the ratio intensity of NMR signal is other observable besides the frequency shift $\lambda$. The surface density of ratio intensity which is generated by unit area of pseudo-random lattices of spin solitons is 
\begin{equation}
n \cdot \frac{|I_{\delta{S_{+}}}|^{2}}{I_{|\delta{S_{+}}|^{2}}}=n \cdot \frac{|\int_{\sigma}\delta{S_{+}}d{\sigma}|^{2}}{\int_{\sigma}|\delta{S_{+}}|^{2}d{\sigma}},
\label{DensityOfRatioIntensity}
\end{equation} 
where $n=D^{-2}$ is the density of 1D nexuses and $\sigma$ is area of  one-half of unit cell of pseudo-random lattice. 

\begin{figure*}
\centerline{\includegraphics[width=0.9\linewidth]{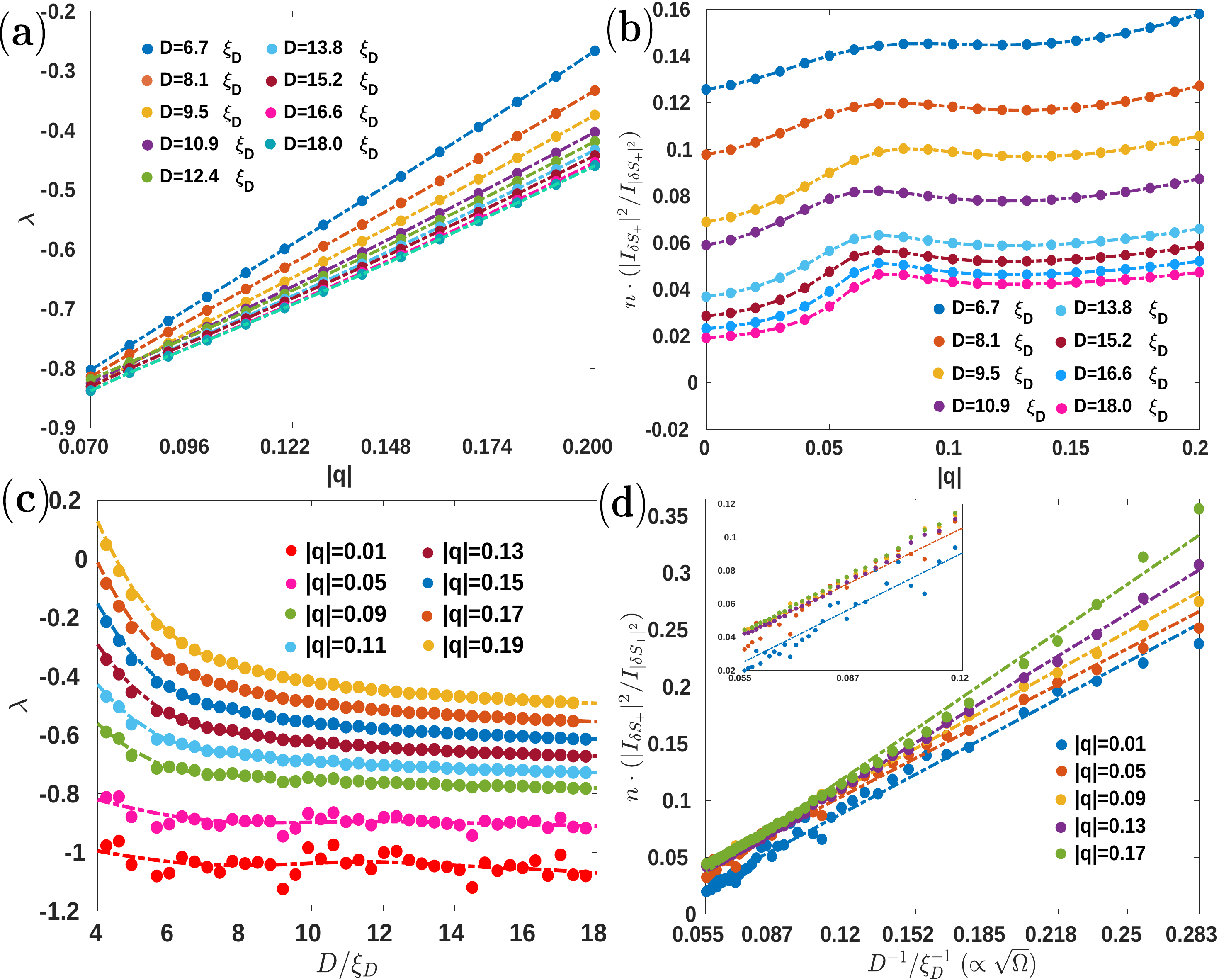}}
\caption{Transverse NMR frequencies shifts $\lambda$ and surface densities of ratio intensity $n \cdot (|I_{\delta{S_{+}}}|^{2}/I_{|\delta{S_{+}}|^{2}})$ of pseudo-random lattices consisting of separable spin solitons (KLS-solitons and solitons). The frequency shifts $\lambda$ are eigenvalues of Eq.~(\ref{NMREigenEquationDimensonless}) with equilibrium textures of $1/4+1/4$ spin solitons in London limit. The surface densities of NMR ratio intensity are calculated by using Eq.~(\ref{DensityOfRatioIntensity}). All scattering dots represent the original numeric results, while colored lines are smoothing spline fittings of these original numeric results. (a) Transverse NMR frequency shifts $\lambda$  increases when $|q|$ increases. This is because only solitons ($|\Delta{\theta}|=\pi-2\theta_{0}$) contribute to the lowest transverse spin dynamic response mode. When $|q|$ increases, $\lambda$ generated by solitons increases, see the details in appendices Sec. \ref{SpinDynamicsSolionAndBigSoliton}. The typical values of $\lambda$ are larger than $-0.9$ when $D\geq10\xi_{D}$. (b) depicts the surface densities of ratio intensity $n \cdot (|I_{\delta{S_{+}}}|^{2}/I_{|\delta{S_{+}}|^{2}})$ of the eigenmodes. (c) depicts the transverse frequency shifts $\lambda$ as function of $D$. (d) The surface densities of ratio intensity $n \cdot (|I_{\delta{S_{+}}}|^{2}/I_{|\delta{S_{+}}|^{2}})$ as function of $1/D \propto \sqrt{\Omega}$. The inset is the magnified plot between $1/D=0.055$ till $1/D=0.120$. \label{NMRLambdaAndRatioIntensitySeparableSoliton}}
\end{figure*}
In Fig.~\ref{SpinDynamicResponseModes}, we demonstrate the modulus of the lowest transverse spin dynamic response modes $|\delta{S_{+}}(\omega)|$ located in the unit cells of pseudo-random lattices of inseparable and separable $2/4$ spin solitons. In the unit cell of pseudo-random lattice consisting of inseparable $2/4$ spin soliton, the lowest spin dynamic response mode locates on the region which is occupied by $\pi$-soliton. While, in the unit cell of pseudo-random lattice consisting of separable $2/4$ spin soliton, the lowest spin dynamic response mode locates on the region which is occupied by soliton ($|\Delta{\theta}|=\pi-2\theta_{0}$). This means the KLS-soliton in the separable spin soliton does not respond the continuous-wave magnetic drive. The transverse NMR frequency shifts $\lambda$ and surface densities of ratio intensity of pseudo-random lattices for inseparable and separable $2/4$ spin solitons are shown in Fig.~\ref{NMRLambdaAndRatioIntensityInseparableSoliton} and Fig.~\ref{NMRLambdaAndRatioIntensitySeparableSoliton} respectively. Let us consider them separately.   

\subsubsection{\label{EigenModeOfInSeparableSoliton}Transverse NMR frequency shifts and surface densities of ratio intensity of pseudo-random lattices consisting of inseparable spin solitons}
The transverse NMR frequency shifts $\lambda$ of pseudo-random lattices of inseparable spin solitons ($\pi$-solitons) exactly coincide with the experimentally observed values in Ref.~\onlinecite{Makinen2019}. As been shown in Fig.~\ref{NMRLambdaAndRatioIntensityInseparableSoliton} (a), the numeric values of $\lambda$ generated by pseudo-random lattices consisting of $\pi$-solitons is around $-1.01$ to $-1.03$ when pseudo-random lattice model is good enough i.e., $D\geq10\xi_{D}$. In this case, the transverse NMR frequency shifts $\lambda$ slightly increase as $|q|$ increasing when $|q|>0.16$. This phenomenon has also been observed in experiment of Ref.~\onlinecite{Makinen2019}. The ratio intensities generated by unit area of pseudo-random lattice consisting of $\pi$-solitons linearly increase when the square root of angular velocity $\sqrt{\Omega} \propto 1/D$ increases, as shown in Fig.~\ref{NMRLambdaAndRatioIntensityInseparableSoliton} (d). This coincides with the $\sqrt{\Omega}$-scaling of satellite intensity observed in the experiment when $T=0.38T_{c}$ ($|q|\approx0.152$) \cite{Makinen2019}. 

In Sec. \ref{EquilibriumSpinTexturesAndFreeEnergy}, we suggested the possible equilibrium state which was observed in experiment is the pseudo-random lattices of $2/4$ inseparable spin solitons of the network of 1D nexus objects. Here we see the results of numeric simulations of transverse NMR spin dynamic response  of this kind of pseudo-random lattices indeed coincide with the experimental observations.   

\subsubsection{\label{EigenModeOfSeparableSoliton}Transverse NMR frequency shifts and surface densities of ratio intensity of pseudo-random lattices consisting of separable spin solitons}
In contrast with pseudo-random lattices consisting of inseparable $2/4$ spin solitons, the transverse NMR frequency shifts of pseudo-random lattices consisting of separable spin solitons strongly deviate from the results of experimental observations, see Fig.~\ref{NMRLambdaAndRatioIntensitySeparableSoliton} (a). $\lambda$ generated by pseudo-random lattice of separable spin solitons increase when $|q|$ increases. This is because only the solitons ($|\Delta{\theta}|=\pi-2\theta_{0}$) of separable spin solitons contribute to the transverse NMR frequency shift, and the frequency shifts $\lambda$ of the soliton ($|\Delta{\theta}|=\pi-2\theta_{0}$) increase as $|q|$ increases, see the details in appendices Sec. \ref{SpinDynamicsSolionAndBigSoliton}. Moreover, the magnitudes of the surface densities of ratio intensity $n \cdot (|I_{\delta{S_{+}}}|^{2}/I_{\delta{S_{+}}^{2}})$ generated by pseudo-random lattices of separable spin solitons are larger than those generated by pseudo-random lattices consisting of inseparable spin solitons, as shown in Fig.~\ref{NMRLambdaAndRatioIntensitySeparableSoliton} (d).

\begin{figure*}
\centerline{\includegraphics[width=0.9\linewidth]{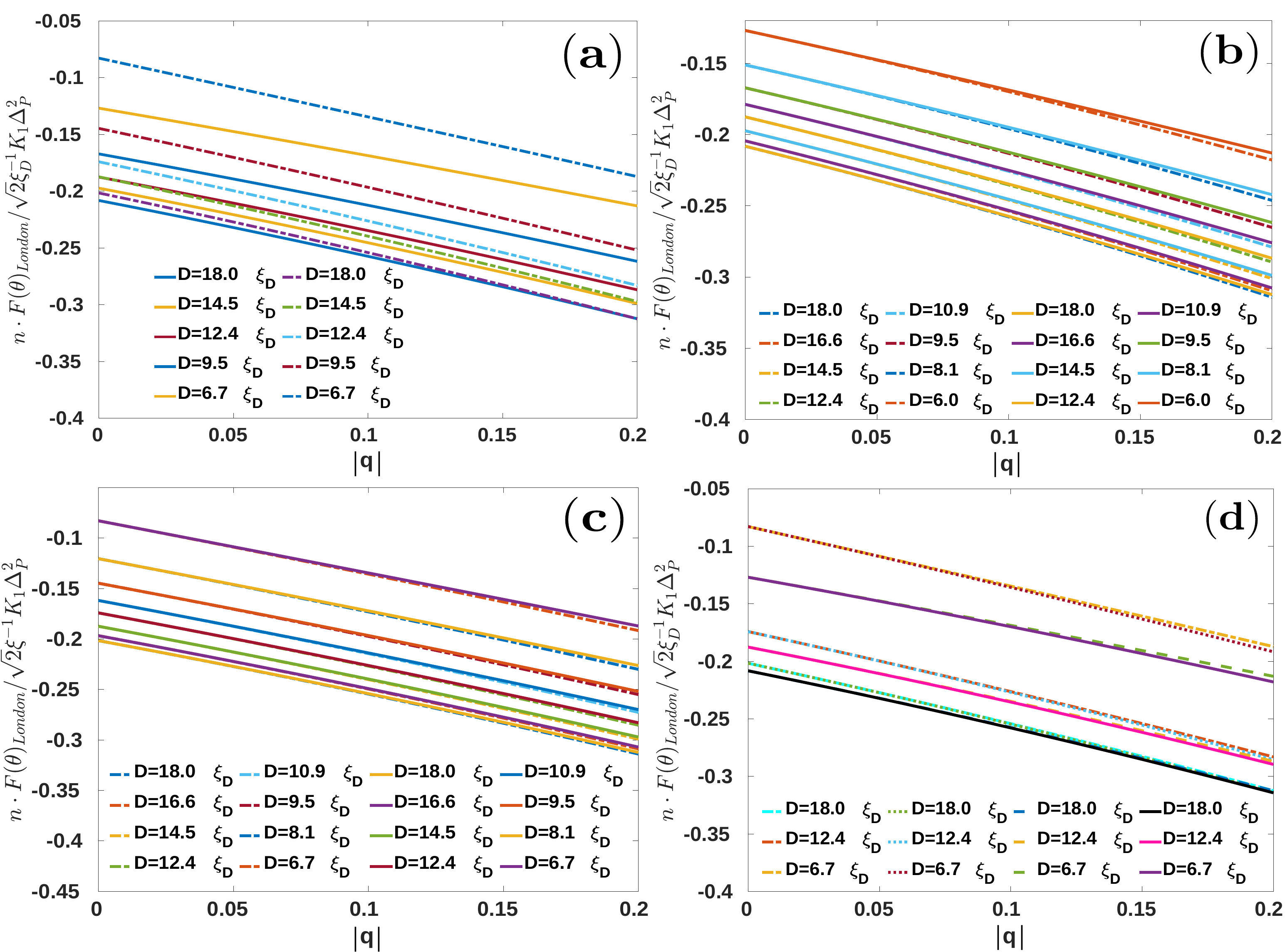}}
\caption{Surface densities of equilibrium London limit free energies of pseudo-random lattices consisting of $2/4$ spin solitons within vacuum state (i) and vacuum state (ii) respective. These two vacuum states are not degenerated with same free energy any more when the direction of KLS domain wall is fixed. (a) The surface densities of London limit free energies of pseudo-random lattices consisting of inseparable spin solitons ($\pi$-solitons) and separable spin solitons (KLS-solitons and solitons) in the vacuum state (ii). This figure shows similar features with Fig.~\ref{FreeEnergyBoxAndDensity}(c). (b) depicts the surface densities of London limit free energies of pseudo-random lattices consisting of inseparable $2/4$ spin solitons of vacuum states (i) and (ii) respectively. The dash-dot lines represent the vacuum state (i) while the solid lines represent the vacuum state (ii). (c) depicts the surface densities of London limit free energies of pseudo-random lattices consisting of separable $2/4$ spin solitons of vacuum states (i) and (ii) respectively. The dash-dot lines represent the vacuum state (i) while the solid lines represent the vacuum state (ii). (d) depicts the surface densities of London limit free energies of pseudo-random lattices in vacuum states (i) and (ii). The solid lines and dash lines represent the pseudo-random lattices consisting of inseparable spin solitons ($\pi$-solitons) within vacuum states (i) and (ii) respectively. The dot lines and dash-dot lines represent the pseudo-random lattices consisting of separable spin solitons (KLS-solitons and solitons) within vacuum states (i) and (ii) respectively. We found the pseudo-random lattices consisting of $2/4$ inseparable spin solitons within vacuum state (i) have lowest equilibrium free energies. \label{DensityOfLondonLimitFreeEnergyPARA2}}
\end{figure*}
\section{\label{DiscreteSymmetry}The mirror symmetry in the presence of KLS domain wall and its breaking} 
As we mentioned before, The London limit free energy $F(\theta)_{London}$ has a mirror symmetry when the coordinates are permuted to each other i.e., $F[\theta(x',y')] = F[\theta(x,y)]$ with $x' = y$ and $y' = x$. This mirror symmetry does not vanish even in the presence of 1D nexus object. As a result, the spin textures of $2/4$ spin solitons have this mirror symmetry as well.

This discrete symmetry originates from the reduction of degenerate space of order parameter by requirement of continuity of order parameter in the presence of KLS domain wall. In order to understand this, we start from the degenerate manifold of PdB which generates from symmetry breaking transition of polar phase vacuum. In this case, $R_{PdB} \cong SO(2)_{S-L} \times \mathbb{Z}_{2}^{S-\Phi}$, in which the nontrivial element of $\mathbb{Z}_{2}^{S-\Phi}$ corresponds to the presence of KLS domain wall \cite{volovik2020}. In the presence of KLS domain wall, the requirement of continuity of order parameter reduces the degenerate space of $\hat{\mathbf{e}}^{1}$ and $\hat{\mathbf{e}}^{2}$ on both sides of domian wall  from $SO(2)_{S-L}$ to (i) $\hat{\mathbf{e}}^{1}$ $\longrightarrow$ $-\hat{\mathbf{e}}^{1}$, while $\hat{\mathbf{e}}^{2}$ keeps its direction and (ii) $\hat{\mathbf{e}}^{2}$ $\longrightarrow$ $-\hat{\mathbf{e}}^{2}$, while $\hat{\mathbf{e}}^{1}$ keeps its direction. The parametrization in Eq.~(\ref{PARA1}), which we used in previous calculations and discussions, corresponds to the vacuum state (i) and the direction of static magnetic field $\mathbf{H}^{(0)}$ is set to parallel with the $\hat{\mathbf{e}}^{2}$. Because the vacuum state (ii) is another possible vacuum state with same free energy of case (i) in the presence of KLS domain wall, the London limit free energy $F(\theta)_{London}$ is invariant when we transform from vacuum state (i) to vacuum state (ii). In our case, the parametrization of vacuum state (ii) is
\begin{align}
\mathbf{\hat{d}} & =\hat{y}cos\theta-\hat{z}sin\theta, \notag \\  
\mathbf{\hat{e}}^{2} & =-\hat{y}sin\theta-\hat{z}cos\theta, \label{PARA2} \\ 
\mathbf{\hat{e}}^{1} & =\hat{x},\,\, \mathbf{H}^{(0)}=H\hat{x}, \notag
\end{align}
and the corresponding dimensionless London limit free energy is 
\begin{align}
\tilde{F}(\theta)_{London}  = 
& {\frac{1}{\xi_{D}}}\int\nolimits_{\Sigma} [\frac{1}{2} (\gamma_{1}+2\gamma_{2}) \partial_{y}\theta \partial_{y}\theta \notag  + \frac{1}{2}\gamma_{1}\partial_{x}\theta \partial_{x}\theta \notag \\ 
& + \frac{1}{\xi_{D}^{2}} 
(-\frac{1}{2}{\gamma_{4}}cos2\theta-\gamma_{3}sin\theta)] d\Sigma \label{FreeEnergyThetaDimensionlessPARA2}
\end{align}
where 
\begin{align}
q & = \frac{\Delta_{\bot1}}{\Delta_{P}},\,\,
 \gamma_{1}=1+|q|^{2},\,\, 
\gamma_{2}=|q|^{2},\,\, \notag \\
& \gamma_{3}=q(1+|q|),\,\,
\gamma_{4}=(1+|q|)^{2},
\end{align}
Comparing Eq.~(\ref{FreeEnergyThetaDimensionlessPARA2}) and Eq.~(\ref{FreeEnergyThetaDimensionless}), we can see the mirror symmetry.

However, this discrete symmetry may be destroyed if the direction of domain wall is fixed in both vacuum states (i) and (ii). In this case, the term containing $\gamma_{3}$ in Eq.~(\ref{FreeEnergyThetaDimensionlessPARA2}) is invariant for both parametrizations, and thus violates this mirror symmetry. As a result, the equilibrium sates of Eq.~(\ref{FreeEnergyThetaDimensionlessPARA2}) and Eq.~(\ref{FreeEnergyThetaDimensionless}) are not identical any more. Then we need to check the equilibrium London limit free energy of these two different equilibrium states. We did the same numeric minimizations of London limit free energy with parameteization Eq.~(\ref{PARA2}) and calculated the surface densities of equilibrium free energies of pseudo-random lattices in vacuum state (ii). The latter can be evaluated as
\begin{align}
n \cdot \tilde{F}(\theta)_{London}|_{(ii)}\, & \sim \,\frac{1}{8}(1+|q|^{2})\frac{\pi^{2}}{\xi_{D}D}  + \frac{\pi^{2}}{4\xi_{D}D}|q|^{2} \notag \\
 & + \, \frac{1}{2} [\tilde{f}_{soc}|_{y>0} \,\, + \tilde{f}_{soc}|_{y<0}]_{q}  \notag \\ 
 & \sim n \cdot \tilde{F}(\theta)_{London}|_{(i)}\, + \frac{\pi^{2}}{4\xi_{D}D}|q|^{2}. \label{FreeEnergyDensityPARA2}
\end{align}
Then we can expect the surface densities of equilibrium London limit free energy of vacuum state (ii) are slightly higher than those of vacuum state (i) when $|q|\leq0.2$. In Fig.~\ref{DensityOfLondonLimitFreeEnergyPARA2}, we show this  
for pseudo-random lattices consisting of inseparable and separable $2/4$ spin solitons respectively. In all cases, the surface densities of London limit free energy of vacuum state (ii) are indeed higher than those of vacuum state (i). 

\section{\label{ConclusionAndDiscussions}conclusions and discussions}
In this work, we discussed the topological origin of the novel 1D nexus objects in PdB phase of nafen-distorted Helium-3 superfluid system. The topological objects named 2D nexus objects which are similar but has higher spatial dimension were predicted in PdB superfluid \cite{volovik2020}. This object is formed by connection between vortex skyrmion of $\hat{\mathbf{d}}$ vector and spin vortices of $\hat{\mathbf{e}}^{1}$ and $\hat{\mathbf{e}}^{2}$ vectors via monopole. Earlier vortex skyrmions formed by phase and orbital degenerate parameters have been suggested and observed in Helium-3 A-phase \cite{Anderson1977,Chechetkin1976,Volovik1977,Seppala1984,Pekola1990}, it also probably be observed in spin and orbital degree of freedom in PdB phase. In contrast to the out of observation of 2D nexus objects, the 1D nexus objects are observed directly in the continuous wave NMR spectrum of the rotating PdB sample \cite{Makinen2019}. There are two reasons making this possible. 

One reason is the pinning effect of nafen-strands. This strong pinning fixes the locations of the HQVs once they appear during cooling down with a given angular velocity. In the limit of low angular velocity i.e., $\Omega \ll \Omega_{c}$, the average distance between pinned HQVs is around hundred microns. As a result, the KLS domain walls attached on the HQVs have very large geometric sizes when the symmetry breaking transition from polar phase to PdB phase occurs. In the spatial regions with length scales $\xi_{D}$, the SOC energy reduces the vacuum manifolds to discrete sets. The reduced vacuum manifolds have spin solitons, which are described by relative homotopy group $\pi_{1}(R_{1}^{H},\tilde{R}^{SOC}_{1})$. Similar process also happens in bulk Helium-3 superfluid and spinor Bose condensate \cite{Kondo1992,Seji2019,Liu2020}. We demonstrated the subgroup $G$ of $\pi_{1}(R_{1}^{H},\tilde{R}_{1}^{SOC})$, which describes the spin solitons with topological invariant $2/4$, is isomorphic to the group $M$ which describes the spin degree of freedom of KLS string wall. This suggests that HQV is 1D nexus which smoothly connect the spin soliton and KLS domain wall. 

The other reason is the textures of $2/4$ spin soltions with length scales $\xi_{D}$ can strongly influence the SOC energy and then modifies the low frequency spin dynamic response of the spin densities under continuous wave drive. This allows us to directly observe the network of KLS string walls, which has short characteristic lengths and high characteristic energies determined by $\Delta_{p}$ and $|q|$, through the easily controllable low energy spin dynamic process.  

In the nafen-distorted Helium-3 superfuid, the 1D nexus objects connect with each other via large size KLS domain wall and  then form network. The $2/4$ spin solitons connected on every KLS string wall form pseudo-random lattices in the absence of coupling between spin solitons. We discussed the equilibrium configurations and the surface densities of equilibrium free energies of two different pseudo-random lattices consisting of spin solitons with topological invariant $2/4$. These two types of pseudo-random lattices correspond to two representations of group $G = \pi_{1}(S_{S}^{1},\tilde{R}_{2})$, the relative homotopy group of $2/4$ spin solitons of 1D nexus objects. Our analysis shows the pseudo-random lattices consisting of inseparable spin solitons are energy favorable. To compare with the experimental observations, we calculated the transverse spin dynamic response under continuous wave drive. The resulted NMR frequency shifts of pseudo-random lattices consisting of inseparable spin solitons exactly coincide with the experimental measurements. The explicit breaking of mirror symmetry in the presence of KLS domain wall is also be considered. 

In the limit of low angular velocity, the pseudo-random lattices models work very well because the randomness of the network of 1D nexus objects doesn't influence the spin textures of spin solitons. Thus we can not find observable effect originated from this randomness. However, when the angular velocity approaches the critic value $\Omega_{c}$, the  coupling between spin solitons can dramatically change the equilibrium spin textures of random lattices of spin solitons. In this case, the random distributions of KLS string wall lead to spin solitons glasses \cite{volovik2019}. Thus we can expect the observable effects of this randomness on the NMR spectrums under high enough angular velocity. Moreover, PdB phase could be a good platform to observe the monopole-antimonople networks because the string monopole is topologically protected by $\pi_{2}$ relative homotopy group \cite{volovik2020}. These kinds of complex networks are predicted in condensed matter system and also in the Grand Unified Theories \cite{Kibble2015,Saurabh2019,Lazarides1980,Shafi2019,Volovik2019d}. The Grand Unified Theories may have a huge variety of networks consisting of monopoles and strings because of their complex symmetry breaking chains \cite{Chakrabortty2018,Chakrabortty2019}. In the absence of magnetic field, the string monopoles in PdB phase may connect to planar solitons with geometric size around $\xi_{D}$ because of the reduction of vacuum manifold by SOC energy. Similar with pseudo-random lattices of spin solitons, these planar solitons may result in observable influence on NMR spectrum.
\begin{acknowledgments}
We especially thank the instructive and inspiring discussions and comments from professor Grigory.~E. Volovik during the process of this work. We  thank Jaakko Nissinen, Vladislav Zavyalov and professor Erkki. V. Thuneberg for important discussions. We also thank professor Vladimir. B. Eltsov, Jere. T. Mäkinen and Juho. Rysti for instructive discussions about experiments of polar distorted B-phase.  This work has been supported by the European
Research  Council  (ERC)  under  the  European  Union’s
Horizon 2020 research and innovation programme (Grant Agreement No. 694248).
\end{acknowledgments}

\appendix
\begin{figure*}
\centerline{\includegraphics[width=1.0\linewidth]{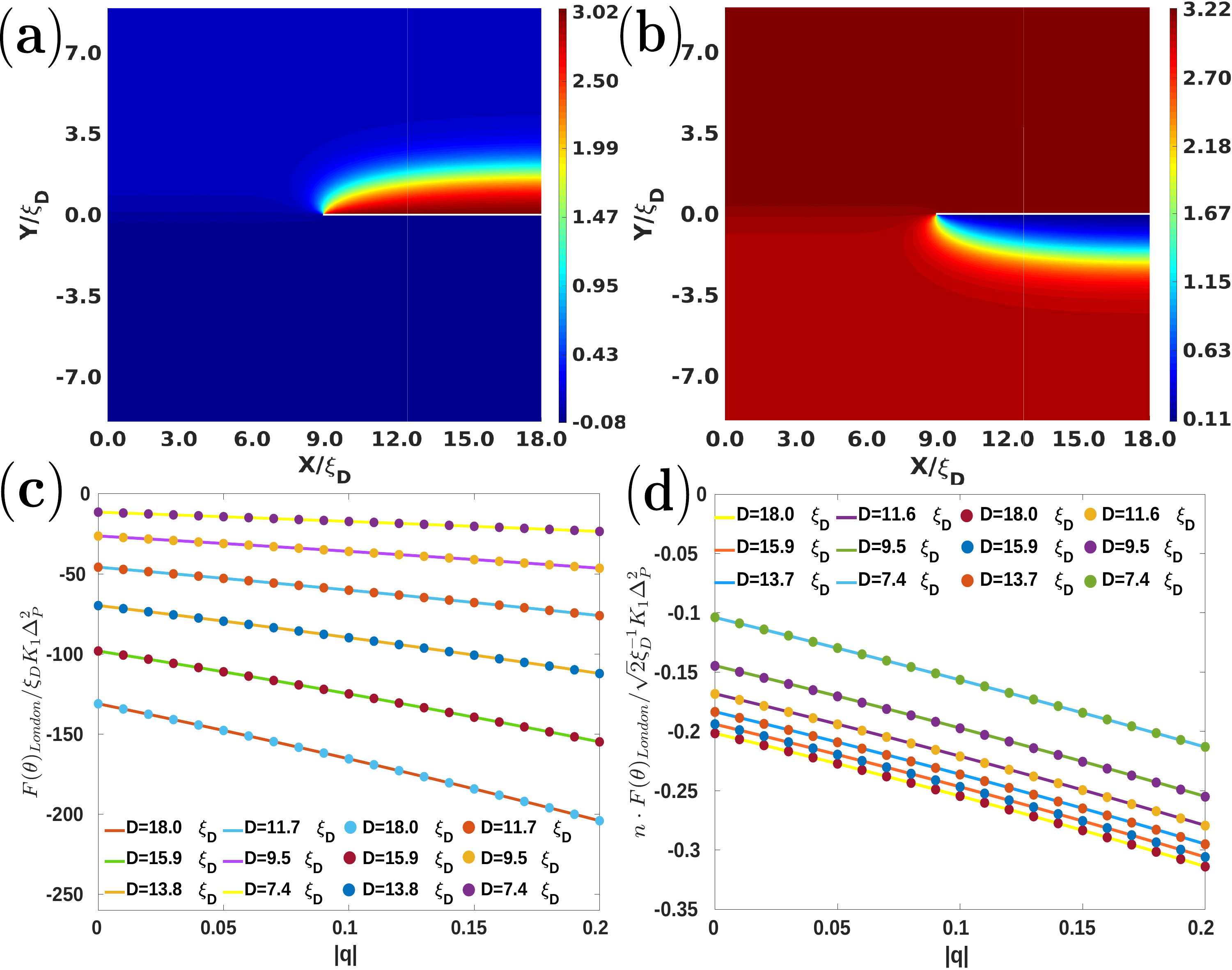}}
\caption{Equilibrium spin textures and equilibrium London limit free energies of one-half unit cell consisting of $1/4+1/4$ separable spin solitons with $\theta_{KLS}=0$ and $\theta_{KLS}=\pi$. The dots represent data of equilibrium configurations with $\theta_{KLS}=0$, while the solid lines represent data of equilibrium configurations with $\theta_{KLS}=\pi$. (a) is the equilibrium spin textures of one-half unit cell with $\theta_{KLS}=0$, $|q|=0.2$ and $D=18\xi_{D}$. (b) is the equilibrium spin textures of one-half unit cell with $\theta_{KLS}=\pi$, $|q|=0.2$ and $D=18\xi_{D}$. (a) and (b) have same $|\Delta{\theta}|=\pi$ and are related by a $\pi$-rotation around $\hat{x}$ axis. (c) depicts the equilibrium London limit free energies  of spin textures with $\theta_{KLS}=0$ and $\theta_{KLS}=\pi$ respectively. (d) depicts the surface densities  of equilibrium London limit free energies of spin textures with $\theta_{KLS}=0$ and $\theta_{KLS}=\pi$ respectively. (c) and (d) demonstrate the pseudo-random lattices consisting of separable spin solitons have same equilibrium London limit free energies for boundary conditions $\theta_{KLS}=0$ and $\theta_{KLS}=\pi$. \label{FiguresOfSeparableSolitonTwoBoundaryConditions}}
\end{figure*}

\section{\label{SeparableSolitionTwoBoundryConditions}Pseudo-random lattices consisting of separable spin solitons with two different domain wall boundary conditions -- $\theta_{KLS}=0$ and $\theta_{KLS} =\pi$}
To show the equivalence of equilibrium configurations of pseudo-random lattices between boundary conditions $\theta_{KLS}=0$ and $\theta_{KLS}=\pi$, we calculated the spin textures of one-half unit cell of the lattices consisting of separable spin solitons with topological invariant $1/4+1/4$ under these two boundary conditions. Based on the resulted equilibrium spin texture, the London limit free energies of one-half unit cell and the surface densities of London limit free energies of the pseudo-random lattices were calculated. In Fig.~\ref{FiguresOfSeparableSolitonTwoBoundaryConditions} (a) and (b), we show the equilibrium spin textures of one-half unit cell with $\theta_{KLS}=0$ and $\theta_{KLS}=\pi$ respectively. We can see these two textures are related by $\pi$-rotation about $x$-axis. They have same London limit free energies and same surface densities of London limit free energy as shown in Fig.~\ref{FiguresOfSeparableSolitonTwoBoundaryConditions} (c) and (d). These information show the equilibrium spin textures of pseudo-random lattices consisting of separable spin solitons with KLS wall boundary conditions $\theta_{KLS}=0$ and $\theta_{KLS}=\pi$ are identical.

\section{The derivation of spin dynamic response equations}
\subsection{\label{For1stOrderEqations} The derivation of the first order dynamic equations of spin densities and degenerate parameters}
Using Eq.~(\ref{GradientEnergy}), Eq.~(\ref{PdBOrderParameter}), Eq.~(\ref{MagneticEnergy}) and Eq.~(\ref{EnergyDensityOfSOC}), All terms of energy densities in hydrodynamic free energy $F_{hydrodynamics}$ are given as
\begin{widetext}
\begin{align}
f_{grad} = & \frac{1}{2}\{K_{1}{\Delta_{P}^{2}}{\partial_{i}{\hat{d}}_{\alpha}}{\partial_{i}{\hat{d}}_{\alpha}} + K_{1}{\Delta_{\bot1}^{2}}{\partial_{i}{\hat{e}}^{1}_{\alpha}}{\partial_{i}{\hat{e}}^{1}_{\alpha}} + K_{1}{\Delta_{\bot2}^{2}}{\partial_{i}{\hat{e}}^{2}_{\alpha}}{\partial_{i}{\hat{e}}^{2}_{\alpha}}  \notag \\ 
& + K_{2}{\Delta_{P}^{2}}{\partial_{j}{\hat{d}}_{\alpha}}{\hat{z}_{j}}{\partial_{i}{\hat{d}}_{\alpha}}{\hat{z}_{i}} + K_{2}{\Delta_{\bot1}^{2}}{\partial_{j}{\hat{e}^{1}}_{\alpha}}{\hat{x}_{j}}{\partial_{i}{\hat{e}^{1}}_{\alpha}}{\hat{x}_{i}} +  K_{2}{\Delta_{\bot2}^{2}}{\partial_{j}{\hat{e}^{2}}_{\alpha}}{\hat{y}_{j}}{\partial_{i}{\hat{e}^{2}}_{\alpha}}{\hat{y}_{i}} \notag \\
& + K_{3}{\Delta_{P}^{2}}{\partial_{i}{\hat{d}}_{\alpha}}{\hat{z}_{i}}{\partial_{j}{\hat{d}}_{\alpha}}{\hat{z}_{j}} + K_{3}{\Delta_{\bot1}^{2}}{\partial_{i}{\hat{e}^{1}}_{\alpha}}{\hat{x}_{i}}{\partial_{j}{\hat{e}^{1}}_{\alpha}}{\hat{y}_{j}} +  K_{3}{\Delta_{\bot2}^{2}}{\partial_{i}{\hat{e}^{2}}_{\alpha}}{\hat{y}_{i}}{\partial_{j}{\hat{e}^{2}}_{\alpha}}{\hat{y}_{j}} \notag \\
& + K_{2}[{\partial_{j}}{\hat{d}_{\alpha}}{\hat{z}_{i}}{\partial_{j}}{\hat{e}^{1}_{\alpha}}{\hat{x}_{j} + {\partial_{j}}{\hat{e}^{1}_{\alpha}}{\hat{x}_{i}}{\partial_{i}}{\hat{d}_{\alpha}}{\hat{z}_{j}}}]{\Delta_{P}}{\Delta_{\bot1}} + K_{2}[{\partial_{j}}{\hat{d}_{\alpha}}{\hat{z}_{i}}{\partial_{i}}{\hat{e}^{2}_{\alpha}}{\hat{y}_{j} + {\partial_{j}}{\hat{e}^{2}_{\alpha}}{\hat{y}_{i}}{\partial_{i}}{\hat{d}_{\alpha}}{\hat{z}_{j}}}]{\Delta_{P}}{\Delta_{\bot2}}  \\
& + K_{2}[{\partial_{j}}{\hat{e}^{1}_{\alpha}}{\hat{x}_{i}}{\partial_{i}}{\hat{e}^{2}_{\alpha}}{\hat{y}_{j} + {\partial_{j}}{\hat{e}^{2}_{\alpha}}{\hat{y}_{i}}{\partial_{i}}{\hat{e}^{1}_{\alpha}}{\hat{x}_{j}}}]{\Delta_{\bot1}}{\Delta_{\bot2}} + K_{3}[{\partial_{i}}{\hat{d}_{\alpha}}{\hat{z}_{i}}{\partial_{j}}{\hat{e}^{1}_{\alpha}}{\hat{x}_{j} + {\partial_{i}}{\hat{e}^{2}_{\alpha}}{\hat{x}_{i}}{\partial_{j}}{\hat{d}_{\alpha}}{\hat{z}_{j}}}]{\Delta_{P}}{\Delta_{\bot1}} \notag \\
& + K_{3}[{\partial_{i}}{\hat{d}_{\alpha}}{\hat{z}_{i}}{\partial_{j}}{\hat{e}^{2}_{\alpha}}{\hat{y}_{j} + {\partial_{i}}{\hat{e}^{2}_{\alpha}}{\hat{y}_{i}}{\partial_{j}}{\hat{d}_{\alpha}}{\hat{z}_{j}}}]{\Delta_{P}}{\Delta_{\bot2}} \notag + K_{3}[{\partial_{i}}{\hat{e}^{1}_{\alpha}}{\hat{x}_{i}}{\partial_{j}}{\hat{e}^{2}_{\alpha}}{\hat{y}_{j} + {\partial_{i}}{\hat{e}^{2}_{\alpha}}{\hat{y}_{i}}{\partial_{j}}{\hat{e}^{1}_{\alpha}}{\hat{x}_{j}}}]{\Delta_{\bot1}}{\Delta_{\bot2}}\}, \notag
\label{DensityOfGrad}
\end{align}
\begin{align}
f_{soc} = & \frac{3g_{D}}{5} \{{\Delta_{P}^{2}}({\hat{d}_{i}{\hat{z}_{i}}})^{2} + {\Delta_{P}}{\Delta_{\bot1}}({\hat{d}_{i}{\hat{z}_{i}}})({\hat{e}^{1}_{j}{\hat{x}_{j}}}) + {\Delta_{P}}{\Delta_{\bot2}}({\hat{d}_{i}{\hat{z}_{i}}})({\hat{e}^{2}_{j}{\hat{y}_{j}}}) + {\Delta_{P}}{\Delta_{\bot1}}({\hat{e}^{1}_{i}{\hat{x}_{i}}})({\hat{d}_{j}{\hat{z}_{j}}}) \notag \\
& + {\Delta_{\bot1}^{2}}({\hat{e}^{1}_{i}{\hat{x}_{i}}})^{2} +  {\Delta_{\bot1}}{\Delta_{\bot2}}({\hat{e}^{1}_{i}{\hat{x}_{i}}})({\hat{e}_{j}{\hat{y}_{j}}}) +  {\Delta_{\bot2}}{\Delta_{\bot1}}({\hat{e}^{2}_{i}{\hat{y}_{i}}})({\hat{d}_{j}{\hat{z}_{j}}}) + {\Delta_{\bot2}}{\Delta_{\bot1}}({\hat{e}^{2}_{i}{\hat{y}_{i}}})({\hat{e}^{1}_{j}{\hat{x}_{j}}}) + {\Delta_{\bot2}^{2}}({\hat{e}^{2}_{i}{\hat{y}_{i}}})^{2} \notag \\
& + {\Delta_{P}^{2}}({\hat{d}_{i}{\hat{z}_{i}}})^{2} + {\Delta_{P}}{\Delta_{\bot1}}({\hat{d}_{i}{\hat{x}_{i}}})({\hat{e}^{1}_{j}{\hat{z}_{j}}}) + {\Delta_{P}}{\Delta_{\bot2}}({\hat{d}_{i}{\hat{z}_{j}}})({\hat{e}^{2}_{j}{\hat{y}_{i}}}) + {\Delta_{P}}{\Delta_{\bot1}}({\hat{e}^{1}_{i}{\hat{x}_{j}}})({\hat{d}_{j}{\hat{z}_{i}}}) + {\Delta_{\bot1}^{2}}({\hat{e}^{1}_{i}{\hat{x}_{i}}})^{2}  \\ 
& +  {\Delta_{\bot1}}{\Delta_{\bot2}}({\hat{e}^{1}_{i}{\hat{x}_{j}}})({\hat{e}^{2}_{j}{\hat{y}_{j}}}) +  {\Delta_{\bot2}}{\Delta_{P}}({\hat{e}^{2}_{i}{\hat{y}_{j}}})({\hat{d}_{j}{\hat{z}_{i}}}) + {\Delta_{\bot1}}{\Delta_{\bot2}}({\hat{e}^{2}_{i}{\hat{y}_{j}}})({\hat{e}^{1}_{j}{\hat{x}_{i}}}) + {\Delta_{\bot2}^{2}}({\hat{e}^{2}_{j}{\hat{y}_{j}}})^{2} \notag \\
& - \frac{2}{3}(\Delta^{2}_{P} + \Delta^{2}_{\bot1} + \Delta^{2}_{\bot2})\}, \notag
\label{DensityOfSOC}
\end{align}
\begin{equation}
f_{H} = -{\gamma}H_{\beta}S_{\beta} + \frac{\gamma^{2}}{2}{\chi_{\bot}^{-1}}[({\hat{d}}_{\beta}S_{\beta})^{2}\delta + S_{\beta}S_{\beta}].
\label{DensityOfMagnatic}
\end{equation}
Then we have all of functional derivatives
\begin{equation}
\frac{\delta{F_{hydrodynamics}}}{\delta{S_{\beta}}}(\mathbf{r}') = \frac{\partial{f_{H}}}{\partial{S_{\beta}}}(\mathbf{r}') = -{\gamma}H_{\beta} + \gamma^{2}{\chi^{-1}_{\bot}}[(S_{\gamma}\hat{d}_{\gamma}){\hat{d}_{\beta}}{\delta} + S_{\beta}],
\label{FunctionalDerivaticveOfS}
\end{equation}
\begin{align}
\frac{\delta{F_{hydrodynamics}}}{\delta{{\hat{d}}_{\beta}}}(\mathbf{r}') = & \frac{\partial{f_{H}}}{\partial{{\hat{d}}_{\beta}}}(\mathbf{r}') + \frac{\partial{f_{soc}}}{\partial{{\hat{d}}_{\beta}}}(\mathbf{r}') - {\partial_{i}}\frac{f_{grad}}{{\partial}{\partial_{i}}{\hat{d}_{\beta}}}(\mathbf{r}') \notag \\
= & {\gamma^{2}}{\chi^{-1}_{\bot}}{\delta}S_{\gamma}{\hat{d}_{\gamma}}S_{\beta} \notag \\
& + \frac{3g_{D}}{5} ( 2{\Delta_{P}^{2}}{\hat{d}_{\gamma}}{\hat{z}_{\gamma}}{\hat{z}_{\beta}} + {\Delta_{P}}{\Delta_{\bot1}}{\hat{z}_{\beta}}{\hat{e}^{1}_{i}}{\hat{x}_{i}} + {\Delta_{P}}{\Delta_{\bot2}}{\hat{z}_{\beta}}{\hat{e}^{2}_{i}}{\hat{y}_{i}} + {\Delta_{P}}{\Delta_{\bot1}}{\hat{z}_{\beta}}{\hat{e}^{1}_{i}}{\hat{x}_{i}} + {\Delta_{P}}{\Delta_{\bot2}}{\hat{z}_{\beta}}{\hat{e}^{2}_{i}}{\hat{y}_{i}} \notag \\
& + 2{\Delta_{P}^{2}}{\hat{d}_{\gamma}}{\hat{z}_{\gamma}}{\hat{z}_{\beta}} + {\Delta_{P}}{\Delta_{\bot1}}{\hat{z}_{i}}{\hat{e}^{1}_{i}}{\hat{x}_{\beta}} + {\Delta_{P}}{\Delta_{\bot2}}{\hat{z}_{i}}{\hat{e}^{2}_{i}}{\hat{y}_{\beta}} + {\Delta_{P}}{\Delta_{\bot1}}{\hat{x}_{\beta}}{\hat{e}^{1}_{i}}{\hat{z}_{i}} + {\Delta_{P}}{\Delta_{\bot2}}{\hat{z}_{i}}{\hat{e}^{2}_{i}}{\hat{y}_{\beta}}) \label{FunctionalDerivaticveOfd} \\
& - \frac{1}{2} ( K_{1}{\Delta^{2}_{P}}2{\partial_{i}}{\partial_{i}}{\hat{d}_{\beta}} + K_{2}{\Delta^{2}_{P}}2{\partial_{i}}{\partial_{j}}{\hat{d}_{\beta}}{\hat{z}_{i}}{\hat{z}_{j}} + 2K_{2}{\Delta_{P}}{\Delta_{\bot1}}{\partial_{i}}{\partial_{j}}{\hat{e}^{1}_{\beta}}{\hat{z}_{j}}{\hat{x}_{i}} + 2K_{2}{\Delta_{P}}{\Delta_{\bot2}}{\partial_{i}}{\partial_{j}}{\hat{e}^{2}_{\beta}}{\hat{z}_{j}}{\hat{y}_{i}} \notag \\
& + 2K_{3}{\Delta^{2}_{P}}{\partial_{i}}{\partial_{j}}{\hat{d}_{\beta}}{\hat{z}_{j}}{\hat{z}_{i}} + 2K_{3}{\Delta_{P}}{\Delta_{\bot1}}{\hat{z}_{i}}{\partial_{i}}{\partial_{j}}{\hat{e}_{\beta}}{\hat{x}_{j}} + 2K_{3}{\Delta_{P}}{\Delta_{\bot1}}{\partial_{i}}{\partial_{j}}{\hat{e}^{2}_{\beta}}{\hat{y}_{j}}{\hat{z}_{i}}), \notag
\end{align}
\begin{align}
\frac{\delta{F_{hydrodynamics}}}{\delta{{\hat{e}^{1}}_{\beta}}}(\mathbf{r}') = & \frac{\partial{f_{soc}}}{\partial{{\hat{e}^{1}}_{\beta}}}(\mathbf{r}') - {\partial_{i}}\frac{f_{grad}}{{\partial}{\partial_{i}}{\hat{e}^{1}_{\beta}}}(\mathbf{r}') \notag \\
= & \frac{3g_{D}}{5}({\Delta_{P}}{\Delta_{\bot1}}{\hat{d}_{\gamma}}{\hat{z}_{\gamma}}{\hat{x}_{\beta}} + {\Delta_{P}}{\Delta_{\bot1}}{\hat{d}_{\gamma}}{\hat{z}_{\gamma}}{\hat{x}_{\beta}} + 2{\Delta^{2}_{\bot1}}{\hat{e}^{1}_{\gamma}}{\hat{x}_{\gamma}}{\hat{x}_{\beta}} \notag \\
& + {\Delta_{\bot1}}{\Delta_{\bot2}}{\hat{e}^{2}_{\gamma}}{\hat{y}_{\gamma}}{\hat{x}_{\beta}} + {\Delta_{P}}{\Delta_{\bot1}}{\hat{d}_{\gamma}}{\hat{x}_{\gamma}}{\hat{z}_{\beta}} + {\Delta_{P}}{\Delta_{\bot1}}{\hat{d}_{\gamma}}{\hat{x}_{\gamma}}{\hat{z}_{\beta}} \label{FunctionalDerivaticveOfe1} \\
& + {\Delta_{\bot1}^{2}}2{\hat{e}^{1}_{\gamma}}{\hat{x}_{\gamma}}{\hat{x}_{\beta}} + {\Delta_{\bot1}}{\Delta_{\bot2}}{\hat{e}^{2}_{\gamma}}{\hat{x}_{\gamma}}{\hat{y}_{\beta}} \notag + {\Delta_{\bot2}}{\Delta_{\bot1}}{\hat{e}^{2}_{\gamma}}{\hat{y}_{\gamma}}{\hat{x}_{\beta}} + {\Delta_{\bot2}}{\Delta_{\bot1}}{\hat{e}^{2}_{\gamma}}{\hat{x}_{\gamma}}{\hat{y}_{\beta}}) \notag  \\
& - \frac{1}{2}(2K_{1}{\Delta^{2}_{\bot1}}{\partial_{i}}{\partial_{i}}{\hat{e}^{1}_{\beta}}{\hat{x}_{j}}{\hat{x}_{j}} + 2K_{2}{\Delta^{2}_{\bot1}}{\partial_{i}}{\partial_{j}}{\hat{e}^{1}_{\beta}}{\hat{x}_{j}}{\hat{x}_{i}} + 2K_{2}{\Delta_{P}}{\Delta_{\bot1}}{\partial_{i}}{\partial_{j}}{\hat{d}_{\beta}}{\hat{x}_{j}}{\hat{z}_{i}} \notag  \\
& + 2K_{2}{\Delta_{\bot1}}{\Delta_{\bot2}}{\partial_{i}}{\partial_{j}}{\hat{e}^{2}_{\beta}}{\hat{x}_{j}}{\hat{y}_{i}} \notag  \\ 
& + 2K_{3}{\Delta_{\bot1}^{2}}{\partial_{i}}{\partial_{j}}{\hat{e}^{1}_{\beta}}{\hat{x}_{j}}{\hat{x}_{i}} + 2K_{3}{\Delta_{P}}{\Delta_{\bot1}}{\partial_{i}}{\partial_{j}}{\hat{d}_{\beta}}{\hat{z}_{j}}{\hat{x}_{i}} + 2K_{3}{\Delta_{\bot1}}{\Delta_{\bot2}}{\partial_{i}}{\partial_{j}}{\hat{e}^{2}_{\beta}}{\hat{y}_{j}}{\hat{x}_{i}}), \notag
\end{align}
\begin{align}
\frac{\delta{F_{hydrodynamics}}}{\delta{{\hat{e}^{2}}_{\beta}}}(\mathbf{r}') = & \frac{\partial{f_{soc}}}{\partial{{\hat{e}^{2}}_{\beta}}}(\mathbf{r}') - {\partial_{i}}\frac{f_{grad}}{{\partial}{\partial_{i}}{\hat{e}^{2}_{\beta}}}(\mathbf{r}') \notag \\
= & \frac{3g_{D}}{5} ({\Delta_{P}}{\Delta_{\bot2}}{\hat{d}_{\gamma}}{\hat{z}_{\gamma}}{\hat{y}_{\beta}} + {\Delta_{P}}{\Delta_{\bot2}}{\hat{e}^{1}_{\gamma}}{\hat{x}_{\gamma}}{\hat{y}_{\gamma}} + {\Delta_{\bot2}}{\Delta_{P}}{\hat{y}_{\beta}}{\hat{d}_{\gamma}}{\hat{z}_{\gamma}} \notag \\
& + {\Delta_{\bot2}}{\Delta_{\bot1}}{\hat{e}^{1}_{\gamma}}{\hat{x}_{\gamma}}{\hat{y}_{\beta}} + 2{\Delta_{\bot2}^{2}}{\hat{e}^{2}_{\gamma}}{\hat{y}_{\gamma}}{\hat{y}_{\beta}} + {\Delta_{P}}{\Delta_{\bot2}}{\hat{d}_{\gamma}}{\hat{y}_{\gamma}}{\hat{z}_{\beta}} \notag \\
& + {\Delta_{\bot1}}{\Delta_{\bot2}}{\hat{e}^{1}_{\gamma}}{\hat{y}_{\gamma}}{\hat{x}_{\beta}} + {\Delta_{\bot2}}{\Delta_{P}}{\hat{d}_{\gamma}}{\hat{y}_{\gamma}}{\hat{z}_{\beta}} + {\Delta_{\bot1}}{\Delta_{\bot2}}{\hat{e}^{1}_{\gamma}}{\hat{y}_{\gamma}}{\hat{x}_{\beta}} + {\Delta^{2}_{\bot2}}{2\hat{e}^{2}_{\gamma}}{\hat{y}_{\gamma}}{\hat{y}_{\beta}}) \label{FunctionalDerivaticveOfe2}  \\
& - \frac{1}{2}(2K_{1}{\Delta^{2}_{\bot2}}{\partial_{i}}{\partial_{i}}{\hat{e}^{2}_{\beta}}{\hat{y}_{j}}{\hat{y}_{j}} + 2K_{2}{\Delta^{2}_{\bot2}}{\partial_{i}}{\partial_{j}}{\hat{e}^{2}_{\beta}}{\hat{y}_{j}}{\hat{y}_{i}} + 2K_{2}{\Delta_{P}}{\Delta_{\bot2}}{\partial_{i}}{\partial_{j}}{\hat{d}_{\beta}}{\hat{y}_{j}}{\hat{z}_{i}} \notag \\
& + 2K_{2}{\Delta_{\bot1}}{\Delta_{\bot2}}{\partial_{i}}{\partial_{j}}{\hat{e}^{1}_{\beta}}{\hat{y}_{j}}{\hat{x}_{i}}  \notag \\
& + 2K_{3}{\Delta_{\bot2}^{2}}{\partial_{i}}{\partial_{j}}{\hat{e}^{2}_{\beta}}{\hat{y}_{j}}{\hat{y}_{i}} + 2K_{3}{\Delta_{P}}{\Delta_{\bot1}}{\partial_{i}}{\partial_{j}}{\hat{d}_{\beta}}{\hat{z}_{j}}{\hat{y}_{i}} \notag \\
& + 2K_{3}{\Delta_{\bot1}}{\Delta_{\bot2}}{\partial_{i}}{\partial_{j}}{\hat{e}^{1}_{\beta}}{\hat{x}_{j}}{\hat{y}_{i}}). \notag
\end{align}
\end{widetext}
Plugging Eq.~(\ref{FunctionalDerivaticveOfS}), Eq.~(\ref{FunctionalDerivaticveOfd}), Eq.~(\ref{FunctionalDerivaticveOfe1}) and Eq.~(\ref{FunctionalDerivaticveOfe2}) into Eq.~(\ref{LiouvilleEquations1}) and Eq.~(\ref{LiouvilleEquations2}), we get Eq.~ (\ref{1stOderEquationsA}) and Eq.~(\ref{1stOderEquationsB}).
\subsection{\label{For2stOrderEqations} The derivation of the second order dynamic response equation of spin densities}
Firstly we take time-derivative to Eq.~(\ref{1stOderEquationsA}) and get
\begin{align}
{\gamma}{\epsilon_{{\alpha}{\beta}{\gamma}}} & ({H^{(0)}_{\beta}}\frac{\partial}{{\partial}t}{\delta{S_{\gamma}}} + \frac{\partial}{\partial{t}}{{\delta}H_{\beta}}{{S^{(0)}_{\gamma}}}) = \notag \\ 
& {\epsilon_{{\alpha}{\beta}{\gamma}}}[-\frac{6g_{D}}{5}{Q^{bd}_{{\beta}j}} ({V^{d(0)}_{j}}\frac{\partial}{\partial{t}}{\delta{V^{b}_{\gamma}}} + \frac{\partial}{\partial{t}}{\delta}{V^{d}_{j}}{V^{b(0)}_{r}}) \notag \\ 
& + {K^{ba}_{ij}} ({\partial_{i}}{\partial_{j}}{V^{b(0)}_{\beta}}{\frac{\partial}{\partial{t}}}{\delta}{V^{a}_{\gamma}} + {\partial_{i}}{\partial_{j}}{\frac{\partial}{\partial{t}}}{\delta{V_{\beta}^{b}}}{V^{a(0)}_{\gamma}})],  \label{TimeDerivativeOf1stOrderEqOfS}
\end{align} 
where
\begin{align}
\frac{\partial}{\partial{t}}{\delta}{V_{\alpha}^{a}} = & \{{H_{\beta}^{(0)}}{V_{\gamma}^{a(0)}}{\gamma} + {\gamma}{H_{\beta}^{(0)}}{\delta}{V_{\gamma}^{a}} + {\gamma}{\delta}{H_{\beta}}{V_{\gamma}^{a(0)}} \notag \\
& - {\gamma^{2}}{\chi^{-1}_{\bot}}{\delta}({S_{\eta}^{(0)}}{V_{\eta}^{3(0)}}) ({V_{\beta}^{3(0)}}{\delta}{V_{\gamma}^{a}} + {V_{\gamma}^{3(0)}}{\delta}{V_{\beta}^{3}}) \notag \\ 
& + {\gamma^{2}}{\chi_{\bot}^{-1}}(S_{\beta}^{(0)}{V_{\gamma}^{a(0)}} + S_{\beta}^{(0)}{\delta}{V_{\gamma}^{a}} + {\delta}{S_{\beta}}{V_{\gamma}^{a(0)}})\}{\epsilon_{{\alpha}{\beta}{\gamma}}}.
\label{DerivativeOfV}
\end{align}
Taking into account the relations:
\begin{equation}
{\gamma}{S_{\beta}^{(0)}}={H_{\alpha}^{(0)}}{\chi_{{\alpha}{\beta}}},\,\,{V^{3(0)}_{\eta}}{S^{(0)}_{\eta}} = {\hat{\mathbf{d}}^{(0)}}{\cdot}{\mathbf{S}^{(0)}} = 0,
\end{equation}
where magnetic susceptibility $\chi_{{\alpha}{\beta}} = {\chi_{\parallel}}{\delta_{\alpha\beta}} - (\chi_{\bot}-\chi_{\bot}){\hat{d}_{\alpha}^{(0)}}{\hat{d}_{\beta}^{(0)}}$,
Eq.~(\ref{DerivativeOfV}) is simplified to
\begin{equation}
\frac{\partial}{\partial{t}}{\delta}{V_{\alpha}^{a}} = {\epsilon_{{\alpha}{\beta}{\gamma}}}{V_{\gamma}^{a(0)}}({\gamma}{\delta}{H_{\beta}^{a(0)}} - \gamma^{2}{\chi^{-1}_{\bot}}{\delta}{S_{\beta}})
\label{DerivativeOfVSimplifed}
\end{equation}
Taking Eq.~(\ref{DerivativeOfVSimplifed}) back into Eq.~(\ref{TimeDerivativeOf1stOrderEqOfS}), we get
\begin{align}
& \frac{\partial^{2}} {\partial{t^{2}}} {\delta}{S_{\alpha}} \notag \\ 
& = {\gamma}{\epsilon_{\alpha\beta\gamma}}(H^{(0)}_{\beta} \frac{\partial}{\partial{t}}{\delta}{S_{\gamma}} + \frac{\partial}{\partial{t}}{\delta}{H_{\beta}}S_{\gamma}^{(0)}) + {\Xi_{{\alpha}{\lambda}}}{\delta}{S_{\lambda}} + {C_{{\alpha}{\eta}}}{\delta}{H_{{\eta}}}. \label{BeforeFourier}
\end{align}
The last step is taking time Fourier transformation for dynamic variables $\delta{S_{\alpha}}$ and $\delta{H_{\beta}}$ as well their derivatives
\begin{align}
{\delta}H_{\beta}(\omega) = & \frac{1}{\sqrt{2\pi}}\int dt {\delta}H_{\beta} e^{-i{\omega}t}, \notag \\
{\delta}S_{\alpha}(\omega) = & \frac{1}{\sqrt{2\pi}}\int dt {\delta}S_{\alpha} e^{-i{\omega}t}, \notag \\
\mathfrak{F}(\frac{\partial}{\partial{t}} & {\delta{H_{\beta}}}) = i\omega {\delta}H_{\beta}(\omega), \notag \\ 
\mathfrak{F}(\frac{\partial}{\partial{t}} & {\delta{S_{\alpha}}}) = i\omega {\delta}S_{\alpha}(\omega), \notag \\
\mathfrak{F}(\frac{\partial^{2}}{\partial{t^{2}}} & {\delta{S_{\alpha}}}) = -\omega^{2} {\delta}S_{\alpha}(\omega)
\end{align}
in Eq.~(\ref{BeforeFourier}), this gives out Eq.~(\ref{2ndOderEquations}).

\subsection{\label{TransverseNMRResponseEquation} The derivation of transverse NMR response equation of $\delta{\mathbf{S}_{+}}$}
In the limit of $|\omega-\omega_{L}| \ll \omega_{L}$ and under parametrization Eq.~(\ref{PARA1}), Eq.~(\ref{2ndOderEquations}) within components form are
\begin{widetext}
\begin{align}
i\omega{\delta}{S_{2}}(\omega) & = - \gamma{S_{3}^{(0)}}{\delta}{H_{1}}(\omega), \notag \\
i\omega{\delta}{S_{1}}(\omega) & = \gamma{H_{2}^{(0)}}{\delta}{S_{3}}(\omega) + \frac{\Xi_{11}}{i\omega}{\delta}{S_{1}}(\omega) 
 + \frac{\Xi_{13}}{i\omega}{\delta}{S_{3}}(\omega) + \frac{C_{31}}{i\omega}{\delta}{H_{1}}(\omega),  \label{ResponseEq1} \\ 
i\omega{\delta}{S_{3}}(\omega) & =  \gamma[S_{2}^{(0)}{\delta}{H_{1}}(\omega)- {H_{2}^{(0)}}{\delta}{S_{1}}(\omega)]
 + \frac{\Xi_{31}}{i\omega}{\delta}{S_{1}}(\omega) + \frac{\Xi_{33}}{i\omega}{\delta}{S_{3}}(\omega) + \frac{C_{31}}{i\omega}{\delta}{H_{1}}(\omega). \notag
\end{align} 
We expand $\omega$ around $\omega_{L}$ as $\omega = \omega_{L} + \epsilon + O^{2}(\epsilon)$, then we get
\begin{equation}
{\delta}S_{1}(\omega) = \frac{{\delta}S_{3}(\omega)}{i(1+\epsilon)}, \,\, {\delta}S_{3}(\omega) = \frac{1}{i(1+\epsilon)}(\frac{\chi_{\bot}}{\gamma}{\delta}{H_{1}}(\omega) - {\delta}S_{1}(\omega)),
\label{RelationBetweenS1S3}
\end{equation}
where $\epsilon = (\omega - \omega_{L})$. By multiplying $i\omega$ and utilizing Eq.~(\ref{RelationBetweenS1S3}), Eq.~ (\ref{ResponseEq1}) can be reorganized as 
\begin{align}
(\omega^{2} - \omega^{2}_{L})(\delta{S_{1}}(\omega) + i \delta{S_{3}}(\omega)) = & i[- {\Xi_{31}}(\frac{{\delta}{S_{1}}}{1+2\epsilon} + i \frac{{\delta}{S_{3}}}{1+2\epsilon}) + \Xi_{13}(\frac{\delta{S_{1}}}{1+2\epsilon} + i{\delta}S_{3})] \notag \\
& + \frac{\Xi_{11}}{1+\epsilon}(\delta{S_{1}} + i {\delta}S_{3}) + \frac{\Xi_{33}}{1+\epsilon}(\delta{S_{1}} + i {\delta}S_{3}) \\
& - (C_{11} +i C_{31})\delta{H_{1}} - \frac{\chi_{\bot}}{\gamma}(\frac{\Xi_{33}}{1+\epsilon} + i\frac{\Xi_{13}}{1+2\epsilon} -i\frac{\Xi_{31}}{1+2\epsilon}) \delta{H_{1}}. \notag
\end{align}
\end{widetext}
In the case of $\epsilon \rightarrow 0$, this gives Eq.~(\ref{TransverseResponseEquation}).

\subsection{\label{SimplifyTheNMREigenEquation} All $\Xi_{{\alpha}{\lambda}}$ terms in Eq.~(\ref{NMREigenEquation})}
\begin{widetext}
By utilizing Eq.~(\ref{XiAndC}) and paramentrization Eq.~(\ref{PARA1}), we have
\begin{align}
\Xi_{11} + \Xi_{33} = & [2c_{1}(K_{1}+K_{2}+K_{3})\Delta^{2}_{\bot2} + c_{1}K_{1}(\Delta^{2}_{\bot1} + \Delta^{2}_{P})]{\partial_{y}}{\partial_{y}}{\delta}S_{+} \notag \\
& + [c_{1}(K_{1}+K_{2}+K_{3})\Delta^{2}_{\bot1} + c_{1}K_{1}(2\Delta^{2}_{\bot2} + \Delta^{2}_{P}] {\partial_{x}}{\partial_{x}}{\delta}S_{+}  \\
& + c_{2}(\Delta_{P} + \Delta_{\bot1})[-(\Delta_{\bot1} + \Delta_{P})cos2{\theta} - 5{\Delta_{\bot2}}sin{\theta}) + c_{2}(\Delta^{2}_{P} + \Delta^{2}_{\bot1} + 4\Delta^{2}_{\bot2}), \notag 
\label{Xi11Xi33}
\end{align}
\begin{align}
\Xi_{13} + \Xi_{31} = & -2c_{1}K_{1}{\Delta^{2}_{P}}{\partial_{i}}{\delta}{S_{+}}{\partial_{i}}{\theta} - 2c_{1}{\Delta^{2}_{\bot}}[K_{1}{\partial_{y}}{\delta}{S_{+}}{\partial_{y}}{\theta} + (K_{1}+K_{2}+K_{3}){\partial_{x}}{\delta}S_{+}{\partial_{x}}{\theta}cos2{\theta}] \notag \\
& - c_{1}\{[K_{1}{\Delta^{2}_{P} + (K_{1} + K_{2} + K_{3}){\Delta^{2}_{\bot1}}}]{\partial_{x}}{\partial_{x}}\theta + K_{1}(\Delta^{2}_{P} + \Delta^{2}_{\bot1}){\partial_{y}}{\partial_{y}}\theta\}, 
\end{align}
where 
\begin{equation}
c_{1} =\frac{\gamma^{2}}{\chi_{\bot}},\,\,\,c_{2} = \frac{6g_{D}{\gamma^{2}}}{5\chi_{\bot}}.
\label{c1c2}
\end{equation}
plugging Eq.~(\ref{Xi11Xi33}), Eq.~(\ref{c1c2}) into Eq.~(\ref{NMREigenEquation}) and multiplying $\tilde{\Omega}^{-2}$ on both sides, we get
\begin{align}
\frac{\omega^{2}-\omega^{2}_{L}}{\tilde{\Omega}^{2}}{\delta}S_{+} = & \{ \frac{5}{6g_{D}} [6K_{1}\rho^{2}_{2} + K_{1}(\rho_{1}^{2} +1)] {\partial_{y}}{\partial_{y}} + \frac{5}{6g_{D}} [3K_{1}\rho^{2}_{1} + K_{1}(2\rho_{2}^{2} +1)] {\partial_{x}}{\partial_{x}} \notag \\
& -i\frac{10}{6g_{D}}[(K_{1} + 3{\rho_{1}^{2}}K_{1}cos2{\theta}){\partial_{x}}\theta{\partial_{x}} -K_{1}(1 + \rho_{1}^{2}) {\partial_{y}}\theta{\partial_{y}}]\} {\delta}S_{+} \label{OriginalResponseEq} \\
& -i\frac{5}{6g_{D}}[K_{1}(1+3{\rho_{1}^{2}}){\partial_{x}}{\partial_{x}}\theta + K_{1}(1 + \rho_{1}^{2}){\partial_{y}}{\partial_{y}}\theta]\delta{S_{+}} \notag \\
& + \{(1 + \rho_{1})[-(1 + \rho_{1})cos2\theta - 5 \rho_{2}sin\theta] + (1 + \rho^{2}_{1} + 4\rho^{2}_{2})\}{\delta}{S_{+}}. \notag
\end{align}
To simplify Eq.~(\ref{OriginalResponseEq}), we need the Lagrangian equation of $\theta$
\begin{align}
\frac{{\delta}{F_{London}}(\theta)}{\delta{\theta}} = & c_{1}\{[K_{1}{\Delta^{2}_{P}} + (K_{1}+K_{2}+K_{3}){\Delta^{2}_{\bot1}}]{\partial_{x}}{\partial_{x}}{\theta} + K_{1}({\Delta^{2}_{P}} + {\Delta^{2}_{\bot1}}){\partial_{y}}{\partial_{y}}{\theta}\} \notag \\ & - c_{2}\{{\Delta_{\bot2}}(\Delta_{P} + \Delta_{\bot1})cos{\theta} + ({\Delta_{P} + {\Delta_{\bot1}}})^{2}sin2{\theta}\} = 0. 
\label{LagrangianEqOftheta}
\end{align}
This equation can be simplified to 
\begin{equation}
(1+3\rho_{1}^{2}){\partial_{x}}{\partial_{x}}\theta + (1+\rho_{1}^{2}){\partial_{y}}{\partial_{y}}\theta = \frac{1}{\xi_{D}^{2}}[(1 + \rho_{1})^{2}sin2{\theta} - (1 + \rho_{1})\rho_{2}cos{\theta}].
\end{equation}
Then Eq.~(\ref{OriginalResponseEq}) can be written as 
\begin{align}
\frac{\omega^{2}-\omega^{2}_{L}}{\tilde{\Omega}^{2}}{\delta}S_{+} = & \{ \frac{5}{6g_{D}} [6K_{1}\rho^{2}_{2} + K_{1}(\rho_{1}^{2} +1)] {\partial_{y}}{\partial_{y}} + \frac{5}{6g_{D}} [3K_{1}\rho^{2}_{1} + K_{1}(2\rho_{2}^{2} +1)] {\partial_{x}}{\partial_{x}} \notag \\
& -i\frac{10}{6g_{D}}[(K_{1} + 3{\rho_{1}^{2}}K_{1}cos2{\theta}){\partial_{x}}\theta{\partial_{x}} -K_{1}(1 + \rho_{1}^{2}) {\partial_{y}}\theta{\partial_{y}}]\} {\delta}S_{+} \label{ReducedlResponseEqAppendices} \\
& -i\frac{5K_{1}}{6g_{D}}{\xi^{-2}_{D}}[(1 + \rho_{1})^{2}sin2{\theta} - (1 + \rho_{1})\rho_{2}cos{\theta}]\delta{S_{+}} \notag \\
& + \{(1 + \rho_{1})[-(1 + \rho_{1})cos2\theta - 5 \rho_{2}sin\theta] + (1 + \rho^{2}_{1} + 4\rho^{2}_{2})\}{\delta}{S_{+}}. \notag
\end{align}
This is Eq.~(\ref{NMREigenEquationDimensonless}).
\end{widetext}
\begin{figure*}
\centerline{\includegraphics[width=0.5\linewidth]{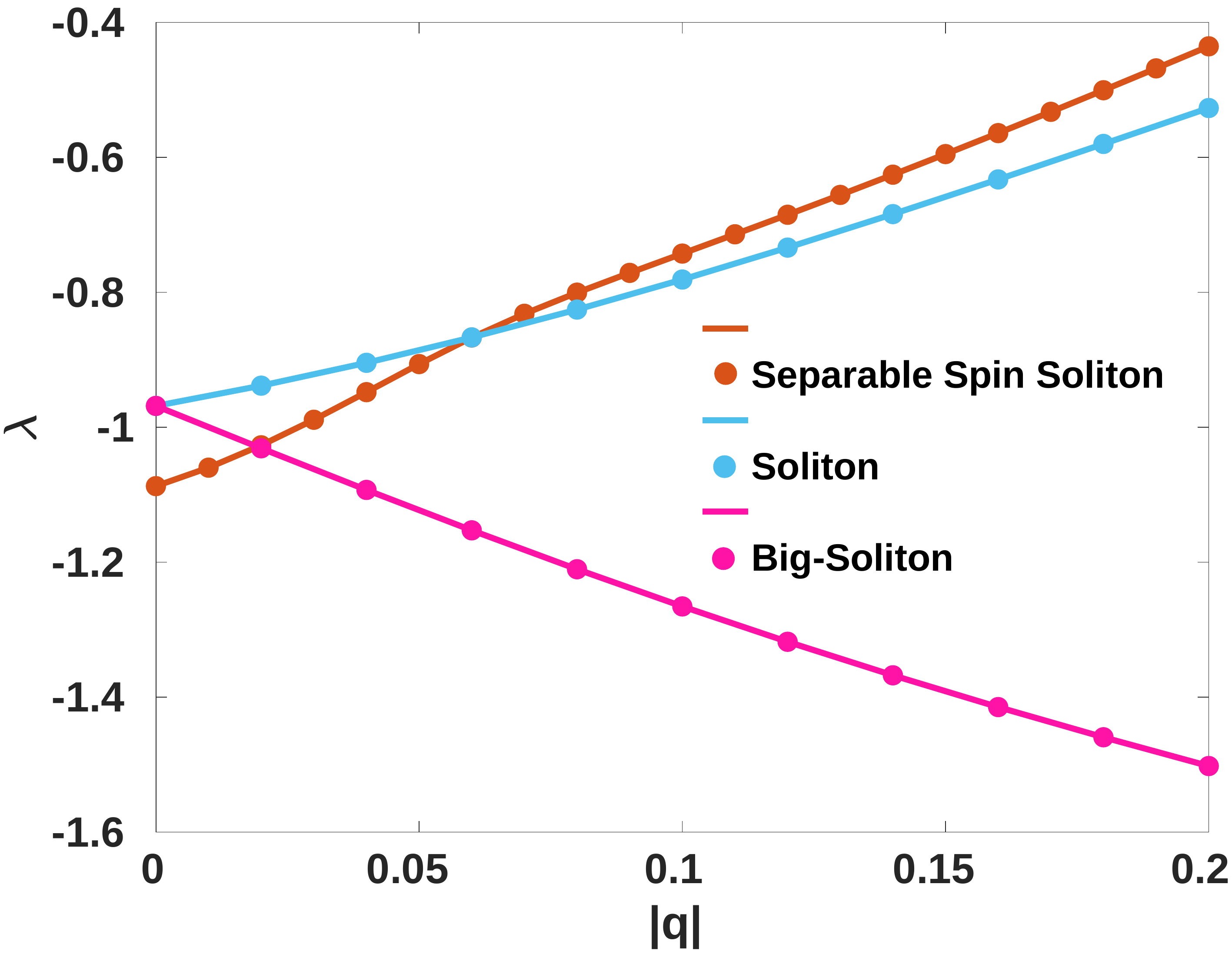}}
\caption{NMR frequency shifts of unit cell of pseudo-random lattices consisting of separable spin soliton, NMR frequency shifts of soliton ($|\Delta{\theta}|=\pi-2\theta_{0}$) and big-soliton ($|\Delta{\theta}|=\pi+2\theta_{0}$). All colored dots are original numeric data, while colored lines are the linear interpolations of numeric data. $D=14.1\xi_{D}$. We found the NMR frequency shifts $\lambda$ of soliton ($|\Delta{\theta}|=\pi-2\theta_{0}$) monotonically  increase when $|q|$ increases. The typical values of $\lambda$ of soliton (blue line) are larger than $-0.7$ when $|q|\geq 0.14$. In contrast, the NMR frequency shifts of big-soliton (pink line) monotonically decrease when $|q|$ increases and the typical values of $\lambda$ in this case are smaller than $-1.35$ when $|q|\geq0.14$. On the other side, the frequency shifts induced by pseudo-random lattices of $2/4$ separable spin solitons (brown lines) monotonically increase as $|q|$ increase. The typical values of $\lambda$ are larger than $-0.65$ when $|q|>0.14$. We find this range of $\lambda$ are very close to those induced by soliton ($|\Delta{\theta}|=\pi-2\theta_{0}$). This is because only the soliton ($|\Delta{\theta}|=\pi-2\theta_{0}$) of $2/4$ separable spin soliton responds to the continuous wave transverse magnetic drive. Moreover, we can see that the frequency shifts of pseudo-random lattices consisting of $2/4$ separable spin solitons is smaller than $-1$ when $|q|=0$. This deviates from the experimental observations of the frequency shifts of spin soliton in polar phase with $|q|=0$ \cite{Autti2016}.  \label{FrequencyShiftSolitonBigSolitonSeparableSoliton}}
\end{figure*}
\section{\label{SpinDynamicsSolionAndBigSoliton}NMR frequency shifts Of soliton ($|\Delta{\theta}|=\pi-2\theta_{0}$) and big-soliton ($|\Delta{\theta}|=\pi+2\theta_{0}$)}
Here we discuss the transverse NMR frequency shifts of soliton ($|\Delta{\theta}|=\pi-2\theta_{0}$) and big-soliton ($|\Delta{\theta}|=\pi+2\theta_{0}$) in the absence of KLS string wall. The frequency shifts $\lambda$ are numeric results of Eq.~ (\ref{NMREigenEquationDimensonless}) with equilibrium spin textures of soliton ($|\Delta{\theta}|=\pi-2\theta_{0}$) and big-soliton ($|\Delta{\theta}|=\pi+2\theta_{0}$) which we got in Sec. \ref{AbsenceOfHQVs}. In Fig.~\ref{FrequencyShiftSolitonBigSolitonSeparableSoliton} we depict the results with $|q|$ from $0.0$ to $0.2$. We found the transverse NMR frequency shift of soliton ($|\Delta{\theta}|=\pi-2\theta_{0}$) is increasing function of $|q|$ while the transverse NMR frequency shift of big-soliton ($|\Delta{\theta}|=\pi+2\theta_{0}$) is decreasing function of $|q|$. When $|q| \geq 0.14$, the typical values of $\lambda$ of soliton and big-soliton are $\lambda \geq -0.7$ and $\lambda \leq -1.3$ respectively. Because the unit cell of pseudo-random lattices of separable spin solitons with topological invariant $1/4+1/4$ contains KLS-soliton and soliton, the transverse NMR frequency shift of unit cell is determined by the equilibrium spin texture of soliton. As are result, $\lambda$ of pseudo-random lattices consisting of separable spin soliton with topological invariant $1/4+1/4$ is very close to those induced by soliton ($|\Delta{\theta}|=\pi-2\theta_{0}$).

\end{document}